\documentclass[twocolumn,appendixfloats]{aastex61}

\usepackage{natbib}

\newcommand{\nineyr}{NG9}

\begin{document}

\title{The NANOGrav 11-year Data Set: High-precision timing of 45 Millisecond Pulsars}
\shorttitle{NANOGrav 11-year Data Set}
\shortauthors{Z.~Arzoumanian et al.}

\author{Zaven Arzoumanian}
\affiliation{Center for Research and Exploration in Space Science and Technology and X-Ray Astrophysics Laboratory, \\NASA Goddard Space Flight Center, Code 662, Greenbelt, MD 20771, USA}
\author{Adam Brazier}
\affiliation{Department of Astronomy, Cornell University, Ithaca, NY 14853, USA}
\author{Sarah Burke-Spolaor}
\affiliation{Department of Physics and Astronomy, West Virginia University, P.O. Box 6315, Morgantown, WV 26506, USA}
\affiliation{Center for Gravitational Waves and Cosmology, West Virginia University, Chestnut Ridge Research Building, Morgantown, WV 26505, USA}
\author{Sydney Chamberlin}
\affiliation{Department of Astronomy and Astrophysics, Pennsylvania State University, University Park, PA 16802, USA}
\author[0000-0002-2878-1502]{Shami Chatterjee}
\affiliation{Department of Astronomy, Cornell University, Ithaca, NY 14853, USA}
\author{Brian Christy}
\affiliation{Department of Mathematics, Computer Science, and Physics, Notre Dame of Maryland University 4701 N Charles Street, Baltimore, MD 21210, USA}
\author{James M. Cordes}
\affiliation{Department of Astronomy, Cornell University, Ithaca, NY 14853, USA}
\author{Neil J. Cornish}
\affiliation{Department of Physics, Montana State University, Bozeman, MT 59717, USA}
\author[0000-0002-2578-0360]{Fronefield Crawford}
\affiliation{Department of Physics and Astronomy, Franklin \& Marshall College, P.O. Box 3003, Lancaster, PA 17604, USA}
\author{H. Thankful Cromartie}
\affiliation{University of Virginia, Department of Astronomy, P.O. Box 400325, Charlottesville, VA 22904, USA}
\author{Kathryn Crowter}
\affiliation{Department of Physics and Astronomy, University of British Columbia, 6224 Agricultural Road, Vancouver, BC V6T 1Z1, Canada}
\author{Megan E. DeCesar}
\altaffiliation{NANOGrav Physics Frontiers Center Postdoctoral Fellow}
\affiliation{Department of Physics, Lafayette College, Easton, PA 18042, USA}
\author{Paul B. Demorest}
\affiliation{National Radio Astronomy Observatory, 1003 Lopezville Road, Socorro, NM 87801, USA}
\author[0000-0001-8885-6388]{Timothy Dolch}
\affiliation{Department of Physics, Hillsdale College, 33 E. College Street, Hillsdale, Michigan 49242, USA}
\author{Justin A. Ellis}
\altaffiliation{NANOGrav Physics Frontiers Center Postdoctoral Fellow}
\affiliation{Department of Physics and Astronomy, West Virginia University, P.O. Box 6315, Morgantown, WV 26506, USA}
\affiliation{Center for Gravitational Waves and Cosmology, West Virginia University, Chestnut Ridge Research Building, Morgantown, WV 26505, USA}
\author{Robert D. Ferdman}
\affiliation{School of Chemistry, University of East Anglia, Norwich, NR4 7TJ, United Kingdom}
\author{Elizabeth C. Ferrara}
\affiliation{NASA Goddard Space Flight Center, Greenbelt, MD 20771, USA}
\author[0000-0001-8384-5049]{Emmanuel Fonseca}
\affiliation{Department of Physics, McGill University, 3600  University Street, Montreal, QC H3A 2T8, Canada}
\author{Nathan Garver-Daniels}
\affiliation{Department of Physics and Astronomy, West Virginia University, P.O. Box 6315, Morgantown, WV 26506, USA}
\affiliation{Center for Gravitational Waves and Cosmology, West Virginia University, Chestnut Ridge Research Building, Morgantown, WV 26505, USA}
\author{Peter A. Gentile}
\affiliation{Department of Physics and Astronomy, West Virginia University, P.O. Box 6315, Morgantown, WV 26506, USA}
\affiliation{Center for Gravitational Waves and Cosmology, West Virginia University, Chestnut Ridge Research Building, Morgantown, WV 26505, USA}
\author{Daniel Halmrast}
\affiliation{Department of Physics, Hillsdale College, 33 E. College Street, Hillsdale, Michigan 49242, USA}
\affiliation{Department of Mathematics, University of California, Santa Barbara, CA 93106, USA}
\author{E. A. Huerta}
\affiliation{NCSA and Department of Astronomy, University of Illinois at Urbana-Champaign, Urbana, IL 61801, USA}
\author{Fredrick A. Jenet}
\affiliation{Center for Gravitational Wave Astronomy, University of Texas-Rio Grande Valley, Brownsville, TX 78520, USA}
\author{Cody Jessup}
\affiliation{Department of Physics, Hillsdale College, 33 E. College Street, Hillsdale, Michigan 49242, USA}
\author{Glenn Jones}
\altaffiliation{NANOGrav Physics Frontiers Center Postdoctoral Fellow}
\affiliation{Department of Physics, Columbia University, New York, NY 10027, USA}
\author{Megan L. Jones}
\affiliation{Department of Physics and Astronomy, West Virginia University, P.O. Box 6315, Morgantown, WV 26506, USA}
\affiliation{Center for Gravitational Waves and Cosmology, West Virginia University, Chestnut Ridge Research Building, Morgantown, WV 26505, USA}
\author[0000-0001-6295-2881]{David L. Kaplan}
\affiliation{Center for Gravitation, Cosmology and Astrophysics, Department of Physics, University of Wisconsin-Milwaukee,\\ P.O. Box 413, Milwaukee, WI 53201, USA}
\author[0000-0003-0721-651X]{Michael T. Lam}
\altaffiliation{NANOGrav Physics Frontiers Center Postdoctoral Fellow}
\affiliation{Department of Physics and Astronomy, West Virginia University, P.O. Box 6315, Morgantown, WV 26506, USA}
\affiliation{Center for Gravitational Waves and Cosmology, West Virginia University, Chestnut Ridge Research Building, Morgantown, WV 26505, USA}
\author{T. Joseph W. Lazio}
\affiliation{Jet Propulsion Laboratory, California Institute of Technology, 4800 Oak Grove Drive, Pasadena, CA 91109, USA}
\author{Lina Levin}
\affiliation{Department of Physics and Astronomy, West Virginia University, P.O. Box 6315, Morgantown, WV 26506, USA}
\affiliation{Center for Gravitational Waves and Cosmology, West Virginia University, Chestnut Ridge Research Building, Morgantown, WV 26505, USA}
\author{Andrea Lommen}
\affiliation{Department of Physics and Astronomy, Haverford College, Haverford, PA, 19041. USA}
\author{Duncan R. Lorimer}
\affiliation{Department of Physics and Astronomy, West Virginia University, P.O. Box 6315, Morgantown, WV 26506, USA}
\affiliation{Center for Gravitational Waves and Cosmology, West Virginia University, Chestnut Ridge Research Building, Morgantown, WV 26505, USA}
\author{Jing Luo}
\affiliation{Center for Gravitational Wave Astronomy, University of Texas-Rio Grande Valley, Brownsville, TX 78520, USA}
\author{Ryan S. Lynch}
\affiliation{Green Bank Observatory, P.O. Box 2, Green Bank, WV 24944, USA}
\author{Dustin Madison}
\affiliation{National Radio Astronomy Observatory, 520 Edgemont Road, Charlottesville, VA 22903, USA}
\author{Allison M. Matthews}
\affiliation{University of Virginia, Department of Astronomy, P.O. Box 400325, Charlottesville, VA 22904, USA}
\author[0000-0001-7697-7422]{Maura A. McLaughlin}
\affiliation{Department of Physics and Astronomy, West Virginia University, P.O. Box 6315, Morgantown, WV 26506, USA}
\affiliation{Center for Gravitational Waves and Cosmology, West Virginia University, Chestnut Ridge Research Building, Morgantown, WV 26505, USA}
\author{Sean T. McWilliams}
\affiliation{Department of Physics and Astronomy, West Virginia University, P.O. Box 6315, Morgantown, WV 26506, USA}
\affiliation{Center for Gravitational Waves and Cosmology, West Virginia University, Chestnut Ridge Research Building, Morgantown, WV 26505, USA}
\author[0000-0002-4307-1322]{Chiara Mingarelli}
\affiliation{Center for Computational Astrophysics, Flatiron Institute, 162 5th Avenue, New York, New York, 10010, USA}
\author[0000-0002-3616-5160]{Cherry Ng}
\affiliation{Department of Physics and Astronomy, University of British Columbia, 6224 Agricultural Road, Vancouver, BC V6T 1Z1, Canada}
\affiliation{Dunlap Institute for Astronomy and Astrophysics, University of Toronto, 50 St. George Street, Toronto, ON M5S 3H4, Canada}
\author[0000-0002-6709-2566]{David J. Nice}
\affiliation{Department of Physics, Lafayette College, Easton, PA 18042, USA}
\author[0000-0001-5465-2889]{Timothy T. Pennucci}
\altaffiliation{NANOGrav Physics Frontiers Center Postdoctoral Fellow}
\affiliation{Department of Physics and Astronomy, West Virginia University, P.O. Box 6315, Morgantown, WV 26506, USA}
\affiliation{Center for Gravitational Waves and Cosmology, West Virginia University, Chestnut Ridge Research Building, Morgantown, WV 26505, USA}
\affiliation{Institute of Physics, E\"{o}tv\"{o}s Lor\'{a}nd University, P\'{a}zm\'{a}ny P. s. 1/A, 1117 Budapest, Hungary}
\affiliation{Hungarian Academy of Sciences MTA-ELTE Extragalatic Astrophysics Research Group, 1117 Budapest, Hungary}
\author[0000-0001-5799-9714]{Scott M. Ransom}
\affiliation{National Radio Astronomy Observatory, 520 Edgemont Road, Charlottesville, VA 22903, USA}
\author{Paul S. Ray}
\affiliation{Space Science Division, Naval Research Laboratory, Washington, DC 20375-5352, USA}
\author{Xavier Siemens}
\affiliation{Center for Gravitation, Cosmology and Astrophysics, Department of Physics, University of Wisconsin-Milwaukee,\\ P.O. Box 413, Milwaukee, WI 53201, USA}
\author[0000-0003-1407-6607]{Joseph Simon}
\affiliation{Jet Propulsion Laboratory, California Institute of Technology, 4800 Oak Grove Drive, Pasadena, CA 91109, USA}
\author[0000-0002-6730-3298]{Ren\'{e}e Spiewak}
\affiliation{Center for Gravitation, Cosmology and Astrophysics, Department of Physics, University of Wisconsin-Milwaukee,\\ P.O. Box 413, Milwaukee, WI 53201, USA}
\affiliation{Centre for Astrophysics and Supercomputing, Swinburne University of Technology, P.O. Box 218, Hawthorn, Victoria 3122, Australia}
\author[0000-0001-9784-8670]{Ingrid H. Stairs}
\affiliation{Department of Physics and Astronomy, University of British Columbia, 6224 Agricultural Road, Vancouver, BC V6T 1Z1, Canada}
\author{Daniel R. Stinebring}
\affiliation{Department of Physics and Astronomy, Oberlin College, Oberlin, OH 44074, USA}
\author{Kevin Stovall}
\altaffiliation{NANOGrav Physics Frontiers Center Postdoctoral Fellow}
\affiliation{National Radio Astronomy Observatory, 1003 Lopezville Road, Socorro, NM 87801, USA}
\author{Joseph K. Swiggum}
\altaffiliation{NANOGrav Physics Frontiers Center Postdoctoral Fellow}
\affiliation{Center for Gravitation, Cosmology and Astrophysics, Department of Physics, University of Wisconsin-Milwaukee,\\ P.O. Box 413, Milwaukee, WI 53201, USA}
\author{Stephen R. Taylor}
\altaffiliation{NANOGrav Physics Frontiers Center Postdoctoral Fellow}
\affiliation{Jet Propulsion Laboratory, California Institute of Technology, 4800 Oak Grove Drive, Pasadena, CA 91109, USA}
\author[0000-0002-4162-0033]{Michele Vallisneri}
\affiliation{Jet Propulsion Laboratory, California Institute of Technology, 4800 Oak Grove Drive, Pasadena, CA 91109, USA}
\author{Rutger van~Haasteren}
\altaffiliation{Currently employed at Microsoft Corporation}
\affiliation{Jet Propulsion Laboratory, California Institute of Technology, 4800 Oak Grove Drive, Pasadena, CA 91109, USA}
\author{Sarah J. Vigeland}
\altaffiliation{NANOGrav Physics Frontiers Center Postdoctoral Fellow}
\affiliation{Center for Gravitation, Cosmology and Astrophysics, Department of Physics, University of Wisconsin-Milwaukee,\\ P.O. Box 413, Milwaukee, WI 53201, USA}
\author{Weiwei Zhu}
\affiliation{National Astronomical Observatories, Chinese Academy of Science, 20A Datun Road, Chaoyang District, Beijing 100012, China}
\affiliation{Max Planck Institute for Radio Astronomy, Auf dem H\"{u}gel 69, D-53121 Bonn, Germany}
 
\collaboration{The NANOGrav Collaboration}
\noaffiliation

\correspondingauthor{David J.~Nice}
\email{niced@lafayette.edu}

\begin{abstract}

We present high-precision timing data over timespans of up to 11 years for 45 millisecond pulsars observed as part of the North American Nanohertz Observatory for Gravitational Waves (NANOGrav) project, aimed at detecting and characterizing low-frequency gravitational waves. The pulsars were observed with the Arecibo Observatory and/or the Green Bank Telescope at frequencies ranging from 327 MHz to 2.3 GHz. Most pulsars were observed with approximately monthly cadence, and six high-timing-precision pulsars were observed weekly.  All were observed at widely separated frequencies at each observing epoch in order to fit for time-variable dispersion delays. We describe our methods for data processing, time-of-arrival (TOA) calculation, and the implementation of a new, automated method for removing outlier TOAs. We fit a timing model for each pulsar which includes spin, astrometric, and (for binary pulsars) orbital parameters; time-variable dispersion delays; and parameters that quantify pulse-profile evolution with frequency. The timing solutions provide three new parallax measurements, two new Shapiro delay measurements, and two new measurements of significant orbital-period variations. We fit models that characterize sources of noise for each pulsar. We find that 11 pulsars show significant red noise, with generally smaller spectral indices than typically measured for non-recycled pulsars, possibly suggesting a different origin.  A companion paper uses these data to constrain the strength of the gravitational-wave background.

\end{abstract}


\section{Introduction}
\label{sec:intro}
High-precision timing of 
millisecond pulsars offers the promise of detecting gravitational waves
with periods of a few years, i.e., 
in the nanohertz (nHz) band of the gravitational-wave spectrum
\citep{BurkeSpolaor2015,Lommen2015}.
An expected signal in this band is
the incoherent superposition of gravitational waves from the cosmic merger
history of supermassive black hole binaries, i.e., a gravitational-wave
background
\citep{Phinney2001,Jaffe2003,Sesana2013}. Its detection is likely within a few years \citep{Taylor2016} depending on the
underlying astrophysics of supermassive black hole binary mergers 
\citep{Kocsis2011,Roedig2012,Sampson2015,Arzoumanian2016,Taylor2017}.
Other possible sources of gravitational waves in this 
band are individual massive binary systems \citep{Arzoumanian2014,Babak2016},
gravitational bursts with memory \citep[e.g.,][]{Seto2009,Madison2014,Zhu2014,Arzoumanian2015},
primordial gravitational waves
from inflation \citep{Grishchuk1976,Grishchuk1977,Starobinsky1980,Lentati2015,Lasky2016}, and gravitational waves
originating from cosmic strings \citep[e.g.][]{Kibble1976,Vilekin1981,Sanidas2012,Arzoumanian2015,Lentati2015}.

Robust detection of nHz gravitational waves requires observing and measuring pulse arrival time series for an 
ensemble of millisecond pulsars; the gravitational-wave signal is manifested as perturbations in the 
arrival time measurements that are
correlated between pulsars,
depending on their relative positions
\citep{Hellings1983,Cornish2013,Taylor2013,Mingarelli2014}.
For this reason, the North American Nanohertz Observatory for Gravitational Waves (NANOGrav) collaboration\footnote{North American Nanohertz Observatory for Gravitational Waves; \url{http://nanograv.org}} has 
undertaken high-precision timing
observations of a large and growing number of millisecond pulsars 
spread across the sky.  Similar programs are being carried out by the
Parkes Pulsar Timing Array \citep{Hobbs2013,Reardon2016} and the European
Pulsar Timing Array \citep{Kramer2013,Desvignes2016}.

Pulsar-timing experiments at nHz frequencies explore
gravitational waves in a band
entirely distinct from other techniques used to explore the gravitational-wave 
spectrum, and hence they are sensitive to a completely different
class of gravitational-wave sources.
For comparison, gravitational waves have been 
detected directly by the LIGO ground-based interferometers 
in the $\sim$100~Hz band
\citep{Abbott2016a,Abbott2016b,Abbott2017a}, and indirectly via binary-pulsar orbital-decay measurements 
in the $\sim100~\mu$Hz band 
\citep[e.g.,][]{Kramer2006,Fonseca2014,Weisberg2016};
proposed space-based detectors will be sensitive
in the $\sim 10^{-2}$~Hz band \citep{Amaro-Seoane2017}.

In addition to gravitational-wave detection,
high-precision pulsar data can be used for a variety of other applications,
including studies of
binary systems and neutron-star masses \citep{Fonseca2016},
measurements of pulsar astrometry and space velocities \citep{Matthews2016},
tests of general relativity \citep{Zhu2015}, and
analysis of the ionized interstellar medium \citep{Lam2016,Levin2016,Jones2017}.

This paper describes NANOGrav data collected over 11~years, our
``11-year Data Set.'' It builds on our previous paper describing
our Nine-year Data Set \citep[][herein \nineyr]{Arzoumanian2015b}.  
This paper is organized as follows.  In \S\ref{sec:obs}, we describe
the observations and data reduction.  In \S\ref{sec:noise}, we characterize
the noise properties of the pulsars.  In \S\ref{sec:astrometry}, we present 
an astrometric analysis of the pulsars, including distance
estimates.
In \S\ref{sec:binary}, we give updated parameters of
those pulsars in our observations that are in binary systems, including
refined measurement of pulsar and companion-star masses. In \S\ref{sec:conclusion}, we summarize our presentation.  In the Appendix, we present
timing residuals and dispersion measure (DM) variations for all pulsars
under observation.  A search for a gravitational-wave background
in these data is presented in a separate paper \citep{Arzoumanian2018b}.

\section{Observations, Data Reduction, and Timing Models}
\label{sec:obs}
The NANOGrav 11-year data set consists of time-of-arrival (TOA)
measurements of 45 pulsars made over time spans of up to 11 years, along with 
a parameterized model fit to the TOAs of each pulsar.

Here, we describe the instrumentation, observations, and data-reduction
procedures applied to produce this data set.  In general, procedures
closely follow those of \nineyr, so
we provide only a brief overview of details already covered in 
\nineyr, highlighting any changes.

The data were collected from 
2004 through the end of 2015.  For the 37 pulsars with data spans
greater than 2.5 years (see Table~\ref{tab:psr_params}), observations
taken though the end of 2013 were previously reported in \nineyr.
This work adds nine new pulsars to
the set; it removes one pulsar (PSR~J1949+3106, which provided relatively
poor timing precision); 
and it extends the time span of all remaining sources by approximately two years.   Five pulsars in \nineyr\ had lengthy
spans of single-receiver observations at their initial years of observations;
for four of these pulsars (PSRs J1853+1303, J1910+1256, J1944+0907, and B1953+29),
we have removed those observations from the present data set because of their susceptibility
to unmodeled variations in DM (see below).  For the
fifth (PSR J1741+1831), we added observations with a second
receiver at those epochs.

\begin{table*}[tp]
\caption{\label{tab:psr_params} Basic pulsar parameters and TOA
statistics}
\begin{center}
{\scriptsize
\begin{tabular}{c|rrrr|rl|rl|rl|rl|rl|r}
\hline
Source & $P$  & $dP/dt$      & DM           & $P_b$ 
         & \multicolumn{10}{|c|}{Median Scaled TOA Uncertainty$^a$ ($\mu$s) / Number of Epochs} 
         & Span \\
       & (ms) & ($10^{-20}$) & (pc~cm$^{-3}$) & (day)  
         & \multicolumn{2}{|c|}{327~MHz}
         & \multicolumn{2}{|c|}{430~MHz}
         & \multicolumn{2}{|c|}{820~MHz}
         & \multicolumn{2}{|c|}{1.4~GHz}
         & \multicolumn{2}{|c|}{2.3~GHz} 
         & (year) \\
\hline
J0023$+$0923 & 3.05 & 1.14 & 14.3 & -& \multicolumn{2}{|c|}{-} & 0.132 & 42 & \multicolumn{2}{|c|}{-} & 0.153 & 50 & \multicolumn{2}{|c|}{-} & 4.4\\
J0030$+$0451 & 4.87 & 1.02 & 4.3 & -& \multicolumn{2}{|c|}{-} & 0.313 & 104 & \multicolumn{2}{|c|}{-} & 0.319 & 115 & \multicolumn{2}{|c|}{-} & 10.9\\
J0340$+$4130 & 3.30 & 0.70 & 49.6 & -& \multicolumn{2}{|c|}{-} & \multicolumn{2}{|c|}{-} & 0.809 & 53 & 1.796 & 52 & \multicolumn{2}{|c|}{-} & 3.8\\
J0613$-$0200 & 3.06 & 0.96 & 38.8 & 1.2& \multicolumn{2}{|c|}{-} & \multicolumn{2}{|c|}{-} & 0.108 & 119 & 0.433 & 115 & \multicolumn{2}{|c|}{-} & 10.8\\
J0636$+$5128 & 2.87 & 0.34 & 11.1 & 0.1& \multicolumn{2}{|c|}{-} & \multicolumn{2}{|c|}{-} & 0.225 & 24 & 0.466 & 24 & \multicolumn{2}{|c|}{-} & 2.0\\
J0645$+$5158 & 8.85 & 0.49 & 18.2 & -& \multicolumn{2}{|c|}{-} & \multicolumn{2}{|c|}{-} & 0.316 & 55 & 0.926 & 56 & \multicolumn{2}{|c|}{-} & 4.5\\
J0740$+$6620 & 2.89 & 1.22 & 15.0 & 4.8& \multicolumn{2}{|c|}{-} & \multicolumn{2}{|c|}{-} & 0.523 & 22 & 0.570 & 24 & \multicolumn{2}{|c|}{-} & 2.0\\
J0931$-$1902 & 4.64 & 0.36 & 41.5 & -& \multicolumn{2}{|c|}{-} & \multicolumn{2}{|c|}{-} & 0.778 & 36 & 1.559 & 35 & \multicolumn{2}{|c|}{-} & 2.8\\
J1012$+$5307 & 5.26 & 1.71 & 9.0 & 0.6& \multicolumn{2}{|c|}{-} & \multicolumn{2}{|c|}{-} & 0.371 & 119 & 0.518 & 124 & \multicolumn{2}{|c|}{-} & 11.4\\
J1024$-$0719 & 5.16 & 1.86 & 6.5 & -& \multicolumn{2}{|c|}{-} & \multicolumn{2}{|c|}{-} & 0.559 & 77 & 0.836 & 78 & \multicolumn{2}{|c|}{-} & 6.2\\
J1125$+$7819 & 4.20 & 0.70 & 11.2 & 15.4& \multicolumn{2}{|c|}{-} & \multicolumn{2}{|c|}{-} & 0.817 & 21 & 1.267 & 24 & \multicolumn{2}{|c|}{-} & 2.0\\
J1453$+$1902 & 5.79 & 1.17 & 14.1 & -& \multicolumn{2}{|c|}{-} & 1.642 & 21 & \multicolumn{2}{|c|}{-} & 2.261 & 23 & \multicolumn{2}{|c|}{-} & 2.4\\
J1455$-$3330 & 7.99 & 2.43 & 13.6 & 76.2& \multicolumn{2}{|c|}{-} & \multicolumn{2}{|c|}{-} & 0.929 & 105 & 1.724 & 103 & \multicolumn{2}{|c|}{-} & 11.4\\
J1600$-$3053 & 3.60 & 0.95 & 52.3 & 14.3& \multicolumn{2}{|c|}{-} & \multicolumn{2}{|c|}{-} & 0.258 & 94 & 0.201 & 100 & \multicolumn{2}{|c|}{-} & 8.1\\
J1614$-$2230 & 3.15 & 0.96 & 34.5 & 8.7& \multicolumn{2}{|c|}{-} & \multicolumn{2}{|c|}{-} & 0.341 & 79 & 0.446 & 91 & \multicolumn{2}{|c|}{-} & 7.2\\
J1640$+$2224 & 3.16 & 0.28 & 18.5 & 175.5& \multicolumn{2}{|c|}{-} & 0.084 & 119 & \multicolumn{2}{|c|}{-} & 0.095 & 130 & \multicolumn{2}{|c|}{-} & 11.1\\
J1643$-$1224 & 4.62 & 1.85 & 62.3 & 147.0& \multicolumn{2}{|c|}{-} & \multicolumn{2}{|c|}{-} & 0.291 & 118 & 0.483 & 117 & \multicolumn{2}{|c|}{-} & 11.2\\
J1713$+$0747 & 4.57 & 0.85 & 15.9 & 67.8& \multicolumn{2}{|c|}{-} & \multicolumn{2}{|c|}{-} & 0.101 & 117 & 0.051 & 326 & 0.030 & 111 & 10.9\\
J1738$+$0333 & 5.85 & 2.41 & 33.8 & 0.4& \multicolumn{2}{|c|}{-} & \multicolumn{2}{|c|}{-} & \multicolumn{2}{|c|}{-} & 0.385 & 53 & 0.385 & 47 & 6.1\\
J1741$+$1351 & 3.75 & 3.02 & 24.2 & 16.3& \multicolumn{2}{|c|}{-} & 0.200 & 45 & \multicolumn{2}{|c|}{-} & 0.213 & 63 & 0.235 & 9 & 6.4\\
J1744$-$1134 & 4.07 & 0.89 & 3.1 & -& \multicolumn{2}{|c|}{-} & \multicolumn{2}{|c|}{-} & 0.113 & 113 & 0.193 & 111 & \multicolumn{2}{|c|}{-} & 11.4\\
J1747$-$4036 & 1.65 & 1.31 & 153.0 & -& \multicolumn{2}{|c|}{-} & \multicolumn{2}{|c|}{-} & 1.094 & 49 & 1.115 & 51 & \multicolumn{2}{|c|}{-} & 3.8\\
J1832$-$0836 & 2.72 & 0.83 & 28.2 & -& \multicolumn{2}{|c|}{-} & \multicolumn{2}{|c|}{-} & 0.606 & 38 & 0.422 & 35 & \multicolumn{2}{|c|}{-} & 2.8\\
J1853$+$1303 & 4.09 & 0.87 & 30.6 & 115.7& \multicolumn{2}{|c|}{-} & 0.390 & 49 & \multicolumn{2}{|c|}{-} & 0.413 & 55 & \multicolumn{2}{|c|}{-} & 4.5\\
B1855$+$09 & 5.36 & 1.78 & 13.3 & 12.3& \multicolumn{2}{|c|}{-} & 0.159 & 101 & \multicolumn{2}{|c|}{-} & 0.154 & 111 & \multicolumn{2}{|c|}{-} & 11.0\\
J1903$+$0327 & 2.15 & 1.88 & 297.5 & 95.2& \multicolumn{2}{|c|}{-} & \multicolumn{2}{|c|}{-} & \multicolumn{2}{|c|}{-} & 0.501 & 58 & 0.497 & 51 & 6.1\\
J1909$-$3744 & 2.95 & 1.40 & 10.4 & 1.5& \multicolumn{2}{|c|}{-} & \multicolumn{2}{|c|}{-} & 0.041 & 113 & 0.090 & 195 & \multicolumn{2}{|c|}{-} & 11.2\\
J1910$+$1256 & 4.98 & 0.97 & 38.1 & 58.5& \multicolumn{2}{|c|}{-} & \multicolumn{2}{|c|}{-} & \multicolumn{2}{|c|}{-} & 0.301 & 67 & 0.326 & 56 & 6.8\\
J1911$+$1347 & 4.63 & 1.69 & 31.0 & -& \multicolumn{2}{|c|}{-} & 0.136 & 22 & \multicolumn{2}{|c|}{-} & 0.131 & 25 & \multicolumn{2}{|c|}{-} & 2.4\\
J1918$-$0642 & 7.65 & 2.57 & 6.1 & 10.9& \multicolumn{2}{|c|}{-} & \multicolumn{2}{|c|}{-} & 0.328 & 110 & 0.548 & 114 & \multicolumn{2}{|c|}{-} & 11.2\\
J1923$+$2515 & 3.79 & 0.96 & 18.9 & -& \multicolumn{2}{|c|}{-} & 0.514 & 36 & \multicolumn{2}{|c|}{-} & 0.568 & 48 & \multicolumn{2}{|c|}{-} & 4.3\\
B1937$+$21 & 1.56 & 10.51 & 71.1 & -& \multicolumn{2}{|c|}{-} & \multicolumn{2}{|c|}{-} & 0.007 & 119 & 0.012 & 197 & 0.007 & 63 & 11.3\\
J1944$+$0907 & 5.19 & 1.73 & 24.3 & -& \multicolumn{2}{|c|}{-} & 0.428 & 44 & \multicolumn{2}{|c|}{-} & 0.475 & 54 & \multicolumn{2}{|c|}{-} & 4.4\\
B1953$+$29 & 6.13 & 2.97 & 104.5 & 117.3& \multicolumn{2}{|c|}{-} & 0.662 & 36 & \multicolumn{2}{|c|}{-} & 0.719 & 47 & \multicolumn{2}{|c|}{-} & 4.4\\
J2010$-$1323 & 5.22 & 0.48 & 22.2 & -& \multicolumn{2}{|c|}{-} & \multicolumn{2}{|c|}{-} & 0.336 & 79 & 0.692 & 79 & \multicolumn{2}{|c|}{-} & 6.2\\
J2017$+$0603 & 2.90 & 0.80 & 23.9 & 2.2& \multicolumn{2}{|c|}{-} & 0.262 & 6 & \multicolumn{2}{|c|}{-} & 0.277 & 54 & 0.283 & 32 & 3.8\\
J2033$+$1734 & 5.95 & 1.11 & 25.1 & 56.3& \multicolumn{2}{|c|}{-} & 0.712 & 20 & \multicolumn{2}{|c|}{-} & 0.716 & 26 & \multicolumn{2}{|c|}{-} & 2.3\\
J2043$+$1711 & 2.38 & 0.52 & 20.7 & 1.5& \multicolumn{2}{|c|}{-} & 0.124 & 75 & \multicolumn{2}{|c|}{-} & 0.139 & 89 & \multicolumn{2}{|c|}{-} & 4.5\\
J2145$-$0750 & 16.05 & 2.98 & 9.0 & 6.8& \multicolumn{2}{|c|}{-} & \multicolumn{2}{|c|}{-} & 0.229 & 95 & 0.494 & 100 & \multicolumn{2}{|c|}{-} & 11.3\\
J2214$+$3000 & 3.12 & 1.47 & 22.5 & 0.4& \multicolumn{2}{|c|}{-} & \multicolumn{2}{|c|}{-} & \multicolumn{2}{|c|}{-} & 0.496 & 53 & 0.464 & 39 & 4.2\\
J2229$+$2643 & 2.98 & 0.15 & 22.7 & 93.0& \multicolumn{2}{|c|}{-} & 0.522 & 21 & \multicolumn{2}{|c|}{-} & 0.527 & 22 & \multicolumn{2}{|c|}{-} & 2.4\\
J2234$+$0611 & 3.58 & 1.20 & 10.8 & 32.0& \multicolumn{2}{|c|}{-} & 0.214 & 20 & \multicolumn{2}{|c|}{-} & 0.214 & 24 & \multicolumn{2}{|c|}{-} & 2.0\\
J2234$+$0944 & 3.63 & 2.01 & 17.8 & 0.4& \multicolumn{2}{|c|}{-} & 0.278 & 4 & \multicolumn{2}{|c|}{-} & 0.280 & 27 & 0.240 & 18 & 2.5\\
J2302$+$4442 & 5.19 & 1.39 & 13.8 & 125.9& \multicolumn{2}{|c|}{-} & \multicolumn{2}{|c|}{-} & 0.992 & 55 & 1.659 & 50 & \multicolumn{2}{|c|}{-} & 3.8\\
J2317$+$1439 & 3.45 & 0.24 & 21.9 & 2.5& 0.071 & 80 & 0.114 & 132 & \multicolumn{2}{|c|}{-} & 0.180 & 76 & \multicolumn{2}{|c|}{-} & 11.0\\
\hline
\multicolumn{5}{r|}{Nominal scaling factor$^b$ (ASP/GASP)} 
  & \multicolumn{2}{|c|}{0.6}
  & \multicolumn{2}{|c|}{0.4}
  & \multicolumn{2}{|c|}{0.8}
  & \multicolumn{2}{|c|}{0.8}
  & \multicolumn{2}{|c|}{0.8}
  & \\
\multicolumn{5}{r|}{Nominal scaling factor$^b$ (GUPPI/PUPPI)} 
  & \multicolumn{2}{|c|}{0.7}
  & \multicolumn{2}{|c|}{0.5}
  & \multicolumn{2}{|c|}{1.4}
  & \multicolumn{2}{|c|}{2.5}
  & \multicolumn{2}{|c|}{2.1}
  & \\
\hline
\end{tabular}

\vspace{0.5em}

{$^a$ For this table, the original TOA uncertainties were scaled by their
bandwidth-time product, $\displaystyle \left( \frac{\Delta \nu}{100~\mathrm{MHz}}
\frac{\tau}{1800~\mathrm{s}} \right)^{1/2}$, to remove variation due to
different instrument bandwidths and integration time.}

\vspace{0.5em}

{$^b$ TOA uncertainties can be rescaled to the nominal full instrumental
bandwidth as listed in Table~1 of \cite{Arzoumanian2015b} by dividing by the
scaling factors given here.}

}

 \end{center}
\end{table*}

Observations were taken using two telescopes, the 305 m William E.
Gordon Telescope of the Arecibo Observatory, and the 100 m Robert C. Byrd
Green Bank Telescope (GBT) of the Green Bank Observatory (formerly the
National Radio Astronomy Observatory).
Pulsars at declinations $0^\circ<\delta<+39^\circ$ were observed with Arecibo, while all
others were observed with the GBT; two sources (PSRs~J1713$+$0747 and B1937$+$21) were observed with
both telescopes.  An approximately monthly
observing cadence was used for most of the observations.
In addition, weekly observations were made for two pulsars at
the GBT beginning in 2013 (PSRs J1713+0747 and J1909$-$3744) and for
five pulsars at Arecibo beginning in 2015
(PSRs J0030+0451, J1640+2224, J1713+0747, J2043+1711, and
J2317+1439).  Observations at Arecibo were temporarily interrupted
in 2007 (telescope painting) and 2014 (earthquake damage).
Observations at GBT were interrupted in 2007 (azimuth-track refurbishment).

\begin{figure*}
\begin{center}
\includegraphics[width=6.0in]{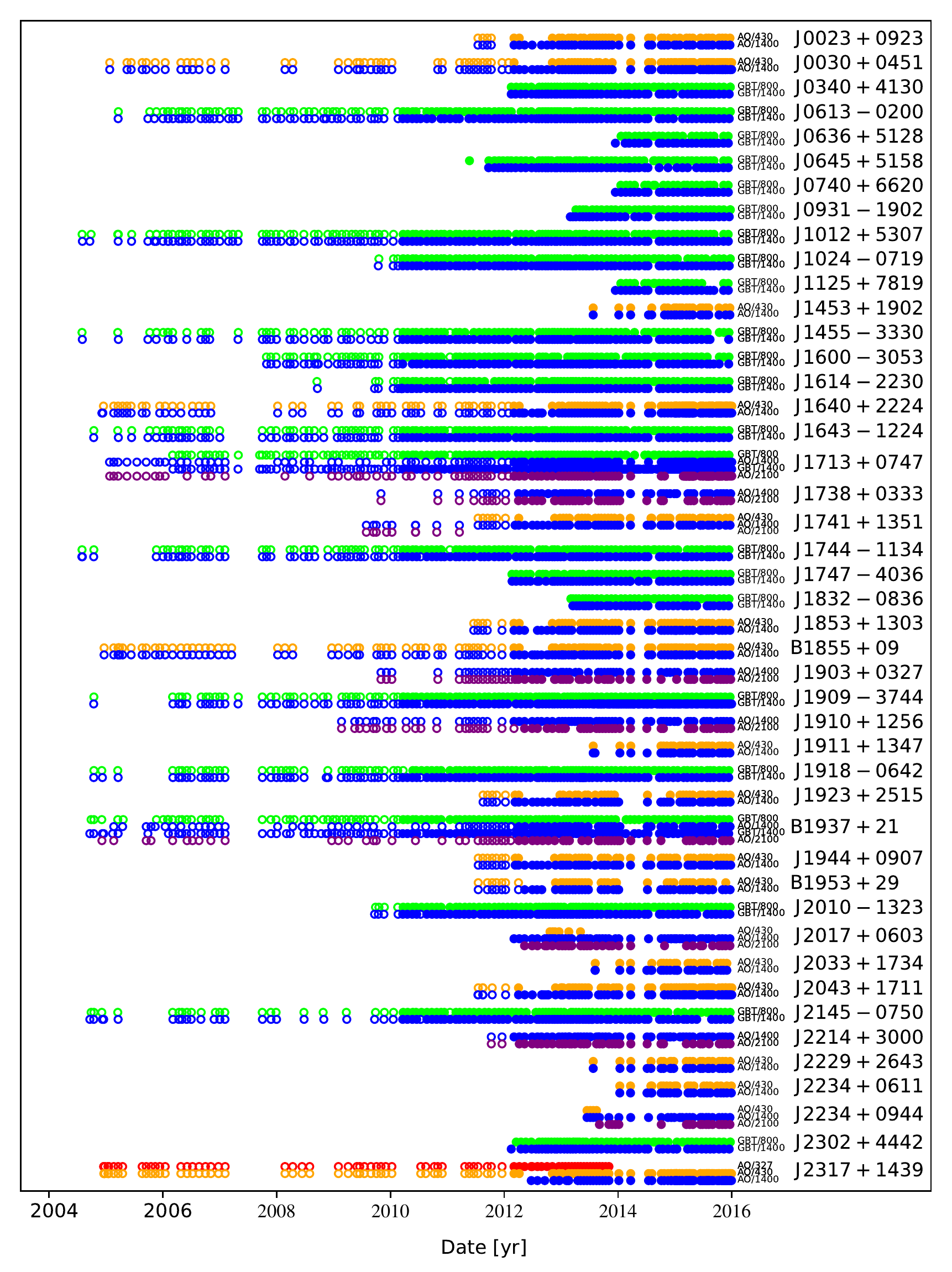}
\caption{\label{fig:obs:epochs}Epochs of all observations in the data
set.  Marker type indicates data-acquisition system: open circles are
ASP or GASP; closed circles are PUPPI or GUPPI.  Colors indicate radio-frequency band, at either telescope: red is 327~MHz; orange is 430~MHz;
green is 820~MHz; blue is 1.4~GHz; and purple is 2.1~GHz.}
\end{center}
\end{figure*}

At most epochs, each pulsar was observed with two separate receiver
systems at widely separated frequencies in order to provide precise DM estimates, as described below.  At the GBT, the 820~MHz and 1.4~GHz receivers were
always used for monthly observations, and the 1.4~GHz receiver alone was
used for weekly observations.  At
Arecibo, two out of four possible receivers, at 327~MHz, 430~MHz, 1.4~GHz, and 2.3~GHz, were chosen
for each pulsar.  For some pulsars, the pair of receivers used has
changed over the course of the project.  In the most recent observations,
we no longer use the 327~MHz system.  See Figure~\ref{fig:obs:epochs},
Table~\ref{tab:psr_params}, and \nineyr\ for more details.

\begin{table*}[tp]
\begin{center}
\caption{\label{tab:timingpar} Summary of timing-model fits}
{\scriptsize
\begin{tabular}{c|r|rrrrrr|cc|ccr|c}
\hline
Source
  & Number
  & \multicolumn{6}{|c|}{Number of Fit Parameters$^a$}
  & \multicolumn{2}{|c|}{RMS$^b$ ($\mu$s)}
  & \multicolumn{3}{|c|}{Red Noise$^c$}
  & Figure \\

  & of TOAs
  & S
  & A
  & B
  & DM
  & FD
  & J
  & Full
  & White
  & $A_{\mathrm{red}}$
  & $\gamma_{\mathrm{red}}$
  & log$_{10}B$
  & Number\\
\hline
J0023+0923 & 8161 & 3 & 5 & 8 & 50 & 1 & 1 & 0.308  & - & - & - & 0.41  & \ref{fig:summary-J0023+0923} \\
J0030+0451 & 5681 & 3 & 5 & 0 & 102 & 1 & 1 & 0.710  & 0.241  & 0.025\phantom{$^d$} & $-$4.0 & 5.29  & \ref{fig:summary-J0030+0451} \\
J0340+4130 & 6475 & 3 & 5 & 0 & 56 & 2 & 1 & 0.454  & - & - & - & $-$0.02  & \ref{fig:summary-J0340+4130} \\
J0613$-$0200 & 11566 & 3 & 5 & 7 & 121 & 2 & 1 & 0.502  & 0.199  & 0.212\phantom{$^d$} & $-$1.2 & 4.10  & \ref{fig:summary-J0613-0200} \\
J0636+5128 & 13699 & 3 & 5 & 6 & 26 & 1 & 1 & 0.611  & - & - & - & 0.57  & \ref{fig:summary-J0636+5128} \\
J0645+5158 & 6370 & 3 & 5 & 0 & 61 & 2 & 1 & 0.180  & - & - & - & 0.01  & \ref{fig:summary-J0645+5158} \\
J0740+6620 & 2090 & 3 & 5 & 6 & 26 & 1 & 1 & 0.190  & - & - & - & 0.09  & \ref{fig:summary-J0740+6620} \\
J0931$-$1902 & 2597 & 3 & 5 & 0 & 39 & 0 & 1 & 0.495  & - & - & - & $-$0.03  & \ref{fig:summary-J0931-1902} \\
J1012+5307 & 16782 & 3 & 5 & 6 & 123 & 1 & 1 & 1.270  & 0.354  & 0.476\phantom{$^d$} & $-$1.5 & 16.20  & \ref{fig:summary-J1012+5307} \\
J1024$-$0719 & 8233 & 4 & 5 & 0 & 82 & 2 & 1 & 0.324  & - & - & - & 0.12  & \ref{fig:summary-J1024-0719} \\
J1125+7819 & 2285 & 3 & 5 & 5 & 25 & 4 & 1 & 0.483  & - & - & - & 0.86  & \ref{fig:summary-J1125+7819} \\
J1453+1902 & 736 & 3 & 5 & 0 & 22 & 0 & 1 & 0.757  & - & - & - & 0.02  & \ref{fig:summary-J1453+1902} \\
J1455$-$3330 & 7526 & 3 & 5 & 6 & 108 & 1 & 1 & 0.571  & - & - & - & 0.04  & \ref{fig:summary-J1455-3330} \\
J1600$-$3053 & 12433 & 3 & 5 & 9 & 106 & 2 & 1 & 0.181  & - & - & - & 0.04  & \ref{fig:summary-J1600-3053} \\
J1614$-$2230 & 11173 & 3 & 5 & 8 & 92 & 2 & 1 & 0.183  & - & - & - & $-$0.05  & \ref{fig:summary-J1614-2230} \\
J1640+2224 & 5945 & 3 & 5 & 8 & 110 & 4 & 1 & 0.382  & - & - & - & 0.00  & \ref{fig:summary-J1640+2224} \\
J1643$-$1224 & 11528 & 3 & 5 & 6 & 122 & 4 & 1 & 3.510  & 0.757  & 1.619$^d$ & $-$1.3 & 28.38  & \ref{fig:summary-J1643-1224} \\
J1713+0747 & 27571 & 3 & 5 & 8 & 209 & 5 & 3 & 0.116  & 0.103  & 0.021\phantom{$^d$} & $-$1.6 & 0.85  & \ref{fig:summary-J1713+0747} \\
J1738+0333 & 4881 & 3 & 5 & 5 & 54 & 1 & 1 & 0.364  & - & - & - & 0.05  & \ref{fig:summary-J1738+0333} \\
J1741+1351 & 3037 & 3 & 5 & 8 & 59 & 2 & 2 & 0.102  & - & - & - & $-$0.02  & \ref{fig:summary-J1741+1351} \\
J1744$-$1134 & 11550 & 3 & 5 & 0 & 116 & 4 & 1 & 0.403  & - & - & - & 1.13  & \ref{fig:summary-J1744-1134} \\
J1747$-$4036 & 6065 & 3 & 5 & 0 & 54 & 1 & 1 & 5.350  & 1.580  & 1.823{$^d$} & $-$1.4 & 4.90  & \ref{fig:summary-J1747-4036} \\
J1832$-$0836 & 3886 & 3 & 5 & 0 & 39 & 0 & 1 & 0.184  & - & - & - & 0.01  & \ref{fig:summary-J1832-0836} \\
J1853+1303 & 2502 & 3 & 5 & 7 & 53 & 0 & 1 & 0.205  & - & - & - & 0.07  & \ref{fig:summary-J1853+1303} \\
B1855+09 & 5618 & 3 & 5 & 7 & 101 & 3 & 1 & 0.796  & 0.482  & 0.069\phantom{$^d$} & $-$3.0 & 6.93  & \ref{fig:summary-B1855+09} \\
J1903+0327 & 3326 & 3 & 5 & 8 & 60 & 1 & 1 & 4.010  & 0.573  & 1.615$^d$ & $-$2.1 & 15.53  & \ref{fig:summary-J1903+0327} \\
J1909$-$3744 & 17373 & 3 & 5 & 9 & 166 & 1 & 1 & 0.187  & 0.070  & 0.042\phantom{$^d$} & $-$1.7 & 23.55  & \ref{fig:summary-J1909-3744} \\
J1910+1256 & 3563 & 3 & 5 & 6 & 67 & 1 & 1 & 0.515  & - & - & - & 0.15  & \ref{fig:summary-J1910+1256} \\
J1911+1347 & 1356 & 3 & 5 & 0 & 25 & 2 & 1 & 0.054  & - & - & - & $-$0.03  & \ref{fig:summary-J1911+1347} \\
J1918$-$0642 & 12505 & 3 & 5 & 7 & 117 & 4 & 1 & 0.297  & - & - & - & 0.01  & \ref{fig:summary-J1918-0642} \\
J1923+2515 & 1944 & 3 & 5 & 0 & 48 & 1 & 1 & 0.229  & - & - & - & $-$0.04  & \ref{fig:summary-J1923+2515} \\
B1937+21 & 14217 & 3 & 5 & 0 & 165 & 5 & 3 & 1.500  & 0.110  & 0.157\phantom{$^d$} & $-$2.8 & 174.46  & \ref{fig:summary-B1937+21} \\
J1944+0907 & 2830 & 3 & 5 & 0 & 53 & 2 & 1 & 0.333  & - & - & - & 0.25  & \ref{fig:summary-J1944+0907} \\
B1953+29 & 2315 & 3 & 5 & 5 & 47 & 2 & 1 & 0.394  & - & - & - & 0.06  & \ref{fig:summary-B1953+29} \\
J2010$-$1323 & 10844 & 3 & 5 & 0 & 88 & 3 & 1 & 0.260  & - & - & - & $-$0.04  & \ref{fig:summary-J2010-1323} \\
J2017+0603 & 2359 & 3 & 5 & 7 & 49 & 0 & 2 & 0.091  & - & - & - & $-$0.12  & \ref{fig:summary-J2017+0603} \\
J2033+1734 & 1511 & 3 & 5 & 5 & 23 & 2 & 1 & 0.500  & - & - & - & 0.08  & \ref{fig:summary-J2033+1734} \\
J2043+1711 & 3241 & 3 & 5 & 7 & 64 & 4 & 1 & 0.119  & - & - & - & $-$0.03  & \ref{fig:summary-J2043+1711} \\
J2145$-$0750 & 10938 & 3 & 5 & 5 & 107 & 2 & 1 & 1.180  & 0.304  & 0.589\phantom{$^d$} & $-$1.3 & 6.34  & \ref{fig:summary-J2145-0750} \\
J2214+3000 & 4569 & 3 & 5 & 5 & 53 & 2 & 1 & 1.330  & - & $^e$ & $^e$ & 6.62  & \ref{fig:summary-J2214+3000} \\
J2229+2643 & 1131 & 3 & 5 & 5 & 21 & 2 & 1 & 0.203  & - & - & - & 0.03  & \ref{fig:summary-J2229+2643} \\
J2234+0611 & 1279 & 3 & 5 & 7 & 23 & 1 & 1 & 0.030  & - & - & - & $-$0.04  & \ref{fig:summary-J2234+0611} \\
J2234+0944 & 3022 & 3 & 5 & 5 & 29 & 2 & 2 & 0.205  & - & - & - & 0.26  & \ref{fig:summary-J2234+0944} \\
J2302+4442 & 6549 & 3 & 5 & 7 & 58 & 3 & 1 & 0.836  & - & - & - & 0.10  & \ref{fig:summary-J2302+4442} \\
J2317+1439 & 5939 & 3 & 5 & 6 & 111 & 5 & 2 & 0.287  & - & - & - & 0.13  & \ref{fig:summary-J2317+1439} \\
\hline
\end{tabular}

\vspace{0.5em}

{$^a$ Fit parameters: S=spin; B=binary; A=astrometry; DM=dispersion measure;
FD=frequency dependence; J=jump}

\vspace{0.5em}

{$^b$ Weighted root-mean-square of epoch-averaged post-fit timing residuals,
calculated using the procedure described in Appendix D of \nineyr.
For sources with red noise, the ``Full'' RMS value includes the red noise
contribution, while the ``White'' RMS does not.}

\vspace{0.5em}

{$^c$ Red-noise parameters: $A_{\mathrm{red}}$ = amplitude of red noise
spectrum at $f$=1~yr$^{-1}$ measured in $\mu$s yr$^{1/2}$;
$\gamma_{\mathrm{red}}$ = spectral index; $B$ = Bayes factor.  See
Eqn.~\ref{eqn:rn_spec} and Appendix~C of \nineyr\ for details.}

\vspace{0.5em}

{$^d$ For these sources, the detected red noise may include contributions from unmodeled
interstellar-medium propagation effects; see
the text for details.}

\vspace{0.5em}

{$^e$ Difficult to model; see the text.}

}

 \end{center}
\end{table*}

Data were recorded using two generations of backend instrumentation. For
approximately the first six years of the project, the older generation ASP and
GASP systems (at Arecibo and Green Bank, respectively) were used, which
allowed for recording up to 64~MHz of bandwidth \citep{Demorest2007}.
During 2010$-$2012 we transitioned to PUPPI and GUPPI, which
can process up to 800~MHz total bandwidth \citep{DuPlain2008,Ford2010}.  ASP and GASP
have now been fully decommissioned, and all new data presented
here were taken using PUPPI and GUPPI.
Detailed lists of frequencies and bandwidths for all receivers
and backends are given in Table~1 of \nineyr.

The raw data produced by the backend instruments are folded pulse
profiles as a function of time, radio frequency, and polarization.
These profiles have 2048 phase bins, a frequency resolution of either 1.5~MHz
(GUPPI/PUPPI) or 4~MHz (ASP/GASP), and a time resolution (subintegration time)
of 1 or 10~s.  

For the ASP and GASP data, we have not calculated new TOAs for the present work; instead,
we use the TOAs and instrumental time offsets (relative to PUPPI and GUPPI) from 
\nineyr.  However, the set of ASP and GASP TOAs included in the present
data set is slightly different from that of \nineyr\ due to the removal of
the long spans of single-frequency observations described above and also due to
a complete reanalysis of all TOAs to eliminate the outliers described below.

For the GUPPI and PUPPI data, in order to ensure consistency,
we re-processed all data, including those in \nineyr.
These data were polarization-calibrated, had
interference-corrupted data segments excised, and were averaged in both
time and frequency using procedures nearly identical to
\nineyr.  In the new analysis, we applied multiple
rounds of interference excision, first on the original full-resolution
uncalibrated data and then again on the calibrated and partially averaged
data set.  Profiles were then integrated in time for up to 30~minutes or 2.5\%
of the orbital period (for binary pulsars), whichever is shorter.
The final frequency
resolution varies from 1.5~to 12.5~MHz depending on receiver
system.  TOAs were generated from these data using standard
procedures.  For pulsars in \nineyr, the same
template profiles were used.  For newly added sources, template profiles
were generated following the procedure described by
\cite{Demorest2013} and \nineyr.  All processing steps
for the profile data were carried out using the PSRCHIVE software
package\footnote{\url{http://psrchive.sourceforge.net}} and our set of
pipeline processing
scripts\footnote{\url{http://github.com/demorest/nanopipe}}.

Following construction of the initial set of TOAs for
each pulsar, the TOAs were examined to remove uninformative
and outlier data points prior to fitting timing models.  This
was done in three steps.
First, as in \nineyr, all times of arrival coming from pulse
profiles with signal-to-noise ratios less than 8 were excluded.  
Second, the sets of TOAs were manually edited to remove outlier points; typically
these were due to data contaminated by radio-frequency interference (RFI).
Third,
the timing data were run
through the automated outlier-identification algorithm described by
\cite{Vallisneri2017}, which estimates 
the probability $p_{i,{\rm out}}$ that each individual TOA is an outlier. This estimate is obtained in full consistence with the Bayesian inference of all pulsar noise parameters.
We removed all TOAs with probability-of-outlier values, $p_{i,{\rm out}}$, greater than 0.1. 
This resulted in the removal of an 221
TOAs across all pulsar data sets. 
The automated outlier algorithm was run last because it was a late
addition to the analysis pipeline.  
For future data sets, we expect to rely primarily on automated,
rather than manual, excision methods.  

In order to robustly measure DM variation on short time scales, 
we group the data from each pulsar into ``epochs'' up
to 6 days long (or 15 days in early ASP/GASP data).  
Because measurements of DM require analyzing arrival
times across a wide range of radio frequencies,
data from any epoch for which the 
fractional bandwidth was less than 10\%
($\nu_{\rm max}/\nu_{\rm min}<1.1$, where $\nu$ is radio frequency)
were excluded from the data set.
This criterion caused us to exclude some data that were used in \nineyr,
particularly long spans of single-receiver data early in 
the data sets of a few pulsars. 

Dispersion measure variations due to the solar wind
can be significant at low solar elongations.  We used a simple test
for the potential significance of such variations within an
observing epoch: we calculated the
expected difference in pulse arrival time within an epoch 
assuming a toy solar wind model in which the 
electron density is $n_e=n_0(r/r_0)^{-2},$ 
where $n_0=5\,{\rm cm}^{-3}$ (a typical value; e.g., \cite{Splaver2005}), $r$ is the distance from the Sun,
and $r_0=1\,{\rm au}$; we then excluded (or split into separate epochs)
data from any epoch in which the model solar-wind time delays varied
by more than 160~ns.    These excluded points are available in
supplementary files along with the data set. 

The TOAs for each pulsar were fit using a
physical timing model using the Tempo\footnote{\url{http://tempo.sourceforge.net}} and
Tempo2\footnote{\url{http://bitbucket.org/psrsoft/tempo2}} timing-analysis software packages.  We
employed a standardized procedure for determining parameters to include in
each pulsar's timing model.  The following parameters
are always included: intrinsic spin and spin-down rate;
five astrometric parameters (position, proper motion,
and parallax); and, for binary pulsars, five
Keplerian orbital parameters, with the orbital
model chosen based on eccentricity, the presence or absence
of relativistic phenomena, etc.
Time-variable
DM was included in the model via a piecewise-constant
model (``DMX'') within each epoch described above.  Arbitrary
constant offsets (``jumps'') were fit between data subsets collected
with different receivers and/or telescopes, with one offset per
receiver/telescope combination.
The following terms were
included in each pulsar's timing model if they were found to be
significant via a $F$-test value of $<0.0027$ (3$\sigma$):
secular evolution of binary parameters, Shapiro delay in binary
systems, and 
frequency-dependent trends due to pulse-profile evolution over frequency (``FD parameters'';
see \nineyr).
In one case, for PSR~J1024$-$0719, a second spin-frequency derivative was included to account for extremely long-period
orbital motion \citep{Bassa2016,Kaplan2016}.  Otherwise, no higher-order spin-frequency
derivatives were fit.  Red and white timing noise were modeled using
procedures described in \S\ref{sec:noise}.
Timing-model best-fit parameter values and uncertainties were
derived using a generalized-least-squares (GLS) fit that makes use of
the noise-model covariance.  See \nineyr\ for
a detailed description of the motivations for, and
implementation of, the procedures used in this timing analysis.
Table~\ref{tab:timingpar} summarizes the timing models for 
individual pulsars.

The timing fits used the JPL DE436 solar system ephemeris 
and the TT(BIPM2016) timescale.

All TOA data and timing models presented here are
included as supplementary material to this paper and are also
publicly available online.\footnote{\url{http://data.nanograv.org}}
Data are given in standard formats compatible with both Tempo and Tempo2.
All data points excised using the procedures described above (outliers,
low signal-to-noise-ratio points, etc.) are
provided in supplementary files along with the data set.

Pulsar-timing models developed from radio observations can be used to
calculate pulse phase as a function of time over the duration of the radio
observations.  This can be useful for purposes such as pulse-period-folding
data collected at other observatories (e.g., photon time tags from high-energy
observatories) made over the same time span as the radio observations.
Because the red noise model described in \S\ref{sec:noise} and included in our
timing models is stochastic, these models are not optimal for precise pulse
phase calculations.  For this reason, we generated a second set of parameter
files, available with the data set, in which red noise (if any) for each
pulsar is modeled as a Taylor expansion in rotation frequency, beyond the
usual rotation frequency and its first derivative (spin-down).

\section{Noise Characterization}
\label{sec:noise}
\subsection{Noise Model}

\begin{figure*}
\begin{center}
\includegraphics[scale=1.0]{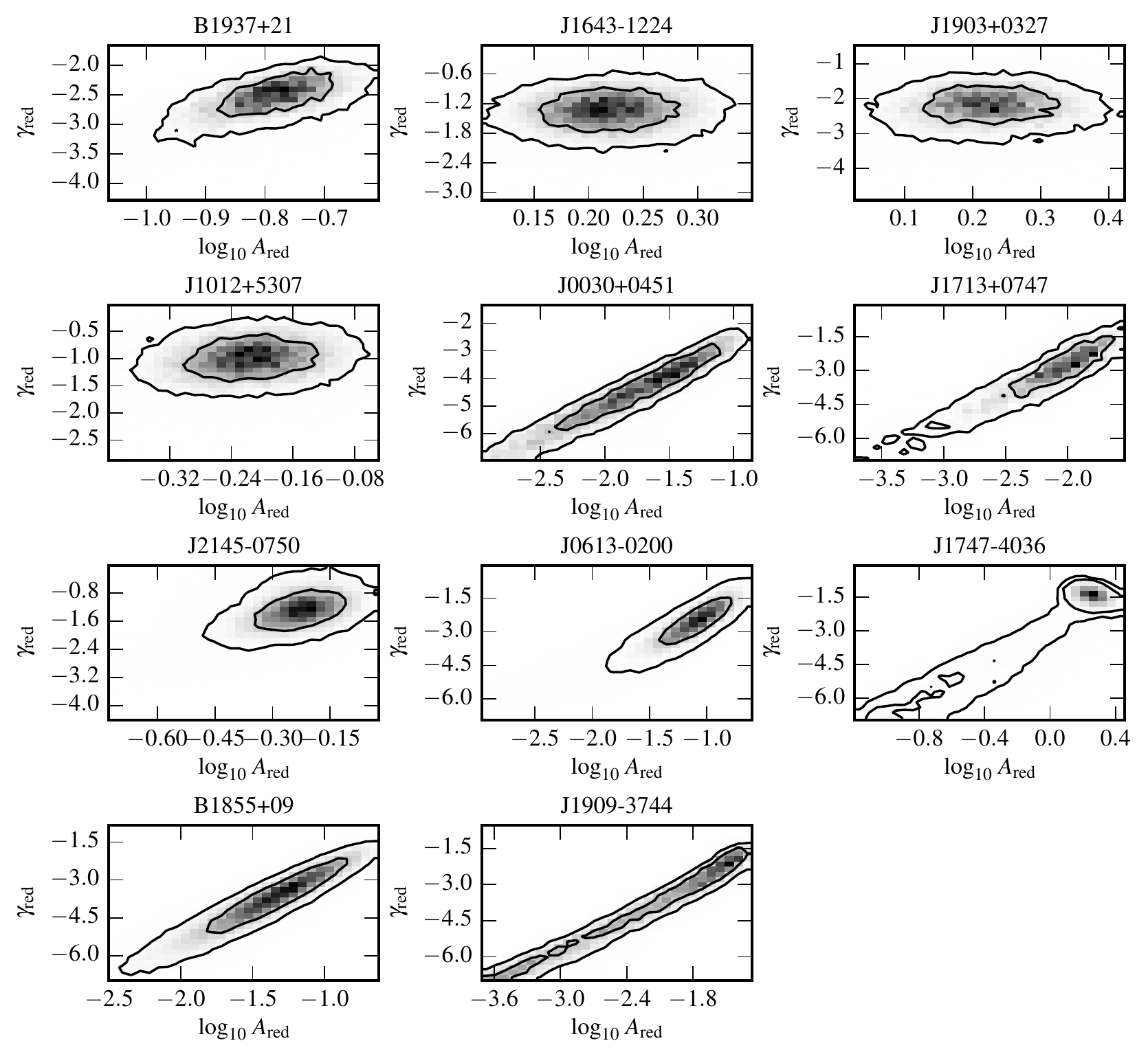}
\caption{2-D posteriors of the amplitude and spectral index of the red-noise parameters for those pulsars with Bayes factors for red noise greater than 100, plus J1713$+$0747. The contours are the 50\% and 90\% credible regions.}
\label{fig:red-post}
\end{center}
\end{figure*}
 \begin{figure}
\begin{center}
\includegraphics[width=\columnwidth]{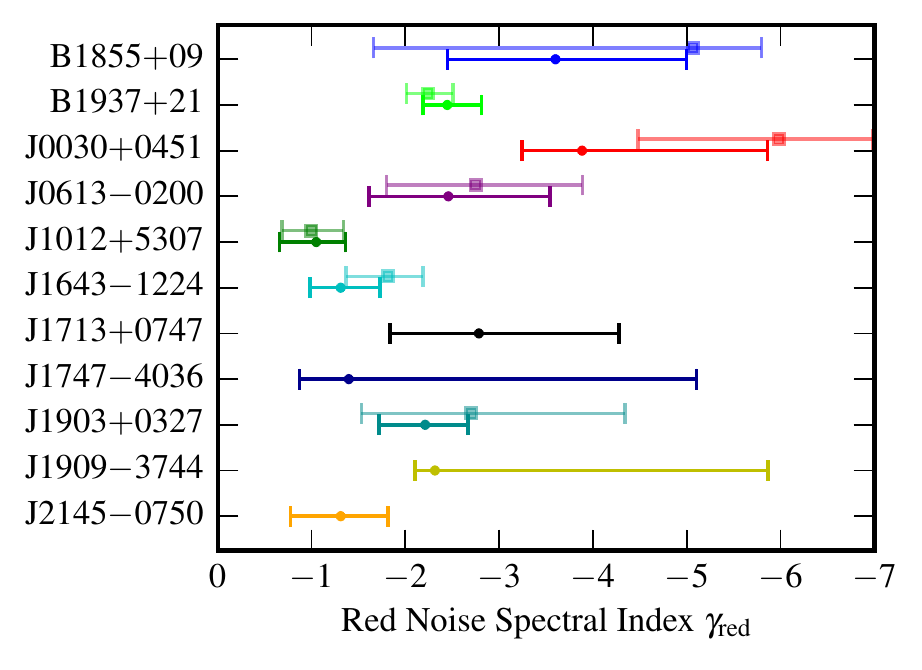}
\caption{Maximum-likelihood estimates of the spectral-index parameters (circles), with 1-$\sigma$ uncertainties (bars). Where applicable, we have plotted the corresponding credible intervals for \nineyr\ with square markers and 1$\sigma$ uncertainties.}
\label{fig:red-comp}
\end{center}
\end{figure}

The noise model used in this analysis is identical to that used in \nineyr ; see that paper for more details. Here, we will qualitatively review the model. In general we model the noise in the residual data as additive Gaussian\footnote{The final noise analysis assumes Gaussian noise after outliers have been removed.} noise, with three white-noise components and one red-noise component, as follows:
\begin{enumerate}

\item EFAC: a multiplication factor on the measured TOA uncertainties, $\sigma_{i}$. We use a separate EFAC parameter, $E_k$, for each pulsar/backend/receiver combination to account for any systematics in the TOA measurement uncertainties.

\item EQUAD: an error term added in quadrature to the TOA uncertainty (before scaling by EFAC). We again use a separate EQUAD parameter, $Q_k$, for each pulsar/backend/receiver combination. This term captures any white noise, in addition to the statistical uncertainties found in the TOA calculations. With this term, the new scaled TOA uncertainty is $\sigma_{i,k}\rightarrow E_{k}(\sigma^{2}_{i,k}+Q_{k}^2)^{1/2}$ for pulsar/backend/receiver combination $k$.

\item ECORR: a short-timescale noise process that is uncorrelated between observing epochs but completely correlated between TOAs obtained simultaneously at different frequencies. This accounts for wideband processes such as pulse jitter \citep[e.g.,][]{lam2016a}.

\item Red noise: a low-frequency stationary Gaussian process that is parameterized by a power-law spectrum of the form
\begin{equation}
P(f) = A_{\rm red}^{2}\left( \frac{f}{f_{\rm yr}} \right)^{\gamma_{\rm red}},
\label{eqn:rn_spec}
\end{equation}
where $A_{\rm red}$ is the amplitude of the red-noise process in units of $\mu\mathrm{s}\,\mathrm{yr}^{1/2}$, $\gamma_{\rm red}$ is the spectral index, and $f_{\rm yr}=1~\mathrm{yr}^{-1}$.
\end{enumerate}

This noise model is incorporated into a joint likelihood containing all timing model parameters, and run through an MCMC inference package\footnote{\url{https://github.com/jellis18/PTMCMCSampler}} and through MultiNest \citep{fhb09} to produce maximum-likelihood parameter estimates and Bayesian evidence for the presence of red noise, respectively. Generally, for those pulsars with a Bayes factor for red noise, $B$, greater than 100,  we included red-noise parameters in the final timing models, while we omitted them from the timing models for other pulsars.
 See Appendix C of \nineyr\ for a complete description of the Bayesian inference model. The red-noise amplitudes, spectral indices, and Bayes factors are given in Table \ref{tab:timingpar}.

\subsection{Noise Analysis}
The noise characteristics of the pulsars broadly match those discussed in \nineyr; we give a brief overview here. From the Bayes-factor analysis described above, we find significant red noise in 11 of the 45 pulsars (Table \ref{tab:timingpar}). PSR J1713$+$0747 is not above our threshold for significant red noise, $B>100$; however, since it is one of our most precisely timed pulsars, and since it does show hints of red noise, $B\sim 7$, we include it in our red-noise analysis. A survey of the red-noise--parameter posterior probability distributions is shown in Figure \ref{fig:red-post},
and 
the 68\% credible intervals of the spectral-index parameter are plotted in Figure \ref{fig:red-comp}.

Figures \ref{fig:red-post} and \ref{fig:red-comp} show that several pulsars have red-noise spectral indices that are well constrained to low values (i.e., $-1$ to $-3$),
 while others are far less constrained and have some non-negligible posterior probability of lying in the full range tested, $-7<\gamma_{\rm red}<0$. 

In Figure~\ref{fig:red-post}, the plots for most pulsars show strong covariance between red-noise amplitude and spectral index.  This arises because the red-noise PSD (power spectral density) is only larger than the white-noise PSD at the lowest frequencies in a given data set, which are typically lower than our fiducial reference frequency of  $f_{\rm yr}=1~\mathrm{yr}^{-1}$.  Extrapolation of the red-noise amplitude from these low frequencies to $f_{\rm yr}$ depends sensitively on the spectral index, hence inducing the large covariance.
In principle, we could use a lower fiducial frequency for the power-law PSD to minimize this covariance,  but the choice of fiducial frequency would be pulsar-dependent and make uniform comparisons complicated.  For this reason, we choose the fiducial reference frequency of 1~$\mathrm{yr}^{-1}$ for all pulsars.

The red-noise spectra that we observe for millisecond pulsars tend to have spectral indices that are shallower than
those seen in canonical (non-millisecond) pulsars,
suggesting different origins of the red noise in these two populations 
\citep{sc10,lcc+17}.
If this behavior is due to a random walk in one of the pulsar-spin parameters, then our data are consistent with random walk in phase\footnote{Random walks in pulsar phase, period, and period derivative lead to underlying power spectral indices of $-2$, $-4$, and $-6$, respectively \citep{sc10}.} as opposed to a random walk in spin-period derivative \citep{lhk+10,sc10}.

In \nineyr\ we suggest that some of the red noise seen in that data set could be due to frequency-dependent propagation effects within the ionized interstellar medium.  One issue is that portions of the \nineyr\ data set only contained observations from a single receiver, inhibiting correction for time-variable DM.  In the present data set we only included observations that have observations over a wide range of frequencies at every epoch, thus ruling out this particular source of apparent red noise. However, even after correction for time-variable DM, non-white frequency-dependent arrival times are still evident in the residual plots for PSRs J1600$-$3053, J1643$-$1224, J1747$-$4036, and J1903+0327, likely indicative of red noise arising from unmodeled propagation effects in the interstellar medium.  We have not attempted to mitigate such effects in the present data set.  Furthermore, 
PSR J1643$-$1224 has been shown to have significant scattering and profile shape variations \citep{slk+16,lkd+17}, which we have not attempted to include in our timing model for this pulsar.  

From inspection of the residual plots, as well as Figure \ref{fig:red-post} and \ref{fig:red-comp}, the red noise falls into two categories: well constrained shallow spectral indices with clearly defined high-frequency residual structure (e.g. PSRs J0613$-$0200 (Figure~\ref{fig:summary-J0613-0200}), J1012$+$5307 (Figure~\ref{fig:summary-J1012+5307}), J1643$-$1224 (Figure~\ref{fig:summary-J1643-1224}), and J2145$-$0750 (Figure~\ref{fig:summary-J2145-0750})) and less constrained steeper spectral indices with clearly defined low-frequency residual structure (e.g. PSRs J0030$+$0451 (Figure~\ref{fig:summary-J0030+0451}), J1713$+$0747 (Figure~\ref{fig:summary-J1713+0747}), B1855+09 (Figure~\ref{fig:summary-B1855+09}), J1903$+$0327 (Figure~\ref{fig:summary-J1903+0327}), J1909$-$3744 (Figure~\ref{fig:summary-J1909-3744})), and B1937$+$21 (Figure~\ref{fig:summary-B1937+21}).

While the power-law noise model seems adequate for this data set, as precision increases and as timing baselines grow we will likely need to use more sophisticated red-noise models such as a spectral model where each PSD component is free to vary or perhaps adaptive techniques like those introduced in \cite{ec16} or \cite{lsc+16}.

For one pulsar, PSR~J2214+3000, the red-noise-detection algorithm indicated the presence of significant noise, but the spectral index could not be easily quantified.  The issue appears to be moderate-level excess noise of unknown origin in mid-to-late 2013.  PSR~J2214+3000 is a black-widow-type binary (short orbital period, very-low-mass companion), but no eclipses are observed \citep{Ransom2011}.  We searched for orbital period variations and for orbital-phase-dependent pulse time delays or DM variations that might indicate variable flow of matter in the system, but found none.  In the appendix, we provide residual plots for this pulsar both with and without the nominal noise model; for parameter fitting, we omitted the red noise model.

\section{Astrometry}
\label{sec:astrometry}
Here, we analyze astrometric measurements in the pulsar-timing models following the procedures used by
\citet[][hereafter M16]{Matthews2016} to analyze astrometry in the \nineyr\ data.
Parallax,
 position, and proper motion were free parameters in the timing model for each pulsar, regardless of statistical significance.  We used ecliptic coordinates for position ($\lambda$, $\beta$) and proper motion ($\mu_\lambda\equiv\dot{\lambda}\cos \beta$, $\mu_\beta$).  For timing parallax, we allowed for both negative and positive values.  Although the former is unphysical, it provides a useful check on our data and an assessment of the veracity of low-significance measurements, as discussed below.

Positions and proper motions in ecliptic coordinates are given in Table~\ref{tab:position_ecliptic}.   We also provide positions
and proper motions in equatorial coordinates ($\alpha$, $\delta$, $\mu_\alpha\equiv\dot{\alpha}\cos\delta$, $\mu_\delta$) in Table~\ref{tab:position_equatorial}.  Because of covariances, the uncertainties in equatorial coordinates tend to be larger than uncertainties in ecliptic coordinates.

All positions and proper motions are relative to the reference frame of the JPL DE436 solar system ephemeris used to reduce these data;
this in turn is aligned with the Second Realization of the International Celestial Reference Frame \citep[ICRF2]{fey15}.

\begin{deluxetable*}{lDDDDc}
\tabletypesize{\scriptsize}
\tablecaption{Positions and Proper Motions in Ecliptic Coordinates
\label{tab:position_ecliptic}}
\tablecolumns{9}
\tablewidth{0pt}
\tablehead{
\multicolumn{1}{c}{Pulsar} &  \multicolumn{2}{c}{$\lambda$}  &  \multicolumn{2}{c}{$\beta$}  &  \multicolumn{2}{c}{$\mu_\lambda\equiv\dot{\lambda}\cos\beta$}  &  \multicolumn{2}{c}{$\mu_\beta$}  & \multicolumn{1}{c}{Epoch} \\ 
 & \multicolumn{2}{c}{} & \multicolumn{2}{c}{} & \multicolumn{2}{c}{(mas\,yr$^{-1}$)} & \multicolumn{2}{c}{(mas\,yr$^{-1}$)} &  \multicolumn{1}{c}{(MJD)}
}
\decimals
\startdata
J0023+0923 & 
9.07039380(1) & 6.30910853(9) & -13.90(3) & -0.3(4) 
 & 56567.000 
\\
J0030+0451 & 
8.910354709(9) & 1.4456968(4) & -5.516(8) & 2.9(4) 
 & 55390.000 
\\
J0340+4130 & 
62.61406187(3) & 21.33447352(8) & -1.33(9) & -3.1(3) 
 & 56675.000 
\\
J0613$-$0200 & 
93.79900756(2) & -25.40713682(4) & 2.13(2) & -10.29(4) 
 & 55413.000 
\\
J0636+5128 & 
96.36314673(2) & 28.24309933(4) & 3.7(2) & -2.0(2) 
 & 57002.000 
\\
J0645+5158 & 
98.05854704(1) & 28.85264208(2) & 2.19(4) & -7.25(6) 
 & 56534.000 
\\
J0740+6620 & 
103.75913772(4) & 44.10249786(4) & -2.7(2) & -32.5(2) 
 & 57017.000 
\\
J0931$-$1902 & 
152.37696636(5) & -31.77672320(8) & -0.6(2) & -4.9(3) 
 & 56864.000 
\\
J1012+5307 & 
133.36109736(3) & 38.75531470(4) & 13.98(3) & -21.50(5) 
 & 55291.000 
\\
J1024$-$0719 & 
160.73435127(1) & -16.04472755(6) & -14.41(3) & -58.0(1) 
 & 56239.000 
\\
J1125+7819 & 
115.6292967(2) & 62.45203072(9) & 16.(1) & 23.6(6) 
 & 57017.000 
\\
J1453+1902 & 
214.2087106(1) & 33.9046168(2) & 4.2(5) & -10.(2) 
 & 56936.000 
\\
J1455$-$3330 & 
231.34753526(5) & -16.0447987(2) & 8.20(5) & 0.5(2) 
 & 55293.000 
\\
J1600$-$3053 & 
244.347677844(6) & -10.07183903(3) & 0.46(1) & -7.16(6) 
 & 55885.000 
\\
J1614$-$2230 & 
245.788293268(7) & -1.2568039(4) & 9.49(1) & -31.3(7) 
 & 56047.000 
\\
J1640+2224 & 
243.98909040(2) & 44.05851688(2) & 4.18(1) & -10.74(2) 
 & 55366.000 
\\
J1643$-$1224 & 
251.08722023(8) & 9.7783313(5) & 5.38(9) & 4.5(5) 
 & 55330.000 
\\
J1713+0747 & 
256.668695241(2) & 30.700360494(4) & 5.267(2) & -3.443(4) 
 & 55391.000 
\\
J1738+0333 & 
264.09491180(2) & 26.88423737(5) & 6.85(5) & 5.4(1) 
 & 56258.000 
\\
J1741+1351 & 
264.364677959(9) & 37.21119874(1) & -8.67(2) & -7.77(2) 
 & 56209.000 
\\
J1744$-$1134 & 
266.11940142(1) & 11.80520111(6) & 19.04(1) & -8.77(6) 
 & 55292.000 
\\
J1747$-$4036 & 
267.5791338(1) & -17.2015403(4) & -1.3(4) & -2.(1) 
 & 56676.000 
\\
J1832$-$0836 & 
278.29200706(1) & 14.59071995(4) & -9.19(5) & -20.7(2) 
 & 56862.000 
\\
J1853+1303 & 
286.25730550(2) & 35.74335095(2) & -1.97(4) & -2.68(6) 
 & 56553.000 
\\
B1855+09 & 
286.86348828(1) & 32.32148622(2) & -3.27(1) & -5.06(2) 
 & 55367.000 
\\
J1903+0327 & 
287.5625787(1) & 25.9379849(2) & -3.7(2) & -5.6(5) 
 & 56258.000 
\\
J1909$-$3744 & 
284.220854447(3) & -15.15551279(1) & -13.863(3) & -34.32(2) 
 & 55339.000 
\\
J1910+1256 & 
291.04141414(3) & 35.10722180(4) & -0.79(5) & -7.25(7) 
 & 56131.000 
\\
J1911+1347 & 
291.71692634(1) & 35.88643155(1) & -3.35(7) & -3.07(8) 
 & 56936.000 
\\
J1918$-$0642 & 
290.31463749(1) & 15.35106180(4) & -7.91(1) & -4.92(5) 
 & 55330.000 
\\
J1923+2515 & 
297.98095097(2) & 46.69620142(3) & -9.74(4) & -12.43(8) 
 & 56583.000 
\\
B1937+21 & 
301.973244534(8) & 42.296752337(9) & -0.018(7) & -0.40(1) 
 & 55321.000 
\\
J1944+0907 & 
299.99545386(2) & 29.89101931(3) & 9.20(4) & -25.10(9) 
 & 56570.000 
\\
B1953+29 & 
309.69134497(5) & 48.68454566(5) & -2.3(1) & -3.7(2) 
 & 56568.000 
\\
J2010$-$1323 & 
301.924487764(9) & 6.49094711(9) & 1.23(2) & -6.4(2) 
 & 56235.000 
\\
J2017+0603 & 
308.26118074(2) & 25.04449436(4) & 2.18(6) & -0.4(2) 
 & 56682.000 
\\
J2033+1734 & 
316.29009241(7) & 35.06284854(8) & -8.6(4) & -7.6(5) 
 & 56945.000 
\\
J2043+1711 & 
318.868484758(6) & 33.96432304(1) & -8.83(1) & -8.49(2) 
 & 56573.000 
\\
J2145$-$0750 & 
326.02461737(4) & 5.3130542(5) & -12.07(5) & -4.2(5) 
 & 55322.000 
\\
J2214+3000 & 
348.80914233(4) & 37.71314985(5) & 17.8(1) & -10.4(2) 
 & 56610.000 
\\
J2229+2643 & 
350.69563878(8) & 33.29017455(6) & -4.3(4) & -4.2(7) 
 & 56937.000 
\\
J2234+0611 & 
342.60523286(1) & 14.07943341(5) & 27.3(1) & -1.1(3) 
 & 57026.000 
\\
J2234+0944 & 
344.11902092(3) & 17.31858876(9) & -6.0(1) & -32.2(4) 
 & 56917.000 
\\
J2302+4442 & 
9.78043764(5) & 45.66543490(5) & -3.3(1) & -4.9(2) 
 & 56675.000 
\\
J2317+1439 & 
356.12940553(2) & 17.68023059(6) & 0.20(1) & 3.74(4) 
 & 54977.000 
\\
\enddata
\tablecomments{Numbers in parentheses are uncertainties in last digit quoted.  Epochs are exact integer dates.}
\end{deluxetable*}
 
\begin{deluxetable*}{lllDDc}
\tabletypesize{\scriptsize}
\tablecaption{Positions and Proper Motions in Equatorial Coordinates
\label{tab:position_equatorial}}
\tablecolumns{7}
\tablewidth{0pt}
\tablehead{
\multicolumn{1}{c}{Pulsar}
          & \multicolumn{1}{c}{$\alpha$} &  \multicolumn{1}{c}{$\delta$}  &  \multicolumn{2}{c}{$\mu_\alpha\equiv\dot{\alpha}\cos\delta$}  &  \multicolumn{2}{c}{$\mu_\delta$}  & \multicolumn{1}{c}{Epoch}\\
 &  & & \multicolumn{2}{c}{(mas\,yr$^{-1}$)} & \multicolumn{2}{c}{(mas\,yr$^{-1}$)} &  \multicolumn{1}{c}{(MJD)}
}
\decimals
\startdata
J0023+0923 & 
 00:23:16.87821(1) & \phs09:23:23.8646(3) & -12.6(2) & -5.8(3)
 & 56567.000 
\\
J0030+0451 & 
 00:30:27.42785(4) & \phs04:51:39.711(1) & -6.2(1) & 0.5(3)
 & 55390.000 
\\
J0340+4130 & 
 03:40:23.28816(1) & \phs41:30:45.2862(3) & -0.5(1) & -3.3(3)
 & 56675.000 
\\
J0613$-$0200 & 
 06:13:43.975825(4) & $-$02:00:47.2372(1) & 1.85(2) & -10.35(4)
 & 55413.000 
\\
J0636+5128 & 
 06:36:04.847128(6) & \phs51:28:59.9609(1) & 3.5(2) & -2.3(2)
 & 57002.000 
\\
J0645+5158 & 
 06:45:59.082079(5) & \phs51:58:14.91290(6) & 1.53(4) & -7.41(6)
 & 56534.000 
\\
J0740+6620 & 
 07:40:45.79492(2) & \phs66:20:33.5593(2) & -10.3(2) & -31.0(2)
 & 57017.000 
\\
J0931$-$1902 & 
 09:31:19.11739(1) & $-$19:02:55.0282(3) & -2.4(2) & -4.4(4)
 & 56864.000 
\\
J1012+5307 & 
 10:12:33.43776(1) & \phs53:07:02.2801(1) & 2.66(3) & -25.50(4)
 & 55291.000 
\\
J1024$-$0719 & 
 10:24:38.667358(6) & $-$07:19:19.5974(2) & -35.29(6) & -48.2(1)
 & 56239.000 
\\
J1125+7819 & 
 11:25:59.8485(1) & \phs78:19:48.7161(3) & 28.3(8) & -1.(1)
 & 57017.000 
\\
J1453+1902 & 
 14:53:45.71922(3) & \phs19:02:12.1270(8) & 0.3(7) & -11.(2)
 & 56936.000 
\\
J1455$-$3330 & 
 14:55:47.97035(2) & $-$33:30:46.3818(6) & 7.98(8) & -2.0(2)
 & 55293.000 
\\
J1600$-$3053 & 
 16:00:51.903178(3) & $-$30:53:49.3919(1) & -0.98(2) & -7.10(6)
 & 55885.000 
\\
J1614$-$2230 & 
 16:14:36.50741(2) & $-$22:30:31.265(1) & 3.8(1) & -32.5(7)
 & 56047.000 
\\
J1640+2224 & 
 16:40:16.745013(3) & \phs22:24:08.82970(6) & 2.08(1) & -11.33(2)
 & 55366.000 
\\
J1643$-$1224 & 
 16:43:38.16189(2) & $-$12:24:58.671(2) & 5.9(1) & 3.7(5)
 & 55330.000 
\\
J1713+0747 & 
 17:13:49.5335505(5) & \phs07:47:37.48838(1) & 4.926(2) & -3.916(4)
 & 55391.000 
\\
J1738+0333 & 
 17:38:53.968032(5) & \phs03:33:10.8893(2) & 7.07(5) & 5.1(1)
 & 56258.000 
\\
J1741+1351 & 
 17:41:31.144731(2) & \phs13:51:44.12188(4) & -8.98(2) & -7.42(2)
 & 56209.000 
\\
J1744$-$1134 & 
 17:44:29.408577(3) & $-$11:34:54.7022(2) & 18.80(1) & -9.29(6)
 & 55292.000 
\\
J1747$-$4036 & 
 17:47:48.71652(4) & $-$40:36:54.784(2) & -1.3(4) & -2.(1)
 & 56676.000 
\\
J1832$-$0836 & 
 18:32:27.592888(3) & $-$08:36:55.0115(1) & -7.97(5) & -21.2(2)
 & 56862.000 
\\
J1853+1303 & 
 18:53:57.318327(4) & \phs13:03:44.05670(7) & -1.65(4) & -2.89(6)
 & 56553.000 
\\
B1855+09 & 
 18:57:36.390442(3) & \phs09:43:17.20167(8) & -2.66(1) & -5.41(2)
 & 55367.000 
\\
J1903+0327 & 
 19:03:05.79256(2) & \phs03:27:19.1851(9) & -3.0(2) & -6.0(5)
 & 56258.000 
\\
J1909$-$3744 & 
 19:09:47.432840(1) & $-$37:44:14.54898(5) & -9.516(4) & -35.77(1)
 & 55339.000 
\\
J1910+1256 & 
 19:10:09.701512(6) & \phs12:56:25.4648(1) & 0.28(5) & -7.29(7)
 & 56131.000 
\\
J1911+1347 & 
 19:11:55.203652(2) & \phs13:47:34.36424(5) & -2.85(6) & -3.54(8)
 & 56936.000 
\\
J1918$-$0642 & 
 19:18:48.032707(3) & $-$06:42:34.8948(2) & -7.15(2) & -5.97(5)
 & 55330.000 
\\
J1923+2515 & 
 19:23:22.492681(4) & \phs25:15:40.59748(9) & -6.96(5) & -14.17(7)
 & 56583.000 
\\
B1937+21 & 
 19:39:38.561253(2) & \phs21:34:59.12518(3) & 0.073(7) & -0.39(1)
 & 55321.000 
\\
J1944+0907 & 
 19:44:09.330945(4) & \phs09:07:23.0118(1) & 14.06(4) & -22.73(9)
 & 56570.000 
\\
B1953+29 & 
 19:55:27.875424(9) & \phs29:08:43.4415(2) & -1.1(1) & -4.2(2)
 & 56568.000 
\\
J2010$-$1323 & 
 20:10:45.921236(5) & $-$13:23:56.0854(3) & 2.59(5) & -6.0(2)
 & 56235.000 
\\
J2017+0603 & 
 20:17:22.705247(5) & \phs06:03:05.5689(2) & 2.22(7) & 0.1(1)
 & 56682.000 
\\
J2033+1734 & 
 20:33:27.51189(2) & \phs17:34:58.4747(3) & -5.9(4) & -9.9(4)
 & 56945.000 
\\
J2043+1711 & 
 20:43:20.881730(1) & \phs17:11:28.91265(3) & -5.72(1) & -10.84(2)
 & 56573.000 
\\
J2145$-$0750 & 
 21:45:50.46014(4) & $-$07:50:18.499(2) & -10.0(2) & -8.0(5)
 & 55322.000 
\\
J2214+3000 & 
 22:14:38.85274(1) & \phs30:00:38.1953(2) & 20.6(1) & -1.3(1)
 & 56610.000 
\\
J2229+2643 & 
 22:29:50.88471(2) & \phs26:43:57.6507(2) & -2.1(6) & -5.7(5)
 & 56937.000 
\\
J2234+0611 & 
 22:34:23.074172(5) & \phs06:11:28.6922(2) & 25.6(2) & 9.4(3)
 & 57026.000 
\\
J2234+0944 & 
 22:34:46.85388(1) & \phs09:44:30.2487(3) & 6.9(2) & -32.0(4)
 & 56917.000 
\\
J2302+4442 & 
 23:02:46.97874(1) & \phs44:42:22.0860(2) & -0.0(1) & -5.9(2)
 & 56675.000 
\\
J2317+1439 & 
 23:17:09.236663(8) & \phs14:39:31.2556(2) & -1.36(2) & 3.49(4)
 & 54977.000 
\\
\enddata
\tablecomments{Numbers in parentheses are uncertainties in last digit quoted.  Epochs are exact integer dates.}
\end{deluxetable*}
 
\subsection{Parallax Measurements with Significant Detections}\label{sec:px_detections}

Measured timing parallax values are listed in Table \ref{tab:px}, along with a selection of previous parallax measurements using
timing and other techniques.  
Of the 45 pulsars, 20 have significant timing parallaxes ($3\sigma$ or greater significance).  Three of these are the first parallax measurements for these sources (PSRs J0740+6620, J2334+0611, and J2234+0944), and many of the others are improvements on previous values.  

For all these pulsars, we calculated distance measurements in the same manner as outlined in M16. In brief, the central value, upper limit, and lower limit given in Table \ref{tab:px} were calculated via $d=\varpi^{-1}$, where $\varpi$ was the 84\%, 50\%, and 16\% point in the measured parallax distribution corresponding to the 16\%, 50\%, and 84\% points in the distance distribution, respectively. This is done to reflect the asymmetry in the hyperbolic distance distribution about the median value.  

Three of the parallax measurements in Table~\ref{tab:px}
(PSRs~J1713+0747, J1741+1351, and J1909$-$3744) are discrepant by 2$\sigma$ or more with previously published values.  
A full investigation into discrepancies is beyond the scope of this work. However, differing treatments of DM among authors may be a contributing factor, as was explored in M16, and differences in noise models may also influence the 
measurements. This is because DM variation and timing noise can both be covariant with the timing signature of the
parallax signal, which is approximately a six-month sinusoidal pattern in pulse arrival times.
 
Two pulsars with both timing and interferometric astrometry are 
discussed further in \S\ref{sec:vlbi}.

\begin{figure}
  \centering
  \includegraphics[scale=0.45]{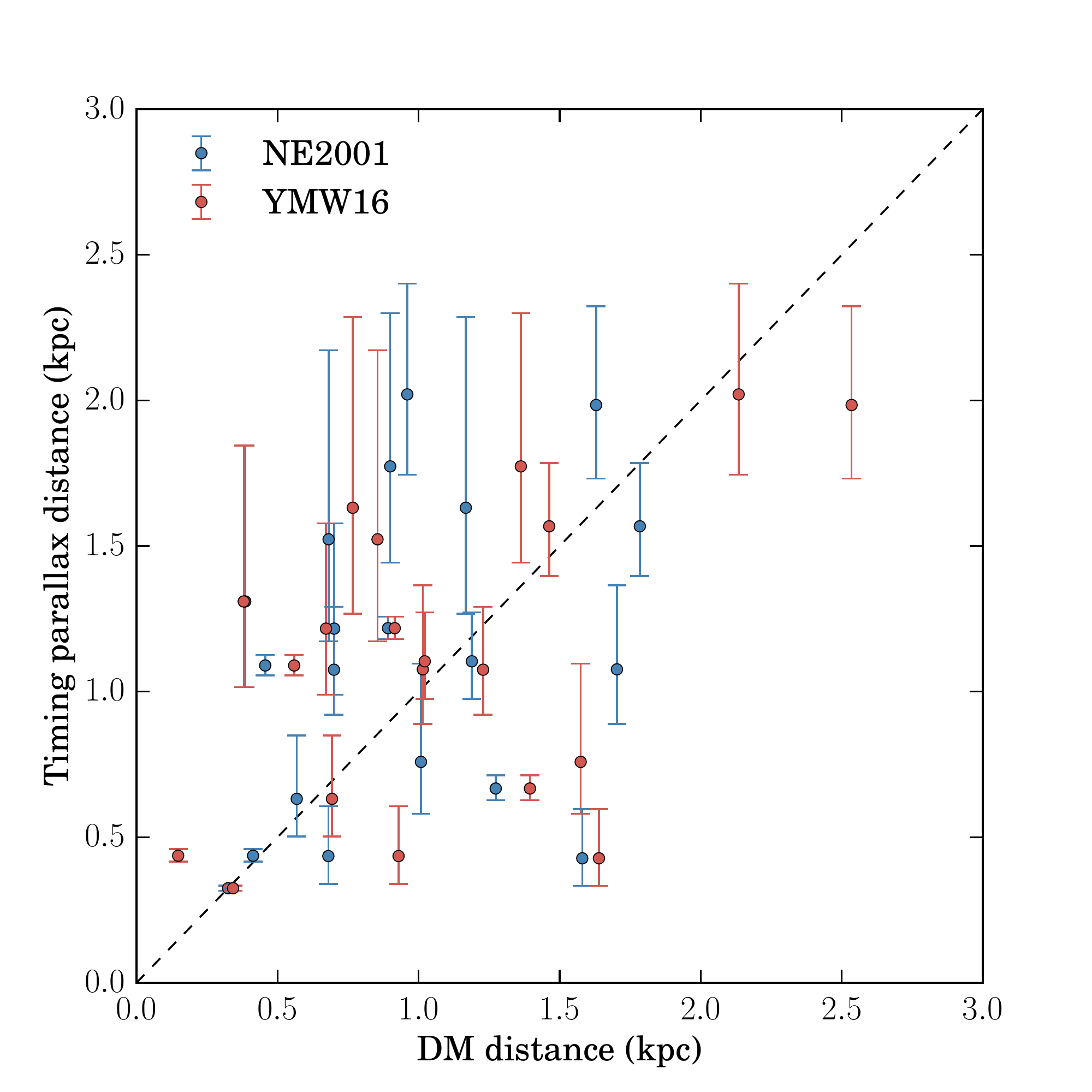}
  \caption{Distances from timing parallax measurements (67\% confidence) versus distances from DM models for pulsars with significant timing parallax detections. The blue points show the NE2001 DM model distances while the red points show the YMW16 model distances. The black dashed line shows a one-to-one relation.\label{fig:dm_vs_px}}
\end{figure}

We compared distances derived from our parallax measurements with distances derived from DMs and models of the Galactic electron-density distribution. For the electron-density models, we used both the NE2001 model \citep{ne2001} and the YMW16 model \citep{ymw16}.  The result of this comparison for both models is given in Figure~\ref{fig:dm_vs_px}. We calculated a simple reduced-chi-square statistic for each DM model, finding $\chi^2\sim 14.0$ for the YMW16 model and $\sim 10.2$ for the NE2001 model.  (The $\chi^2$ statistic accounted only for uncertainties in the parallax distances, not in the electron-density models.)  The similarity in these $\chi^2$ values suggests the two models are comparable in their ability to predict distances to pulsars based on DM values, at least in the distance regime probed by millisecond pulsar timing parallax measurements (within $\sim 2.5$~kpc of the Sun).

\subsection{Parallax Upper Limits}

The remaining 25 pulsars had timing parallax measurements with significance of less than $3\sigma$.  Of these, 22 had positive parallaxes and only 3 had negative parallaxes in the formal timing fits.  This is strong evidence that most of the non-detections have a true low-level parallax signal; otherwise, there would be comparable numbers of negative and positive parallax measurements.  None of the negative parallaxes had large significance.

We used 95\%-confidence upper limits on parallax to compute 95\%-confidence lower limits on distance for these pulsars. The details of this calculation are outlined in M16. We have not attempted to correct for Lutz--Kelker bias \citep{Verbiest2010}, as we do not believe we have enough prior information on the spatial and luminosity distribution of the millisecond pulsar population to accurately correct for this bias (see M16).  For the lower limits on distance, our 95\% limit remains a conservative approach, in the sense that compensation for Lutz--Kelker bias would only push these limits higher.

Among the pulsars with parallax upper limits, a few have
significant discrepancies with previous measurements 
(PSRs~J0636+5128, J1455$-$3330, J1640+1224, B1937+21).  As with the pulsars discussed in
\S\ref{sec:px_detections}, the source of these discrepancies are not clear, but
may result from differences in DM variation models or red-noise models.

\subsection{Distance Constraints from Rotation and Orbital Period Derivatives}\label{sec:pdotpbdot}

Observed pulsar rotation period derivatives are a combination of intrinsic
pulsar spin-down and kinematic terms due to the acceleration and transverse
motion of the pulsar relative to the Sun \citep{shk70,nt95}.  By assuming that
the pulsar is losing rotational energy, so that it has a positive rotation
period derivative, an upper limit can be placed on the pulsar distance.  See M16
\S 4.3 for details.  
Table~\ref{tab:px} lists such ``$\dot{P}$ distance''
constraints, which we calculated following the procedures of M16.  In the table,
we omitted distance limits greater than 10~kpc as physically uninteresting, and
we did not calculate a constraint for PSR J1024$-$0719, as its observed rotation
period derivative is biased by orbital motion \citep{Kaplan2016}.

Similarly, in binary pulsars, observed orbital period derivatives can be used to
place upper limits on distances.  Measured orbital period derivatives are a
combination of intrinsic orbital period changes, due to relativistic orbital
decay or other phenomena, and acceleration and
transverse motion of the binary system \citep{Damour1991}.  Following the procedures of M16 \S 4.2,
we calculated orbital period derivative distance constraints for binary pulsars
in this work.  
To estimate the relativistic orbital period derivatives
in these calculations, we followed M16 and assumed negligible relativistic
orbital decay from wide binaries; we used masses and orbital geometry
constraints from the present work for tight binaries; and we used independently
determined mass and geometry constraints for one pulsar.   
We omitted pulsars
with weakly constrained proper motions (less than 5$\sigma$ measurement in
either component) and pulsars likely to be black widow systems, in which
intrinsic orbital period variability can arise from non-relativistic
sources.
Table~\ref{tab:px} lists the ``$\dot{P_{\rm b}}$ distance'' of each
pulsar for which we could derive distance measurements or
interesting limits (less than 10~kpc) using this method.

Throughout the table, measurements are 67\% confidence, but limits are a more conservative 95\% confidence.

The distance constraints found by these methods are consistent with distances measured by timing parallax.  In one case, PSR J1909$-$3744, the orbital period derivative distance, $1.103\pm 0.011$~kpc, is more precise than the parallax distance, $1.09^{+0.04}_{-0.03}$~kpc.

\startlongtable
\begin{deluxetable*}{lcclcccc}
\tablewidth{0pt}
\tabletypesize{\scriptsize}
\tablecaption{Timing Parallax Measurements and Distance Estimates\label{tab:px}}
\tablehead{\colhead{PSR} & \colhead{Timing Parallax} & \colhead{Distance} & \multicolumn{3}{c}{Selected Previous Measurements} &\colhead{$\dot{P}$ Distance} & \colhead{$\dot{P_{\rm b}}$ Distance}  \\[1pt]
\cline{4-6}
 & \colhead{(mas)} & \colhead{(kpc)} & \colhead{Parallax} & \colhead{Reference \rule{0pt}{10pt}} & \colhead{Type$^{1}$} & \colhead{(kpc)} & \colhead{(kpc)}
\\
}
\startdata
\multicolumn{8}{c}{Timing Parallax Detections ($>3\sigma$) and Distances} \\[2pt]
\cline{1-8}
J0023+0923    &  \phm{$-$}$\phn0.9(2) \phn$  & \phm{ }\phm{${>}$}$1.1^{+0.2\phn\phn}_{-0.2\phn\phn}\phn\phn $    & \phm{$-$}$0.4(3)$   & \cite{Matthews2016}  & T       &$ {<}       7.3$   &   \nodata  \\  
J0030+0451    &  \phm{$-$}$\phn3.08(8)$      & \phm{ }\phm{${>}$}$0.325^{+0.009}_{-0.009}$                     & \phm{$-$}$3.3(2)$   & \cite{Matthews2016}  & T       &   \nodata         &   \nodata  \\    
J0613$-$0200  &  \phm{$-$}$\phn0.9(2) \phn$  & \phm{ }\phm{${>}$}$1.1^{+0.3\phn\phn}_{-0.2\phn\phn}\phn\phn$     & \phm{$-$}$0.9(1)$   & \cite{Reardon2016}   & T       &     \nodata        &   1.8(8)   \\  
              &                              &                                                               & \phm{$-$}$1.3(1)$   & \cite{Desvignes2016} & T       &                   &            \\
              &                              &                                                               & \phm{$-$}$0.9(2)$   & \cite{Matthews2016}  & T       &                   &            \\    
J0645+5158    &  \phm{$-$}$\phn0.8(2) \phn$  & \phm{ }\phm{${>}$}$1.2^{+0.4\phn\phn}_{-0.2\phn\phn}\phn\phn$     & \phm{$-$}$1.3(3)$   & \cite{Matthews2016}  & T       &$ {<}       3.4$   &   \nodata  \\  
J0740+6620    &  \phm{$-$}$\phn2.3(6) \phn$  & \phm{ }\phm{${>}$}$0.4^{+0.2\phn\phn}_{-0.1\phn\phn}\phn\phn$     & \phm{$-$}\nodata    & \nodata              & \nodata &$ {<}       1.7$   &   \nodata  \\      
J1024$-$0719  &  \phm{$-$}$\phn0.8(2) \phn$  & \phm{ }\phm{${>}$}$1.3^{+0.5\phn\phn}_{-0.3\phn\phn}\phn\phn$     & \phm{$-$}$0.8(1)$   & \cite{Bassa2016}     & T       &\nodata            &   \nodata  \\     
              &                              &                                                               & \phm{$-$}$0.6(3)$   & \cite{Matthews2016}  & T       &                   &            \\    
J1600$-$3053  &  \phm{$-$}$\phn0.50(7)$      & \phm{ }\phm{${>}$}$2.0^{+0.3\phn\phn}_{-0.3\phn\phn}\phn\phn$     & \phm{$-$}$0.64(7)$  &\cite{Desvignes2016}  & T       &\nodata            &   \nodata  \\ 
              &                              &                                                               & \phm{$-$}$0.34(9)$   & \cite{Matthews2016}  & T       &                   &            \\    
J1614$-$2230  &  \phm{$-$}$\phn1.5(1) \phn$  & \phm{ }\phm{${>}$}$0.67^{+0.05\phn}_{-0.04\phn}\phn$             & \phm{$-$}$1.5(1) $  & \cite{Matthews2016}  & T       &$ {<}       1.3$   &   0.85(11) \\  
              &                              &                                                               & \phm{$-$}$1.30(9)$  & \cite{Guillemot2016} & T       &                   &            \\
J1713+0747    &  \phm{$-$}$\phn0.82(3)$      & \phm{ }\phm{${>}$}$1.22^{+0.04\phn}_{-0.04\phn}\phn$             & \phm{$-$}$0.84(9)$  & \cite{Reardon2016}   & T       &\nodata            &   \nodata  \\   
              &                              &                                                               & \phm{$-$}$0.90(3)$  & \cite{Desvignes2016} & T       &                   &            \\
              &                              &                                                               & \phm{$-$}$0.85(3)$  & \cite{Matthews2016}  & T       &                   &            \\
              &                              &                                                               & \phm{$-$}$0.95(6)$  & \cite{Chatterjee2009}& V       &                   &            \\
J1741+1351    &  \phm{$-$}$\phn0.6(1) \phn$  & \phm{ }\phm{${>}$}$1.8^{+0.5\phn\phn}_{-0.3\phn\phn}\phn\phn$     & \phm{$-$}$0.93(4)$  & \cite{Espinoza2013}  & T       &\nodata            &   \nodata  \\  
              &                              &                                                               & \phm{$-$}$0.0(5)$   & \cite{Matthews2016}  & T       &                  &            \\
J1744$-$1134  &  \phm{$-$}$\phn2.3(1) \phn$  & \phm{ }\phm{${>}$}$0.44^{+0.02\phn}_{-0.02\phn}\phn$             & \phm{$-$}$2.38(8)$  & \cite{Desvignes2016} & T       &$ {<}       1.9$   &   \nodata  \\ 
              &                              &                                                               & \phm{$-$}$2.53(7)$  & \cite{Reardon2016}   & T       &                   &            \\     
              &                              &                                                               & \phm{$-$}$2.4(1)$   & \cite{Matthews2016}  & T       &\nodata           &   \nodata  \\        
B1855+09       &  \phm{$-$}$\phn0.6(2) \phn$  & \phm{ }\phm{${>}$}$1.6^{+0.7\phn\phn}_{-0.4\phn\phn}\phn\phn$     & \phm{$-$}$0.7(3)$  & \cite{Desvignes2016} & T       &\nodata            & \nodata    \\  
              &                              &                                                               & \phm{$-$}$0.5(3)$   & \cite{Reardon2016}   & T       &                   &            \\      
              &                              &                                                               & \phm{$-$}$0.3(2)$   & \cite{Matthews2016}  & T       &\nodata        &   \nodata  \\    
J1909$-$3744  &  \phm{$-$}$\phn0.92(3)$      & \phm{ }\phm{${>}$}$1.09^{+0.04\phn}_{-0.03\phn}\phn$             & \phm{$-$}$0.810(3)$ & \cite{Reardon2016}   & T       &$ {<}       1.4$   &  1.103(11) \\   
              &                              &                                                               & \phm{$-$}$0.87(2)$  & \cite{Desvignes2016} & T       &                   &            \\   
              &                              &                                                               & \phm{$-$}$0.94(3)$  & \cite{Matthews2016}  & T       &                  &   \nodata  \\
J1918$-$0642  &  \phm{$-$}$\phn0.9(1) \phn$  & \phm{ }\phm{${>}$}$1.1^{+0.2\phn\phn}_{-0.1\phn\phn}\phn\phn$     & \phm{$-$}$1.1(2)$   & \cite{Matthews2016}  & T       &\nodata            &   ${<}8.2$ \\   
J2043+1711    &  \phm{$-$}$\phn0.64(8)$      & \phm{ }\phm{${>}$}$1.6^{+0.2\phn\phn}_{-0.2\phn\phn}\phn\phn$     & \phm{$-$}$0.8(2)$   & \cite{Matthews2016}  & T       &$ {<}       7.7$   &   ${<}5.1$ \\   
J2145$-$0750  &  \phm{$-$}$\phn1.6(4) \phn$  & \phm{ }\phm{${>}$}$0.6^{+0.2\phn\phn}_{-0.1\phn\phn}\phn\phn$     & \phm{$-$}$1.63(4)$  & \cite{Deller2016}    & V       &$ {<}       4.7$   &   ${<}1.3$ \\    
              &                              &                                                               & \phm{$-$}$1.3(2)$   & \cite{Matthews2016}  & T       &\nodata           &   \nodata  \\
J2214+3000    &  \phm{$-$}$\phn2.3(7) \phn$  & \phm{ }\phm{${>}$}$0.4^{+0.2\phn\phn}_{-0.1\phn\phn}\phn\phn$     & \phm{$-$}$1.7(9)$   & \cite{Guillemot2016} & T       &$ {<}       4.9$   &   \nodata  \\
              &                              &                                                               & \phm{$-$}$1\phm{.}(1)$   & \cite{Matthews2016}  & T       &\nodata           &   \nodata  \\ 
J2234+0611    &  \phm{$-$}$\phn0.7(2) \phn$  & \phm{ }\phm{${>}$}$1.5^{+0.6\phn\phn}_{-0.4\phn\phn}\phn\phn$     & \phm{$-$}\nodata    & \nodata              & \nodata &$ {<}       1.8$   &   \nodata  \\
J2234+0944    &  \phm{$-$}$\phn1.3(4) \phn$  & \phm{ }\phm{${>}$}$0.8^{+0.3\phn\phn}_{-0.2\phn\phn}\phn\phn$     & \phm{$-$}\nodata    & \nodata              & \nodata &$ {<}       2.2$   &   \nodata  \\
J2317+1439    &  \phm{$-$}$\phn0.50(8)$      & \phm{ }\phm{${>}$}$2.0^{+0.4\phn\phn}_{-0.3\phn\phn}\phn\phn$     & \phm{$-$}$0.7(2)$   & \cite{Matthews2016}  & T       &  \nodata          &   ${<}6.4$  \\
\cline{1-8}
\multicolumn{8}{c}{Timing Parallax Non-detections ($<3\sigma$) and Distance Lower Limits} \\[2pt]
\cline{1-8}
J0340+4130    &  \phm{$-$}$\phn0.7(4)\phn$   & \phm{ }${>}0.7 \phn\phn $\phm{$^{+0.009}$} & \phm{$-$}$0.7(7)$       & \cite{Matthews2016} & T       &\nodata            &   \nodata  \\
J0636+5128    &  \phm{$-$}$\phn0.9(3)\phn$   & \phm{ }${>}0.7 \phn\phn $\phm{$^{+0.009}$} & \phm{$-$}$4.9(6)$       & \cite{Stovall2014}  & \nodata &\nodata            &   \nodata  \\
J0931$-$1902  &  \phm{$-$}$\phn1.2(9)\phn$   & \phm{ }${>}0.4 \phn\phn $\phm{$^{+0.009}$} & \phm{$-$}$8\phm{.}(8)$  & \cite{Matthews2016} & T       &\nodata   &   \nodata  \\
J1012+5307    &  \phm{$-$}$\phn1.3(4)\phn$   & \phm{ }${>}0.5 \phn\phn $\phm{$^{+0.009}$} & \phm{$-$}$0.7(2)$       & \cite{Desvignes2016}& T       &$ {<}       2.1$   &   1.2(2)   \\
J1125+7819    &  \phm{$-$}$11.(8)\phn$      & \phm{ }${>}0.04    \phn $\phm{$^{+0.009}$} & \phm{$-$}\nodata        & \nodata             & \nodata &$ {<}       1.0$   &   \nodata  \\
J1453+1902    &  \phn$-3\phm{.}(2)\phn\phn$  & \phm{ }${>}0.6 \phn\phn $\phm{$^{+0.009}$} & \phm{$-$}\nodata        & \nodata             & \nodata &\nodata          &   \nodata  \\
J1455$-$3330  &  \phn$-0.1(4) \phn$          & \phm{ }${>}1.4 \phn\phn $\phm{$^{+0.009}$} & \phm{$-$}$1.0(2)$       & \cite{Guillemot2016}& T       &\nodata          &   \nodata  \\
              &                              &                                           & \phm{$-$}$0.2(6)$       & \cite{Matthews2016} & T       &   &            \\    
J1640+2224    &  \phm{$-$}$\phn0.2(4) \phn$  & \phm{ }${>}1.0 \phn\phn $\phm{$^{+0.009}$} & $-1.0(6)$               & \cite{Matthews2016} & T       &$ {<}       3.4$            &   \nodata  \\
J1643$-$1224  &  \phm{$-$}$\phn0.7(9) \phn$  & \phm{ }${>}0.4 \phn\phn $\phm{$^{+0.009}$} & \phm{$-$}$1.2(2)$       & \cite{Desvignes2016}& T       &\nodata            &   \nodata  \\
              &                              &                                           & \phm{$-$}$1.3(2)$       & \cite{Reardon2016}  & T       &                  &            \\
              &                              &                                           & \phm{$-$}$0.7(6)$       & \cite{Matthews2016}  & T       &                  &            \\              
J1738+0333    &  \phm{$-$}$\phn0.4(3) \phn$  & \phm{ }${>}1.2 \phn\phn $\phm{$^{+0.009}$} & \phm{$-$}$0.68(5)$      & \cite{Freire2012}   & T       & \nodata       &   \nodata  \\
              &                              &                                           & \phm{$-$}$0.4(5)$       & \cite{Matthews2016}  & T       &                  &            \\              
J1747$-$4036  &  \phm{$-$}$\phn0\phm{.}(1)\phn\phn$  & \phm{ }${>}0.4 \phn\phn $\phm{$^{+0.009}$} & $-0.4(7)$       & \cite{Matthews2016} & T       & \nodata &   \nodata  \\
J1832$-$0836  &  \phm{$-$}$\phn0.4(1) \phn$  & \phm{ }${>}1.7 \phn\phn $\phm{$^{+0.009}$} & \phm{$-$}$5\phm{.}(5)$  & \cite{Matthews2016} & T       & $ {<}       2.4$ &   \nodata  \\
J1853+1303    &  \phm{$-$}$\phn0.4(2) \phn$  & \phm{ }${>}1.2 \phn\phn $\phm{$^{+0.009}$} & \phm{$-$}$1.0(6)$       & \cite{Gonzalez2011} & T       & \nodata &   \nodata  \\
              &                              &                                           & \phm{$-$}$0.1(5)$       & \cite{Matthews2016}  & T       &                  &            \\              
J1903+0327    &  \phm{$-$}$\phn0.2(9) \phn$  & \phm{ }${>}0.5 \phn\phn $\phm{$^{+0.009}$} & \phm{$-$}$0.4(8)$       & \cite{Matthews2016} & T       & \nodata &   \nodata  \\
J1910+1256    &  \phn$-0.4(4)\phn$           & \phm{ }${>}2.2 \phn\phn $\phm{$^{+0.009}$} & \phm{$-$}$1.4(7)$       & \cite{Desvignes2016}& T       & \nodata &   \nodata  \\
              &                              &                                           & $-0.3(7)$               & \cite{Matthews2016}  & T       &                  &            \\              
J1911+1347    &  \phm{$-$}$\phn0.4(2) \phn$  & \phm{ }${>}1.4 \phn\phn $\phm{$^{+0.009}$} & \phm{$-$}\nodata        & \nodata             & \nodata &\nodata &   \nodata  \\
J1923+2515    &  \phm{$-$}$\phn1.2(4) \phn$  & \phm{ }${>}0.5 \phn\phn $\phm{$^{+0.009}$} & \phm{$-$}$2\phm{.}(1)$  & \cite{Matthews2016} & T       &$ {<} \     5.0$ &   \nodata  \\
B1937+21       &  \phm{$-$}$\phn0.15(8)$      & \phm{ }${>}3.4 \phn\phn $\phm{$^{+0.009}$} & \phm{$-$}$0.22(8)$      & \cite{Desvignes2016}& T       &\nodata &   \nodata  \\
              &                              &                                           & \phm{$-$}$0.5(2)$       & \cite{Reardon2016}  & T       & &            \\
              &                              &                                           & \phm{$-$}$0.1(1)$       & \cite{Matthews2016}  & T       &                  &            \\              
J1944+0907    &  \phm{$-$}$0.5(3) $          & \phm{ }${>}1.1 \phn\phn $\phm{$^{+0.009}$} & \phm{$-$}$0.0(4)$       & \cite{Matthews2016} & T       &$ {<}       2.0$&   \nodata  \\
B1953+29       &  \phm{$-$}$\phn0\phm{.}(1) \phn\phn$  & \phm{ }${>}0.5 \phn\phn $\phm{$^{+0.009}$} & $-4\phm{.}(2)$ & \cite{Matthews2016} & T       &\nodata&   \nodata  \\
J2010$-$1323  &  \phm{$-$}$\phn0.3(1)\phn $  & \phm{ }${>}1.7 \phn\phn $\phm{$^{+0.009}$} & \phm{$-$}$0.1(2)$       & \cite{Matthews2016} & T       &\nodata &   \nodata  \\
J2017+0603    &  \phm{$-$}$\phn0.4(2)\phn $  & \phm{ }${>}1.3 \phn\phn $\phm{$^{+0.009}$} & \phm{$-$}$1.2(5)$       & \cite{Guillemot2016}& T       &\nodata &   \nodata  \\
              &                              &                                           & \phm{$-$}$0.4(3)$       & \cite{Matthews2016}  & T                          &   \        \\              
J2033+1734    &  \phm{$-$}$\phn0\phm{.}(1) \phn\phn$  & \phm{ }${>}0.5 \phn\phn $\phm{$^{+0.009}$} & \phm{$-$} \nodata & \nodata          & \nodata &$ {<}       9.2$ &   \nodata  \\
J2229+2643    &  \phm{$-$}$\phn0.8(6)\phn $  & \phm{ }${>}0.5 \phn\phn $\phm{$^{+0.009}$} & \phm{$-$} \nodata & \nodata                   & \nodata &  \nodata        &   \nodata  \\
J2302+4442    &  \phm{$-$}$\phn0\phm{.}(1) \phn\phn $  & \phm{ }${>}0.5 \phn\phn $\phm{$^{+0.009}$} & ${<}2.5$ & \cite{Guillemot2016}     & T       &\nodata &   \nodata  \\ 
              &                              &                                           & \phm{$-$}$2\phm{.}(2)$       & \cite{Matthews2016}  & T                          &            \\              
\enddata
\tablenotetext{1}{Timing measurement designated by ``T'', VLBI measurement by ``V.''}
\tablecomments{Values in parentheses denote the $1\sigma$ uncertainty in the preceding digit(s).}
\end{deluxetable*}

\subsection{Timing and Interferometric Astrometry}\label{sec:vlbi}

\cite{Deller2016} presented a comparison of astrometric measurements made via
very long baseline interferometry (VLBI) with measurements made using pulsar-timing
and noted some discrepancies.  Here, we discuss PSRs~J1713+0747 and J2145$-$0750,
the two pulsars in their discussion that are part of our work.
The timing and interferometric measurements were made using
different coordinate systems and different epochs of position measurements.
To facilitate comparison, we re-ran our timing analysis of these two
pulsars using equatorial coordinates and the
position epochs used by the interferometry analyses.  The results 
are summarized in Table~\ref{tab:vlbi_comparison}.  

For positions, a complete comparison requires careful accounting for the absolute reference
frame of each measurement, a subject that is beyond the scope of the present paper.
The published interferometric position analyses incorporated uncertainties in the tie to the 
ICRF2 reference frame, whereas our timing analyses simply report the
formal uncertainties relative to the reference frame of the ephemeris used
(DE436) without attempting to account for the accuracy with which its 
frame is tied to ICRF2; hence the smaller uncertainties on most timing position-parameter values.  In any case, the positions differ by, at most, a little 
more than $2\sigma$.

\begin{figure*}
\begin{center}
\includegraphics[width=3.50in]{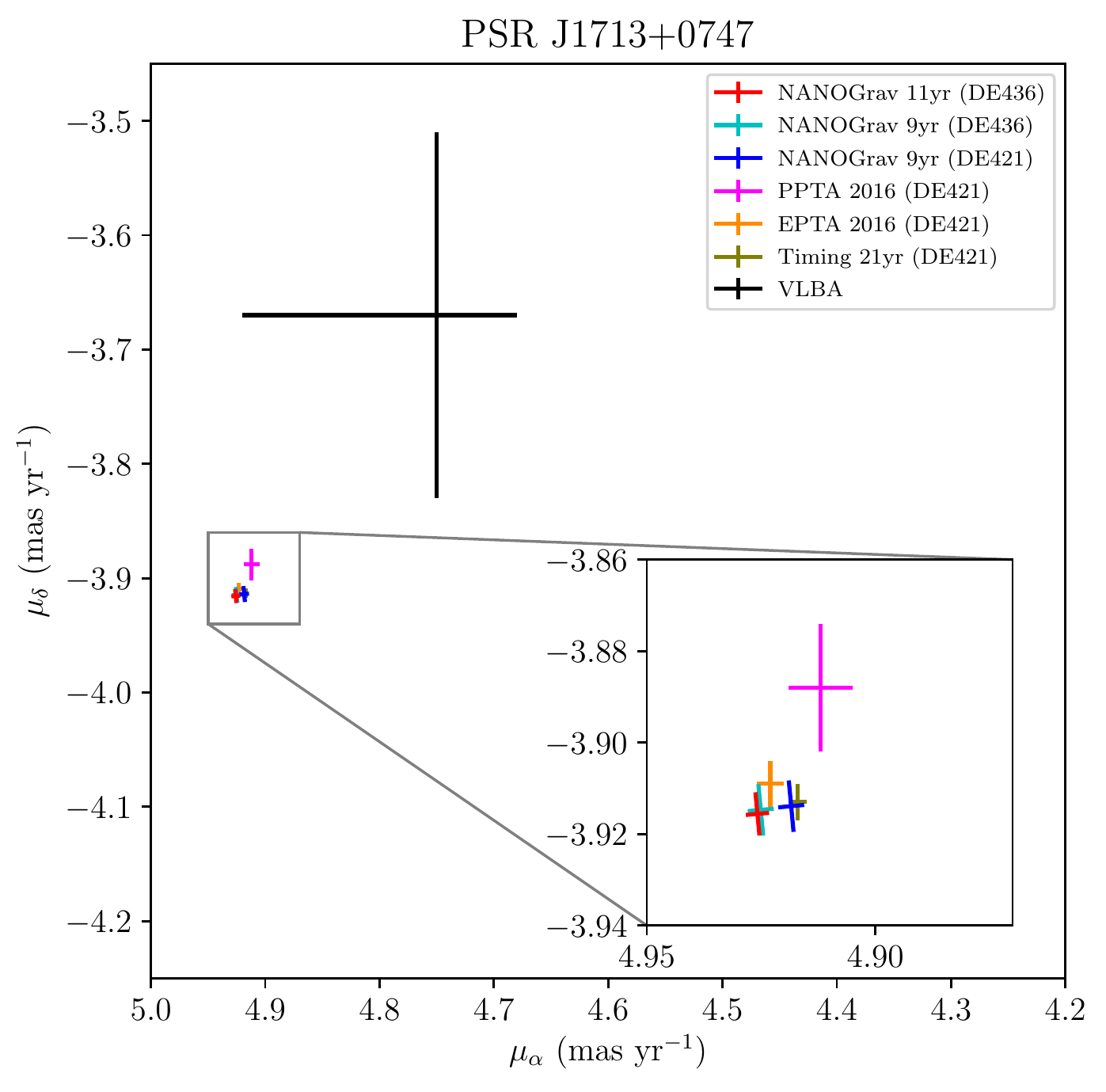}
\includegraphics[width=3.50in]{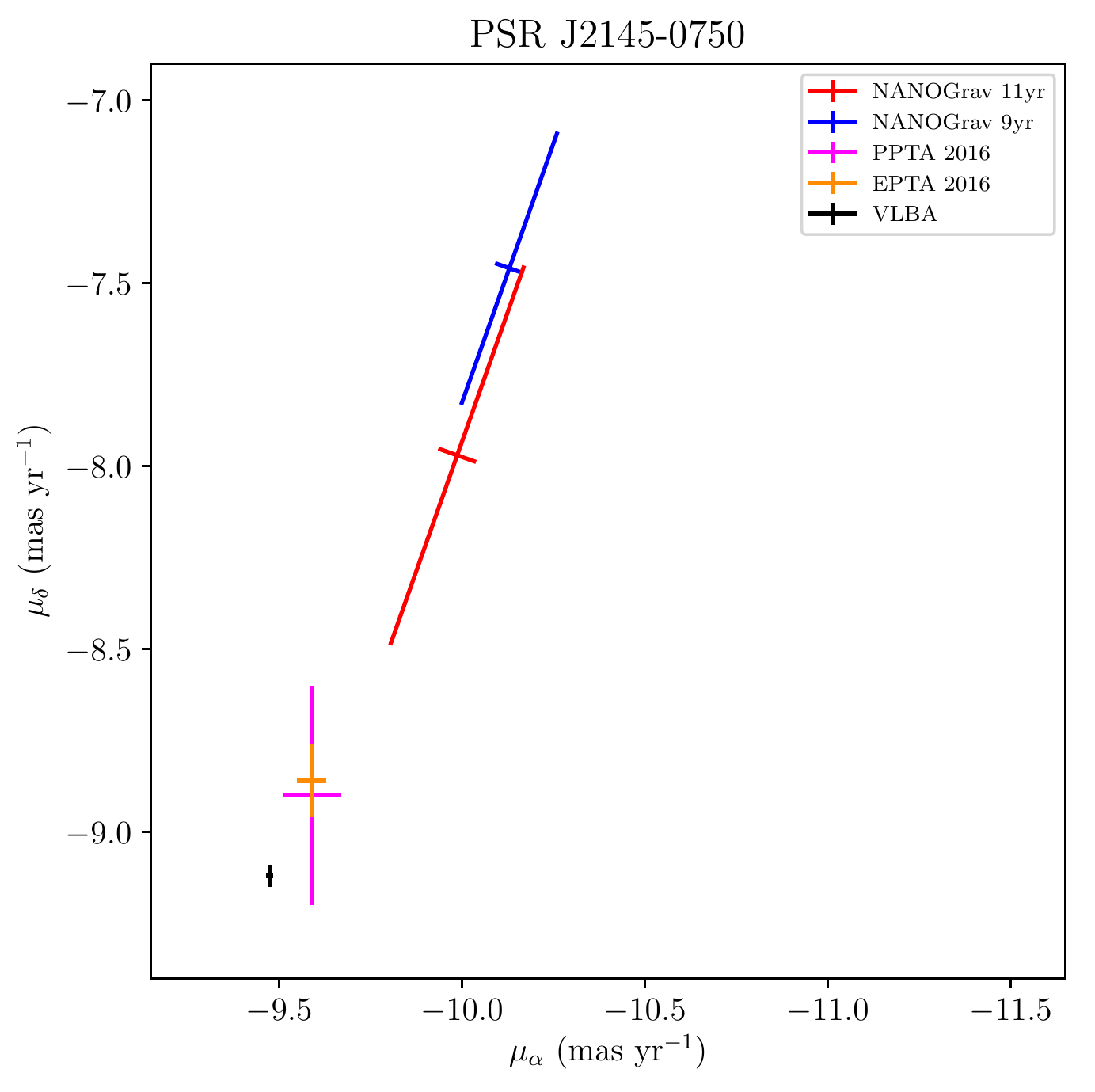}
\caption{\label{fig:propmo}Comparison of proper motion
measurements for PSRs J1713+0747 (left panel)
and J2145-0750 (right panel).  All uncertainties are 1$\sigma$.  For PSR J1713+0747 timing proper motions,
choice of solar system ephemeris is specified (DE421 or DE436);
this choice significantly affects the reported measurements, as can be seen
by comparing the NANOGrav 9-year values calculated using DE421 and DE436.  For J2145$-$0745, 
the choice of solar system ephemeris had a negligible influence compared to the
measurement uncertainties.
Measurements in the plots are from the following sources:
NANOGrav 11yr: this paper (ecliptic coordinate analysis); 
NANOGrav 9yr (DE421): \cite{Matthews2016};
NANOGrav 9yr (DE436): \cite{Matthews2016} re-analyzed for this paper using the JPL DE436 Ephemeris; 
PPTA 2016: \cite{Reardon2016}; 
EPTA 2016: \cite {Desvignes2016}; 
Timing 21yr: \cite{Zhu2015};
VLBA for J1713+0747: \cite{Chatterjee2009}; 
VLBA for J2145-0750: \cite{Deller2016}.
}
\end{center}
\end{figure*}

For proper motions, Figure~\ref{fig:propmo} gives a comparison of
interferometric values, our measured timing values, and other published timing
values for each pulsar.  For PSR~J1713+0747, the VLBI uncertainties are much
larger than the timing uncertainties.  Given the uncertainties in the plot,
there are no noteworthy disagreements between measurements.  As shown in the plot,
a marginal disagreement between the \nineyr\ value reported in M16 (measured
relative to ephemeris DE421) and our value (measured relative to ephemeris
DE436) is eliminated by reprocessing the \nineyr\ data using DE436.  This
highlights the importance of ephemeris reference frame choice at the level of
10 $\mu$as\,yr$^{-1}$. For PSR~J2145$-$0750, \cite{Deller2016} noted a $3\sigma-5\sigma$
discrepancy between their very-high-precision interferometric
proper motion and the \nineyr\ values reported in M16.  As shown in
Figure~\ref{fig:propmo}, that discrepancy is significantly reduced in our new
data set, which has a measured proper motion closer to the interferometric
value and which also has a larger uncertainty.  We suspect that the
improvement in the timing proper-motion accuracy, as well as its larger
uncertainty, is due to the adoption of a red-noise timing model for this
pulsar in the present work, whereas in \nineyr\ the noise was assumed to be
white.

For parallaxes, the timing and VLBI measurements for PSR~J2145$-$0750 agree within the uncertainties,
while the PSR~J1713+0747 measurements show marginal disagreement (2$\sigma$,
taking into account uncertainties in both measurements).  The cause of this
small disagreement is not known.  We note that, as Table~\ref{tab:px}
shows, several independent timing parallax measurements have been made for
PSR~J1713+0747, and all such measurements are less than the VLBI parallax value.

\begin{deluxetable*}{lllllll} 
\tablecaption{Comparison of VLBI and Timing Astrometric Parameters
\label{tab:vlbi_comparison}}
\tablehead{
\multicolumn{1}{c}{Measurement} &
\multicolumn{1}{c}{$\alpha$} &
\multicolumn{1}{c}{$\delta$} &
\multicolumn{1}{c}{$\mu_{\alpha}$} &
\multicolumn{1}{c}{$\mu_{\delta}$} &
\multicolumn{1}{c}{$\varpi$}  &
Reference
\\
\multicolumn{1}{c}{Technique}  & & & \multicolumn{1}{c}{(mas\,yr$^{-1}$)} & \multicolumn{1}{c}{(mas\,yr$^{-1}$)} & \multicolumn{1}{c}{(mas)}  
}
\startdata
\multicolumn{7}{c}{\rule{0pt}{12pt}PSR J1713+0747; Epoch MJD 52275.000} \\[2pt]
\tableline
VLBI           & 17:13:49.5306(1)             & 07:47:37.519(2)\phn\phn\phn & 4.75$^{+0.17}_{-0.07}$ & $-3.67^{+0.16}_{-0.15}$ & 0.95$^{+0.06}_{-0.05}$ & \cite{Chatterjee2009} \\
Timing         & 17:13:49.5308237(5)          & 07:47:37.51969(15)          & 4.926(2)               & $-3.916(4)$             & 0.82(3)                & This paper \\[2pt]
\tableline
\multicolumn{7}{c}{\rule{0pt}{12pt}PSR J2145-0751; Epoch MJD 56000.000} \\[2pt]
\cline{1-7}
\tableline
VLBI           &  21:45:50.4588(1)      & $-$07:50:18.513(2) & \phn $-$9.46(5)    &  $-$9.08(6)     & 1.63(4)    & \cite{Deller2016} \\
Timing         &  21:45:50.45895(4)     & $-$07:50:18.516(2) &      $-$10.0(2)    &  $-$8.0(5)      & 1.6(4)     & This paper  \\
\enddata
\tablecomments{Numbers in parentheses are uncertainties in the last digit quoted.  Epochs are exact integer dates.}
\end{deluxetable*}
 
\section{Binary pulsars}
\label{sec:binary}

Of the 45 pulsars analyzed in this paper, 31 are in binary systems. To analyze them, we followed procedures similar to those used by \cite{Fonseca2016} (herein F16) to analyze the NG9 data.  The binary systems were parameterized using five Keplerian orbital elements, along with any significant post-Keplerian orbital elements as described below.  Two binary-timing models were used (along with small variations).  The choice of binary model was based on the eccentricity of the orbit.

For eccentric binary systems, we used the ``DD" binary model \citep{Damour85, Damour86, Damour1992}.  Its Keplerian parameters are: orbital period, $P_{\rm b}$; semi-major axis projected onto the line of sight, $x=a_{\rm p}\sin i$, where $a_p$ is the pulsar orbit semi-major axis and $i$ is inclination; eccentricity, $e$; argument of periastron, $\omega$; and epoch of periastron passage, $T_0$.   The DD model can include secular variations in the Keplerian parameters, due to relativistic or geometric effects.  A variation on the DD Model (``DDK'') includes orbital-annual and secular terms due to proper motion \citep{Kopeikin95, Kopeikin96}; this was used for PSR J1713+0747.

For nearly circular systems, the periastron parameters ($\omega$, $T_0$) are highly covariant, making the DD model numerically unsuitable.  In such cases, we used the small-$e$ expansion (``ELL1") binary model \citep{Lange2001}, which parameterizes the orbit by: $P_{\rm b}$ and $x$ as in the DD model; two Laplace-Lagrange parameters, $\eta = e\sin\omega$, and $\kappa = e\cos\omega$; and the epoch of ascending-node passage, $T_{\rm asc}$.  The ELL1 model also allows for the fitting of post-Keplerian parameters.

We used a statistical criterion to determine which binary parameterization (DD or ELL1) to use: if the weighted root-mean-square timing residual for a given pulsar is less than $xe^2$, then the DD model is used to parameterize the orbital motion;  otherwise, the ELL1 model is used. The implementation of this criterion led us to change the binary models used for three pulsars (PSRs J1853+1303, B1855+09, and J2145$-$0750) from DD (used in \nineyr) to ELL1.  This did not lead to any significant changes in their physical-parameter estimates.

We tested for the significance of secular variations in Keplerian orbital elements $P_{\rm b}$, $x$, and $\omega$ for all binary pulsars using the $F$-test described in \S\ref{sec:obs}.  Table~\ref{tab:postKeppar} lists all such parameters with significant values.

We tested for the significance of the Shapiro delay using
the orthometric parameterization of the Shapiro delay in the DD/ELL1 timing models \citep{Freire2010}. For low-inclination, ELL1 orbits, the orthometric representation allows for a Fourier decomposition of the TOA residuals across orbital phase for improved detection of variations from the Shapiro delay.  In such cases, the Shapiro delay is approximately parameterized by the third and fourth harmonic amplitudes of the Fourier spectrum ($h_3$ and $h_4$, respectively). For eccentric systems, or ELL1 systems with high orbital inclinations, $h_3$ and the harmonic ratio $\varsigma = h_4 / h_3$ are more appropriate Shapiro-delay parameters, and the exact expressions for the timing delay are used to calculate the Shapiro delay.

Table~\ref{tab:postKeppar} lists the best-fit values of the orthometric Shapiro-delay parameters for all binary pulsars, whether or not the parameters were significant.
We measured significant ($h_3 > 3\sigma$) Shapiro-delay signals in sixteen systems.  This includes the first measurements of significant Shapiro timing delays in PSRs J0740+6620 and J1853+1303, and confirms previous Shapiro-delay measurements for the other 14 pulsars.

In the data parameter files accompanying this paper, we include traditional Shapiro-delay parameters (companion mass, $m_{\rm c}$; and sine of orbital inclination, $\sin i$) for systems in which two significant Shapiro-delay parameters could be measured, and we include orthometric parameter $h_3$ for systems in which only one Shapiro delay could be measured.  We also include any significant measurements of secular variations in Keplerian orbital elements.

\begin{figure}
    \centering
    \includegraphics[scale=0.4]{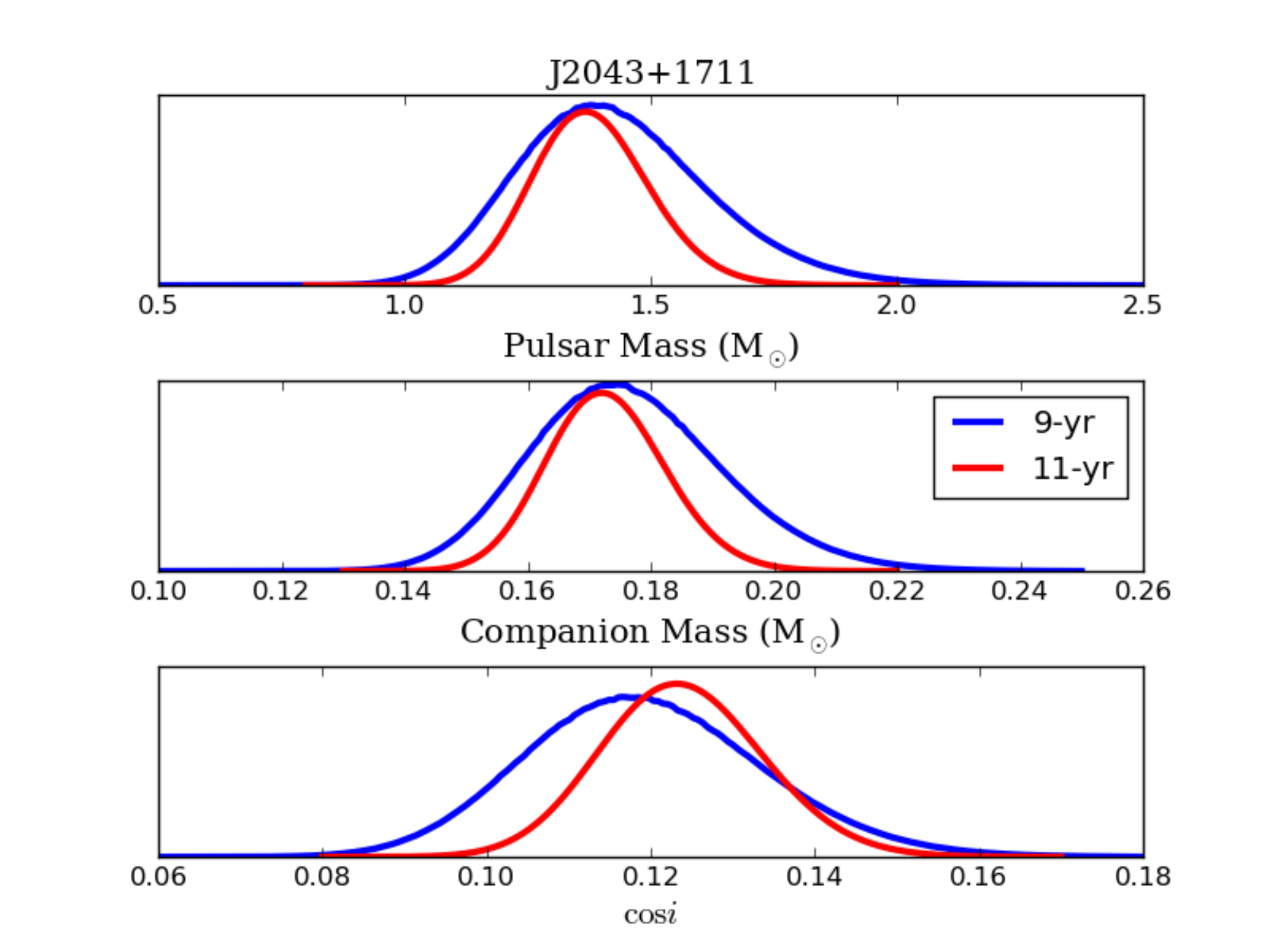}
    \caption{Posterior probability density functions for the Shapiro-delay parameters measured in the J2043+1711 binary system, computed from a $(m_{\rm c}$, $\cos i)$ grid of $\chi^2$ values. The pulsar mass was derived by translation of the ($m_{\rm c}$, $\cos i$) map to the ($m_{\rm p}$, $\cos i$) space using the mass function for this system, and then integrating the ($m_{\rm p}$, $\cos i$) map over all $\cos i$ values to obtain a PDF for $m_{\rm p}$.\label{fig:2043mass}}
\end{figure}

We obtained posterior probability distributions and credible intervals for the mass and geometric parameters of the most significant Shapiro-delay signals (i.e., $h_3 > 10\sigma$), using the PAL2\footnote{\url{https://github.com/jellis18/PAL2}} Bayesian inference suite. The results from PAL2 MCMC simulations are shown in Table \ref{tab:SDresults}. The statistical significance in several sets of these measurements has improved since they were previously studied by F16. For example, the precision of the pulsar mass for PSR J2043+1711, $m_{\rm p} = 1.38^{+0.12}_{-0.13}\textrm{ M}_\odot$, has improved by a factor of two (Figure~\ref{fig:2043mass}).

Three binary systems -- those of PSRs~ J1600$-$3053, J1903+0327, and J2234+0611 -- show a significant time variation in their periastron arguments ($\dot{\omega}$). Previous studies using other data sets interpreted the observed $\dot{\omega}$ values for PSR J1903+0327 \citep{Freire2011} and PSR J2234+0611 \citep{Antoniadis2016} as being due to general relativistic orbital precession,  while F16 used the \nineyr\ data set to conclude the same for the $\dot{\omega}$ of PSR~J1600$-$3053.
 For PSR J1600$-$3053, the precision in the $\dot{\omega}$ measurement has improved by a factor of $\sim1.7$ since \nineyr, consistent with the expected improvement in measurability of post-Keplerian variations over time \citep{Damour1992}; based on this scaling relation, the significance of $\dot{\omega}$ will reach $10\sigma$ by $\sim$2020. The constrained estimates of the Shapiro-delay parameters, have similarly improved, allowing us to measure $m_{\rm p} = 2.3^{+0.7}_{-0.6}\textrm{ M}_\odot$ for this pulsar.

Five binary systems show significant variations of their orbital periods over time ($\dot{P}_{\rm b}$). For three of these systems -- PSRs J1012+5307, J1614$-$2230, and J1909$-$3744 -- previous analyses by \citet{Desvignes2016} and F16 showed that the dominant mechanism for the observed variations is relative acceleration between the solar system barycenter and the binary systems (see \S\ref{sec:pdotpbdot}). We also measured large orbital-period variations in PSRs J0023+0923 and J0636+5128 for the first time; the timing solution for PSR~J0023+0932 fits four significant time derivatives of the orbital frequency $n_{\rm b} = 2\pi/P_{\rm b}$, though only the derived $\dot{P}_{\rm b}$ is shown in Table \ref{tab:postKeppar}. A sixth pulsar, PSR~J0613$-$0200, shows a marginally significant $\dot{P}_{\rm b} = 0.06(2)\times10^{-12}$; this was not included in our fit files, as it did not pass the $F$-test used
for parameter significance. 
 However, its value is consistent with the estimate made by \citet{Desvignes2016}, and is in agreement with the predicted value from various sources of relative acceleration. 

For PSRs~J0023+0923 and J0636+5128, short periods ($\sim$ 3 hr) and very-low minimum companion masses ($m_{\rm c, min}\sim 0.05\textrm{ M}_\odot$), suggest these are black widow systems, in which torques produced from tidal interactions with an oblate companion can cause orbital variability \citep[e.g.][]{Applegate1994}. Recent simulations of long-term variability in black widow systems have shown that such behavior will not significantly impact the detection of nHz-frequency GWs \citep{Bochenek2015}. However, long-term timing of the PSR~J2051$-$0827 black widow system illustrated apparent variations in $x$, the projected
semi-major axis, that may also need to be accounted for in such systems \citep{Shaifullah2016}.

Many of our binary pulsars exhibit significant changes in their projected semi-major axes over time ($\dot{x}$). The dominant mechanism for these observed variations is the change in apparent inclination of the orbital plane due to proper motion of the system \citep{Kopeikin96}. F16 used the observed Shapiro delay and $\dot{x}$ in the PSR~J1741+1351 system to directly estimate a value for the longitude of ascending node ($\Omega$), along with the masses and system inclination. We find that $\Omega = 330^\circ\pm 30^\circ$, consistent with the measurement of $317^\circ\pm 35^\circ$ in F16.

\begin{deluxetable*}{lllllllcc}
    \centering
    \tabletypesize{\scriptsize}
    \tablecaption{Secular Variations and Shapiro-Delay Parameters in Binary Systems\label{tab:postKeppar}}
    \tablewidth{0in}
    \tablehead{\colhead{PSR} & \colhead{$\dot{\omega}$ (deg yr$^{-1}$)} & \colhead{$\dot{x}$ (10$^{-12}$)} & \colhead{$\dot{P}_{\rm b}$ (10$^{-12}$)} & \colhead{$h_3$ ($\mu$s)} & \colhead{$h_4$ ($\mu$s)} & \colhead{$\varsigma$} & \colhead{Detection of $\Delta_{\rm S}$?} & \colhead{Span (yr)}}
    \startdata
J0023+0923\tablenotemark{a} & \ldots & \ldots & 2.8(2) & \phn 0.05(3) & \ldots & \ldots & N & \phn 4.4 \\
J0613$-$0200 & \ldots & \ldots & 0.054(18)  & \phn 0.27(3) & \ldots & 0.71(6) & Y & 10.8 \\
J0636+5128 & \ldots & \ldots & 2.5(3) & \phn 0.00(3) & \phn 0.00(5) & \ldots & N & \phn 2.0 \\
J0740+6620 & \ldots & \ldots & \ldots & \phn 0.95(16) & \phn 0.0(3) & \ldots & Y & \phn 2.0 \\
J1012+5307 & \ldots & \ldots & 0.081(16) & \phn0.02(7) & \phn 0.05(10) & \ldots & N & 11.4 \\
J1125+7819 & \ldots & $-$0.36(11) & \ldots & \phn 0.0(1.5) & $-$1.2(1.8) & \ldots & N & \phn 2.0 \\
J1455$-$3330 & \ldots & $-$0.020(4) & \ldots & \phn 0.30(15) & \ldots & 0.5(3) & N & 11.4 \\
J1600$-$3053 & 0.0052(14) & $-$0.0040(6) & \ldots & \phn 0.34(2) & \ldots & 0.63(5) & Y & \phn 8.1 \\
J1614$-$2230 & \ldots & \ldots & 1.7(2) & \phn 2.32(1) & \ldots & 0.9862(2) & Y & \phn 7.2 \\
J1640+2224 & \ldots & \phn 0.0135(9) & \ldots & \phn 0.44(5) & \ldots & 0.58(11) & Y & 11.1 \\
J1643$-$1224 & \ldots & $-$0.054(5) & \ldots & $-$0.018(12) & \ldots & 0.91(17) & N & 11.2 \\
J1713+0747 & \ldots & \phn 0.00645(11) & \ldots & \phn 0.54(3) & \ldots & 0.73(1) & Y & 10.9 \\
J1738+0333 & \ldots & \ldots & \ldots & $-$0.03(8) & $-$0.05(9) & \ldots & N & \phn 6.1 \\
J1741+1351 & \ldots & $-$0.005(1) & \ldots & \phn 0.45(3) & \ldots & 0.76(6) & Y & \phn 6.4 \\
J1853+1303\tablenotemark{b} & \ldots & \phn 0.013(2) & \ldots & \phn 0.26(6) & 0.10(6) & \ldots & Y & \phn 4.5 \\
B1855+09 & \ldots & \ldots & \ldots & \phn 1.07(4) & \ldots & 0.966(5) & Y & 11.0 \\
J1903+0327 & 0.0002403(5) & \ldots & \ldots & \phn 2.5(3) & \ldots & 0.88(6) & Y & \phn 6.1 \\
J1909$-$3744 & \ldots & $-$0.00040(13) & 0.502(5) & \phn 0.847(5) & \ldots & 0.940(1) & Y & 11.2 \\
J1910+1256\tablenotemark{b} & \ldots & $-$0.023(4) & \ldots & \phn 0.02(16) & \ldots & $-$0.3(8) & N & \phn 6.8 \\
J1918$-$0642 & \ldots & \ldots & \ldots & \phn 0.86(2) & \ldots & 0.911(7) & Y & 11.2 \\
B1953+29\tablenotemark{b} & \ldots & \phn 0.011(3) & \ldots & $-$0.00(1) & \ldots & $-$0.8(1.0) & N & \phn 4.4 \\
J2017+0603 & \ldots & \ldots & \ldots & \phn 0.38(6) & \ldots & 0.7(1) & Y & \phn 3.8 \\
J2033+1734 & \ldots & \ldots & \ldots & \phn 1.0(4) & \ldots & $-$0.4(5) & N & \phn 2.3 \\
J2043+1711 & \ldots & \ldots & \ldots & \phn 0.585(18) & \ldots & 0.884(9) & Y & \phn 4.5 \\
J2145$-$0750 & \ldots & \phn \ldots & \ldots & \phn 0.17(7) & \ldots & 0.7(3) & N & 11.3 \\
J2214+3000 & \ldots & \ldots & \ldots & \phn0.11(17) & 0.0(2) & \ldots & N & \phn 4.2 \\
J2229+2643 & \ldots & \ldots & \ldots & $-0.2(4)$ & \ldots & 0.1(9) & N & \phn 2.4 \\
J2234+0611 & 0.000871(16) & $-$0.041(11) & \ldots & \phn0.10(7) & \ldots & 0.96(7) & N & \phn 2.4 \\ 
J2234+0944 & \ldots & \ldots & \ldots & $-$0.17(12) & \phn0.21(11) & \ldots & N & \phn 2.0 \\ 
J2302+4442 & \ldots & \ldots & \ldots & \phn 1.5(3) & \ldots & 0.55(15) & Y & \phn 3.8 \\
J2317+1439 & \ldots & \ldots & \ldots & \phn 0.20(3) & \ldots & 0.55(13) & Y & 11.0  \\
    \enddata
    \tablecomments{Values in parentheses denote the $1\sigma$ uncertainty in the preceding digit(s), as determined from TEMPO.}
    \tablenotetext{a}{Four derivatives in orbital frequency ($n_{\rm b} = 2\pi/P_{\rm b}$) were fitted; the Shapiro delay $h_4$ is not currently implemented as a fit parameter in this particular TEMPO binary model, and is therefore not fit for.}
    \tablenotetext{b}{Early single-frequency ASP data removed for this data release.}
\end{deluxetable*}

\begin{deluxetable*}{llll}
    \tabletypesize{\small}
    \tablecaption{Pulsar-binary Component Masses and Inclination Angles}
    \tablewidth{0in}
    \tablehead{\multicolumn{1}{c}{PSR} & Pulsar Mass (M$_{\odot})$ & Companion Mass (M$_{\odot})$ & System Inclination}
    \startdata
    J1600$-$3053 & $2.5^{+0.9}_{-0.7}$ & $0.34^{+0.09}_{-0.07}$ & $62^{+3}_{-3}$ \\
    J1614$-$2230 & $1.908^{+0.016}_{-0.016}$ & $0.493^{+0.003}_{-0.003}$ & $89.204^{+0.014}_{-0.014}$ \\
    J1713$+$0747\tablenotemark{a} & $1.35^{+0.07}_{-0.07}$ & $0.292^{+0.011}_{-0.011}$ & $71.8^{+0.5}_{-0.6}$ \\ 
    J1741$+$1351\tablenotemark{a} & $1.14^{+0.43}_{-0.25}$ & $0.22^{+0.05}_{-0.04}$ & $73^{+3}_{-4}$ \\ 
    B1855$+$09 & $1.37^{+0.13}_{-0.10}$ & $0.244^{+0.014}_{-0.012}$ & $88.0^{+0.3}_{-0.4}$ \\ 
    J1903$+$0327\tablenotemark{a} & $1.666^{+0.010}_{-0.012}$ & $1.033^{+0.011}_{-0.008}$ & $77^{+2}_{-2}$ \\ 
    J1909$-$3744 & $1.48^{+0.03}_{-0.03}$ & $0.208^{+0.002}_{-0.002}$ & $86.47^{+0.10}_{-0.09}$ \\ 
    J1918$-$0642 & $1.29^{+0.10}_{-0.09}$ & $0.231^{+0.010}_{-0.010}$ & $84.7^{+0.4}_{-0.5}$ \\ 
    J2043$+$1711 & $1.38^{+0.12}_{-0.13}$ & $0.173^{+0.010}_{-0.010}$ & $83.0^{+0.6}_{-0.6}$ \\ 
    \enddata
    \label{tab:SDresults}
    \tablecomments{All estimates were made using the ``traditional" ($m_{\rm c}$, $\sin i$) parameterization of the Shapiro delay. All uncertainties reflect 68.3\% credible intervals.}
    \tablenotetext{a}{One or more observed secular variations were used as constraints for the masses and/or geometry.}
\end{deluxetable*}
  
\section{Summary and Conclusions}
\label{sec:conclusion}
We have presented the timing data and analysis for 45 millisecond pulsars observed for time spans of up to 11 years from
the NANOGrav timing program. We outlined the analysis procedure used to calculate TOAs and fit these TOAs to models including
spin, astrometric, and binary (if necessary) parameters, along with a parameterized noise model for each pulsar.
The timing and noise analysis methods used for the 11-year data set are nearly identical to those 
described in our previous (9-year) data set paper, \cite{Arzoumanian2015b}. However, we made several improvements to the initial stages of preparing the TOAs for fitting. We incorporated a more sophisticated, automated analysis to identify outlier TOAs. We also excluded data with insufficient radio-frequency coverage to fit an accurate DM. In addition, we adjusted the criterion for DM-epoch determination for data for which a solar wind model predicted large delays. These improvements provided greater immunity to corruption of timing- and noise-model parameters due to instrumental effects or unmodeled dispersive delays. 

In general, timing solutions are comparable with NG9, with reduced uncertainties on timing-derived parameters. We measured several timing parameters for the first time in the 11-year data set. We measured parallax of 20 pulsars, 3 for the first time (PSRs J0740+6620,
J2334+0611, and J2234+0944). We measured Shapiro delay for PSRs J0740+6620 and J1853+1303 for the first time ever, and improved the Shapiro-delay measurements, and hence mass estimates, for an additional 14 pulsars. Large orbital-period variations have been measured for two pulsars (PSRs J0023+0923 and
J0636+5128) for the first time. We attribute these variations to torques produced by tidal interactions with their low-mass companions. 

Our noise analysis indicates that 11 pulsars show evidence for significant red noise, i.e., noise with power peaking at lower frequencies. Because we have excluded data with insufficient frequency coverage for reliable dispersion-delay measurements, and those with large predicted unmodeled solar-wind delays, in most cases we cannot attribute this red noise to interstellar-medium effects. 
All of the pulsars with well constrained red-noise spectral indices show low values, ranging from $-1$ to $-3$, indicating flatter spectra than observed for normal pulsars, for which red noise has been attributed to intrinsic spin noise. If the red noise is intrinsic, this may suggest a different origin in millisecond pulsars. Most importantly, the spectral indices are less steep than the $-11/3$ predicted for the stochastic gravitational-wave background, indicating that we should continue to gain in sensitivity as the time span of our data set grows.  

NANOGrav is committed to publicly releasing its timing data on a regular basis.  This data set is the third NANOGrav release.
As with previously released data sets,
the data described in this paper
are being used to constrain the presence of gravitational waves due to a stochastic background of supermassive black hole binaries
\citep{Arzoumanian2018b}.
They will also be used to search for single (or continuous-wave) sources and for burst sources. Future papers will detail these
analyses and their astrophysical implications. These data represent a significant increase in sensitivity over the nine-year set, which contained 37 pulsars.
As we continue to add pulsars to the array, and as the total time span lengthens, our sensitivity to gravitational waves will grow, with a detection expected within the next 5--6 years \citep{Taylor2016}.
 
\acknowledgements

{\it Author contributions.}  
The alphabetical-order author list reflects the broad variety of contributions
of authors to the NANOGrav project.  Some specific contributions to this
paper, particularly incremental work between \nineyr\ and the present work, are
as follows.
ZA, KC, PBD, TD, EF, RDF, EF, PAG, CJ, GJ, DH, MTL,
LL, DRL, RSL, MAM, CN, DJN, TTP, SMR, RS, IHS, KS,
JKS, and WZ
each made at least 20 hours of observations for this project.
MED, EF, MJ, MTL, DRL, MAM, CN, DJN, TTP, PSR, SMR, and IHS
generated and checked timing solutions for individual pulsars.
PBD, MED, JAE, RDF, MTL, CN, DJN, and IHS
developed and refined procedures and computational tools for the timing 
pipeline.
RvH, MV, and JAE implemented the TOA outlier-detection procedure.  
PBD wrote observing proposals, coordinated Green Bank observations,
performed calibration and TOA generation, coordinated the
data flow, developed the data files for public release,
and contributed substantially to the text.
JAE developed and refined the algorithms and software implemented
used for the noise model.  He wrote substantial amounts of the text, 
and contributed several tables and figures.
DJN coordinated the development of the data set and the writing of 
this paper, co-authored observing proposals, chaired the NANOGrav Timing
Working Group, undertook some of the astrometric analysis, and wrote portions of the text.
IHS coordinated the Arecibo observations.
EF wrote observing proposals, assisted in coordination of Arecibo observations, and undertook the analysis and write-up of binary pulsars in \S\ref{sec:binary}.
AM undertook the analysis and write-up of parallax and distance measurements in \S\ref{sec:astrometry}.

 The NANOGrav project receives support from National Science Foundation
(NSF) Physics Frontiers Center award number 1430284.  
Pulsar research at UBC is supported by an NSERC Discovery Grant and by the 
Canadian Institute for Advanced Research.
The Arecibo Observatory is operated by SRI International under a cooperative
agreement with the NSF (AST-1100968), and in alliance with Ana G.
M\'{e}ndez-Universidad Metropolitana, and the Universities Space
Research Association. 
The Green Bank Observatory is a facility of the National Science Foundation operated under cooperative agreement by Associated Universities, Inc.
Part of this research was carried out at the Jet Propulsion Laboratory, California Institute of
Technology, under a contract with the National Aeronautics and Space Administration.
The Flatiron Institute is supported by the Simons Foundation.
The Dunlap Institute is funded by an endowment established by the David Dunlap family and the University of Toronto.
JAE was partially supported by NASA Einstein Fellowship grant PF4-150120.
RvH was supported by NASA Einstein Fellowship grant PF3-140116.
WWZ is supported by the CAS Pioneer Hundred Talents Program and the Strategic Priority Research Program of the Chinese Academy of Sciences, Grant No.\ XDB23000000. 

{

\sloppy

\software{PSRCHIVE \citep{Hotan2004}, \\
Nanopipe \citep{nanopipe2018}, Tempo \citep{tempo2015}, Tempo2 \cite{Edwards2006,Hobbs2006}, PAL2 \citep{pal2}, PTMCMCSampler \cite{ptmcmcsampler}, MultiNest \citep{fhb09}}

}
 
\appendix

\section{Daily Averaged Residuals}

\label{sec:resid}
This appendix includes plots of residual time series and DM variations
for each pulsar in our data set.

As described in \S\ref{sec:obs}, each observation in our data set produced a
large number of arrival times representing data collected simultaneously over
a range of radio-frequency bands.  The top panel of each figure in this
appendix shows the residual arrival time (observed minus computed) for every
arrival time measurement in the data set for a given pulsar.  Points
in the plots are
colored based on receiver frequency, and the predominant data-collection
instrument over any given time period is indicated at the top of each plot,
with vertical dashed lines indicating times at which data-collection
instruments changed.

The models used for residual plots include the effect of red noise,
but any red noise corresponding to a linear or quadratic trend has
been removed, as it is completely covariant with pulsar rotation frequency
and frequency derivative in the timing model, and hence 
would be absorbed by fits for these quantities.

Daily-average residuals for each receiver are shown in the second panel 
for each pulsar.  These were computed using the procedure described
in Appendix~D of \nineyr.

For pulsars with whose timing models include red noise 
(Table~\ref{tab:timingpar}), the third panel of the figure 
shows whitened residuals, which were calculated by subtracting the red-noise model from the daily-average residuals.

The bottom panel of each figure shows the variation in DM
for each pulsar.  These are presented as 
${\rm DMX}_i \equiv {\rm DM}_i - {\rm DM}_{\rm average}$, where
${\rm DM}_i$ is the DM at epoch $i$ and 
${\rm DM}_{\rm average}$ is the average DM
over the entire data set for the pulsar.
Lengths of epochs are described in \S\ref{sec:obs}, and
are typically 6 days or less, except in the earliest data.
Subtracting the average value is advantageous because it
allows us to remove the uncertainty in ${\rm DM}_{\rm average}$
(which arises due to covariance with the FD parameters described
in \S\ref{sec:obs}) from the uncertainties in ${\rm DMX}_i$
shown in the figures.

\bibliographystyle{aasjournal}

\begin{figure*}[p]
\centering
\includegraphics[scale=1.0]{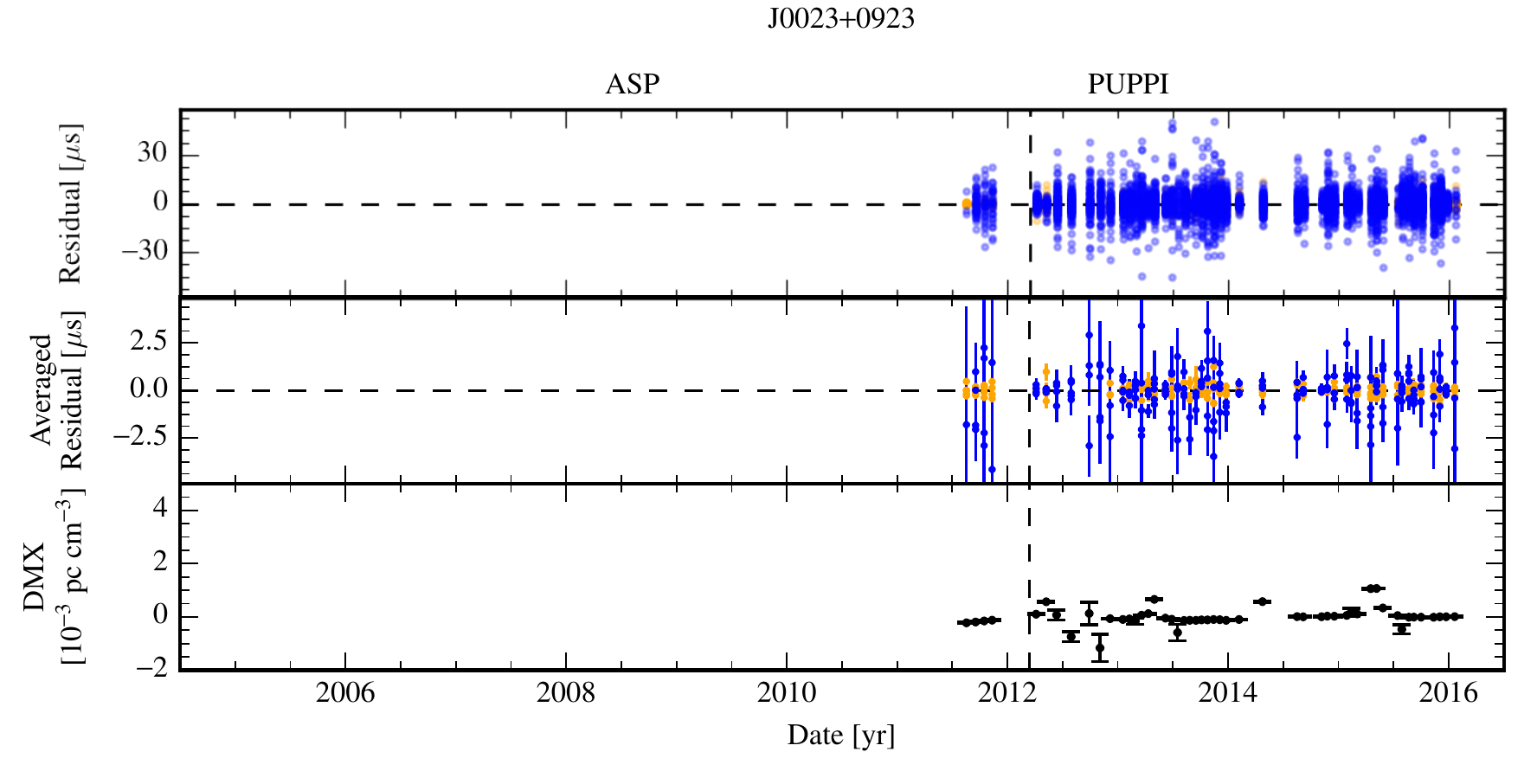}
\caption{Timing summary for PSR J0023+0923. Colors are: Blue: 1.4 GHz, Purple: 2.1 GHz, Green: 820 MHz, Orange: 430 MHz, Red: 327 MHz. In the top panel, individual points are semi-transparent; darker regions arise from the overlap of many points.}
\label{fig:summary-J0023+0923}
\end{figure*}

\begin{figure*}[p]
\centering
\includegraphics[scale=1.0]{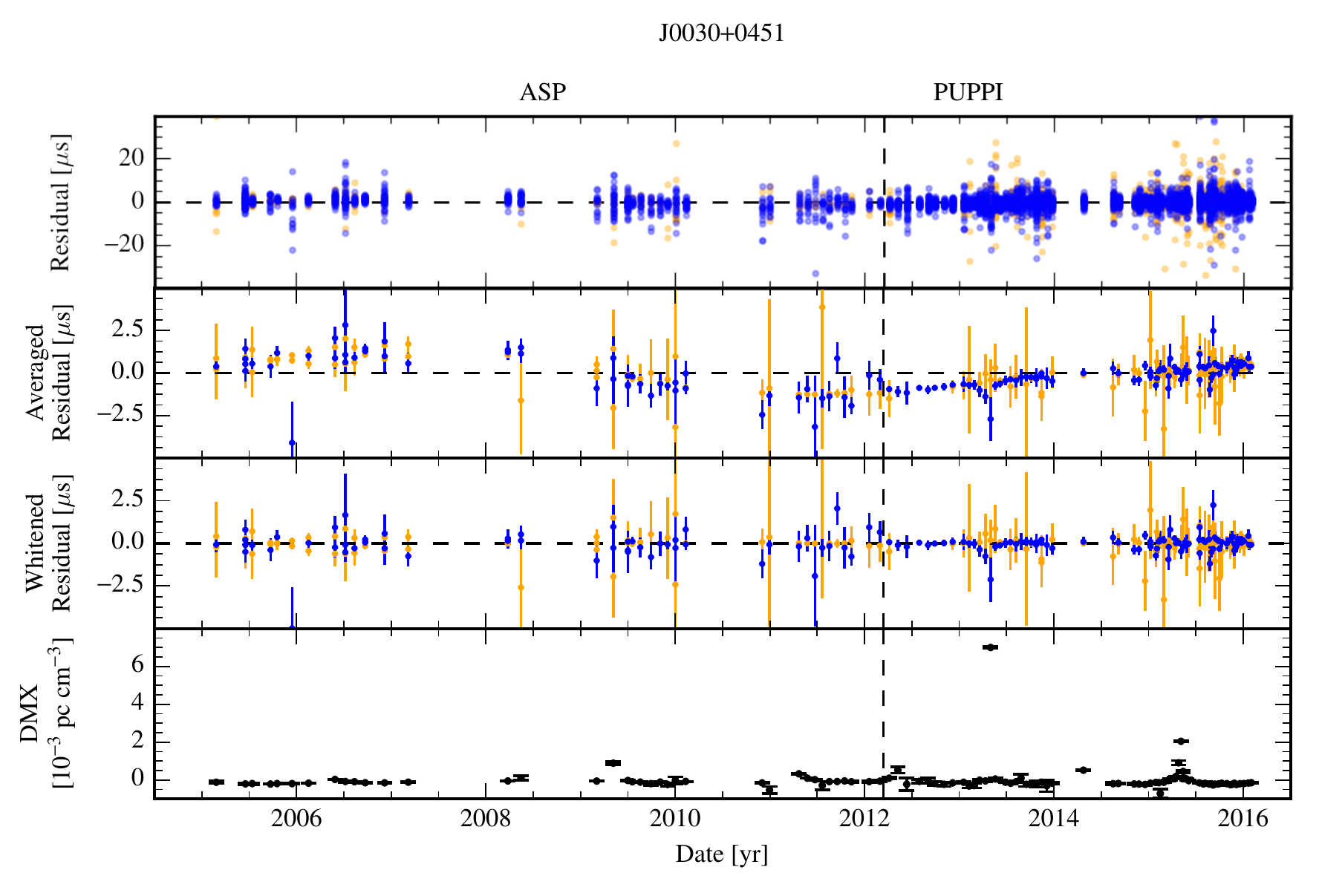}
\caption{Timing summary for PSR J0030+0451. Colors are: Blue: 1.4 GHz, Purple: 2.1 GHz, Green: 820 MHz, Orange: 430 MHz, Red: 327 MHz. In the top panel, individual points are semi-transparent; darker regions arise from the overlap of many points.}
\label{fig:summary-J0030+0451}
\end{figure*}

\begin{figure*}[p]
\centering
\includegraphics[scale=1.0]{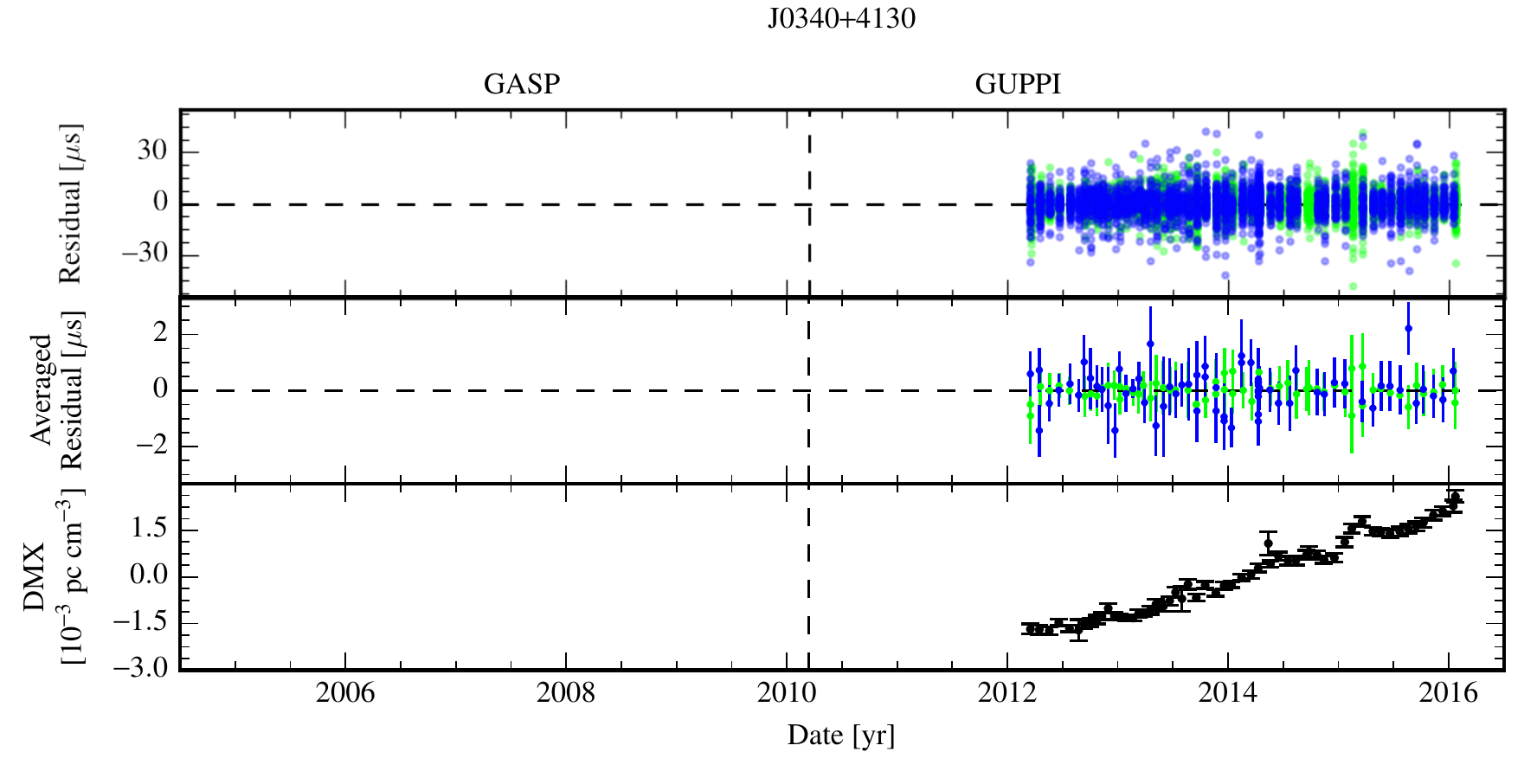}
\caption{Timing summary for PSR J0340+4130. Colors are: Blue: 1.4 GHz, Purple: 2.1 GHz, Green: 820 MHz, Orange: 430 MHz, Red: 327 MHz. In the top panel, individual points are semi-transparent; darker regions arise from the overlap of many points.}
\label{fig:summary-J0340+4130}
\end{figure*}

\begin{figure*}[p]
\centering
\includegraphics[scale=1.0]{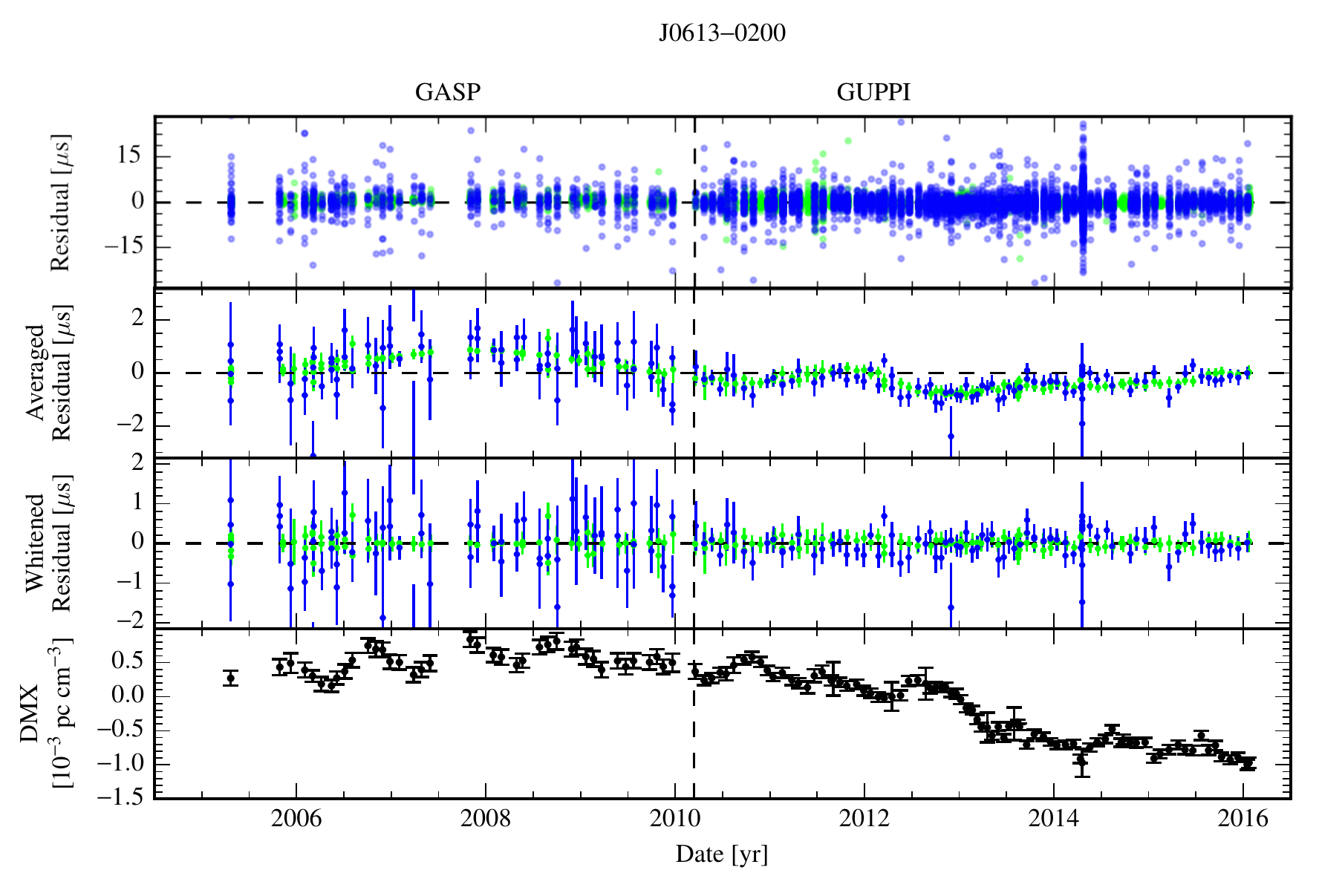}
\caption{Timing summary for PSR J0613-0200. Colors are: Blue: 1.4 GHz, Purple: 2.1 GHz, Green: 820 MHz, Orange: 430 MHz, Red: 327 MHz. In the top panel, individual points are semi-transparent; darker regions arise from the overlap of many points.}
\label{fig:summary-J0613-0200}
\end{figure*}

\begin{figure*}[p]
\centering
\includegraphics[scale=1.0]{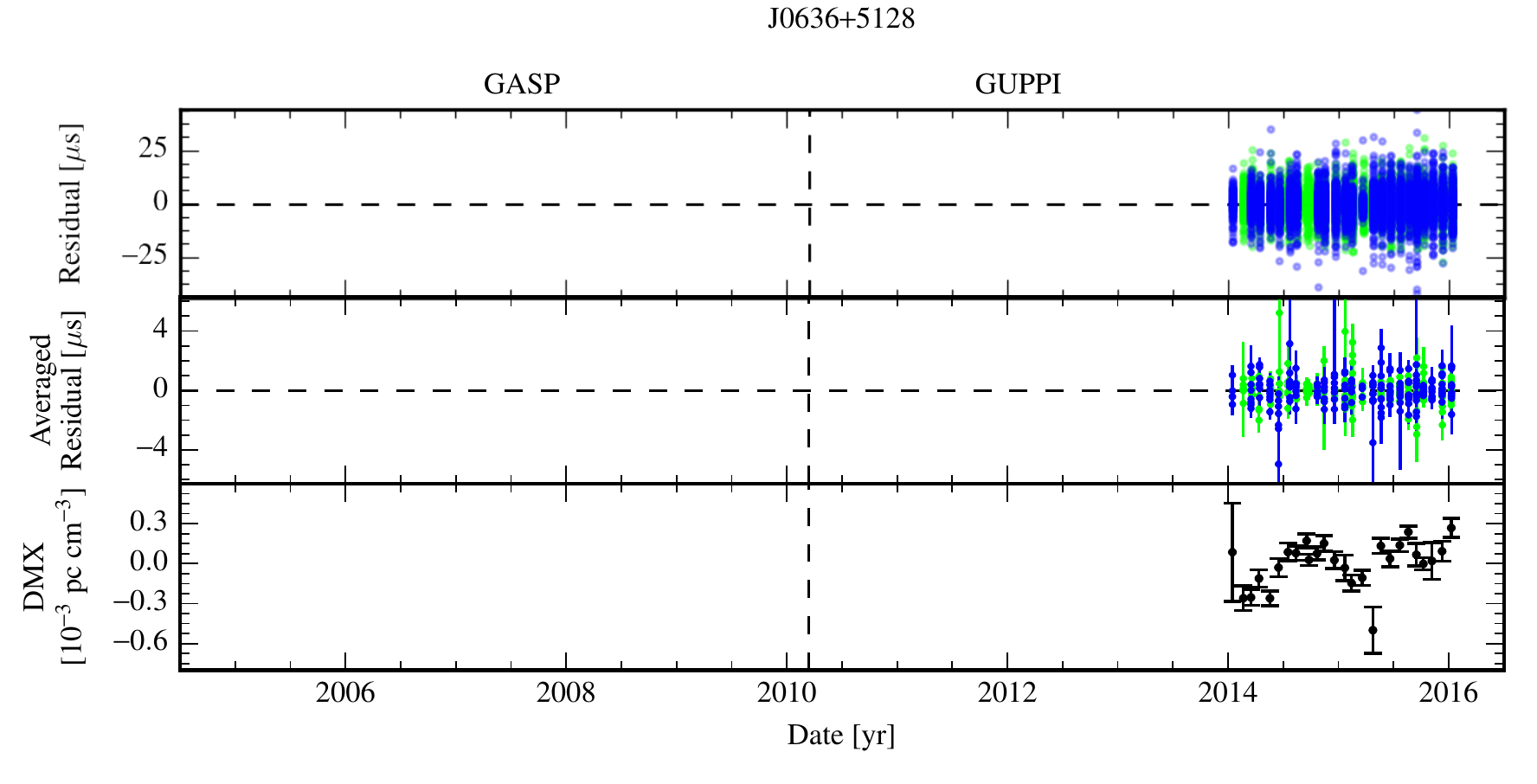}
\caption{Timing summary for PSR J0636+5128. Colors are: Blue: 1.4 GHz, Purple: 2.1 GHz, Green: 820 MHz, Orange: 430 MHz, Red: 327 MHz. In the top panel, individual points are semi-transparent; darker regions arise from the overlap of many points.}
\label{fig:summary-J0636+5128}
\end{figure*}

\begin{figure*}[p]
\centering
\includegraphics[scale=1.0]{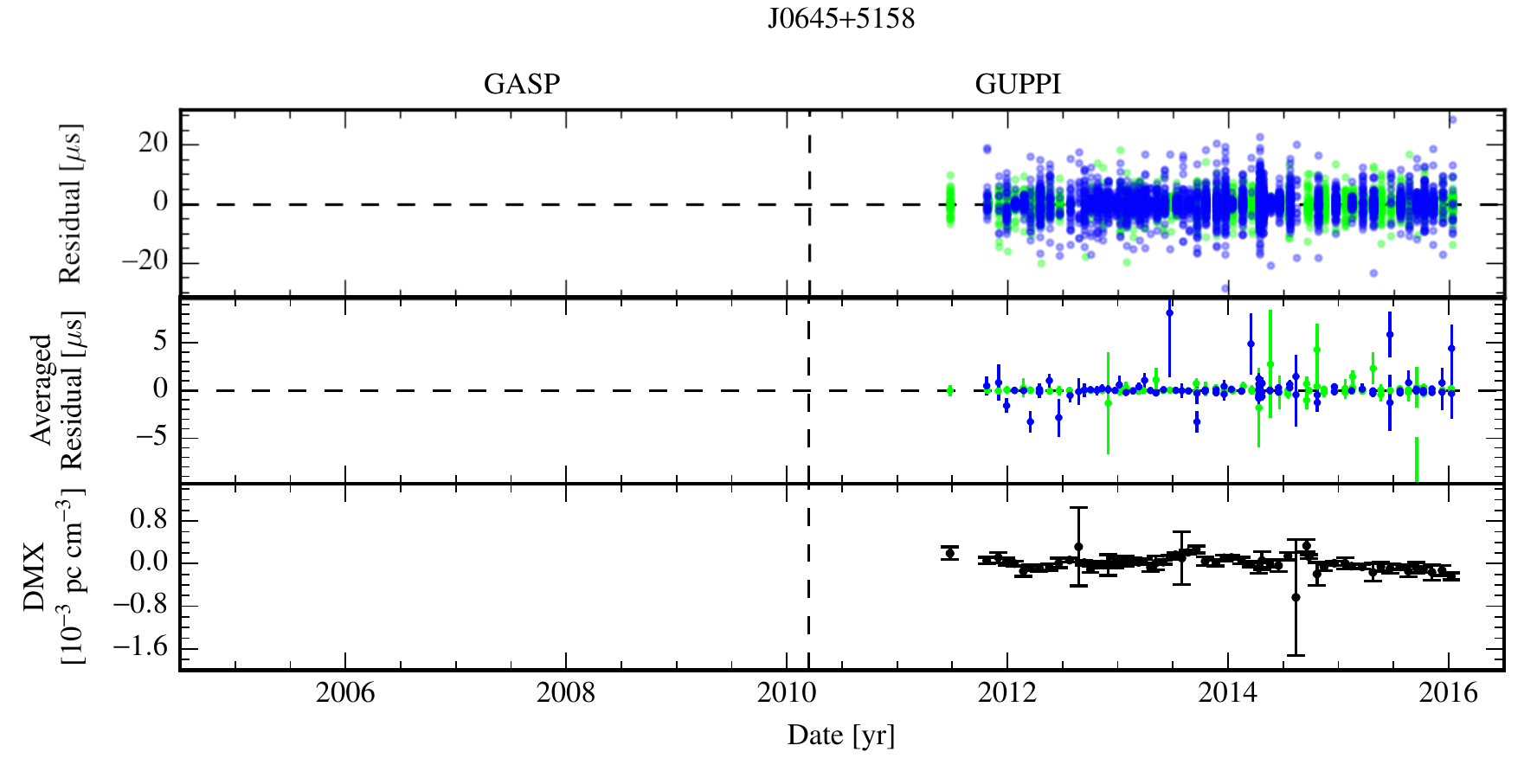}
\caption{Timing summary for PSR J0645+5158. Colors are: Blue: 1.4 GHz, Purple: 2.1 GHz, Green: 820 MHz, Orange: 430 MHz, Red: 327 MHz. In the top panel, individual points are semi-transparent; darker regions arise from the overlap of many points.}
\label{fig:summary-J0645+5158}
\end{figure*}

\begin{figure*}[p]
\centering
\includegraphics[scale=1.0]{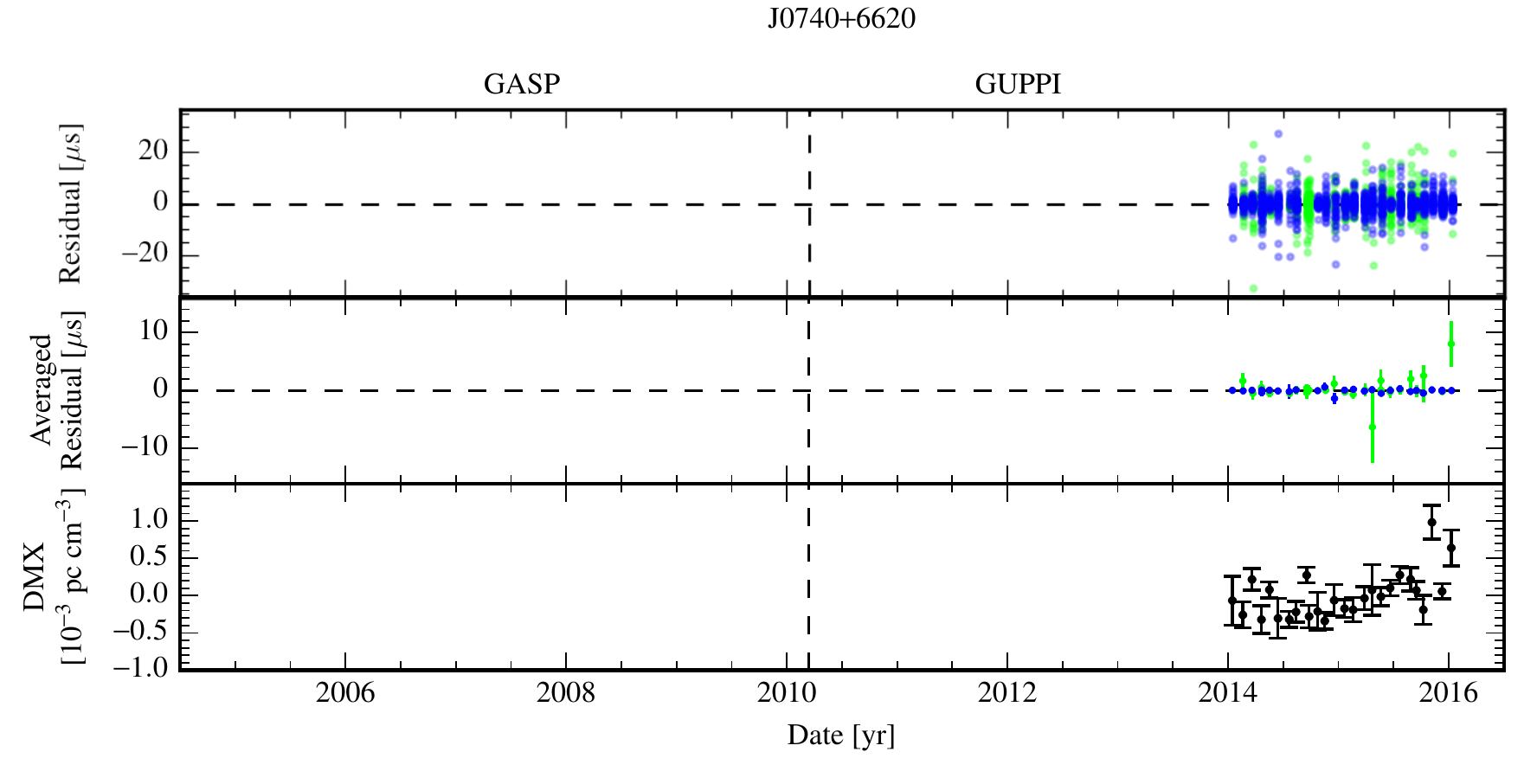}
\caption{Timing summary for PSR J0740+6620. Colors are: Blue: 1.4 GHz, Purple: 2.1 GHz, Green: 820 MHz, Orange: 430 MHz, Red: 327 MHz. In the top panel, individual points are semi-transparent; darker regions arise from the overlap of many points.}
\label{fig:summary-J0740+6620}
\end{figure*}

\begin{figure*}[p]
\centering
\includegraphics[scale=1.0]{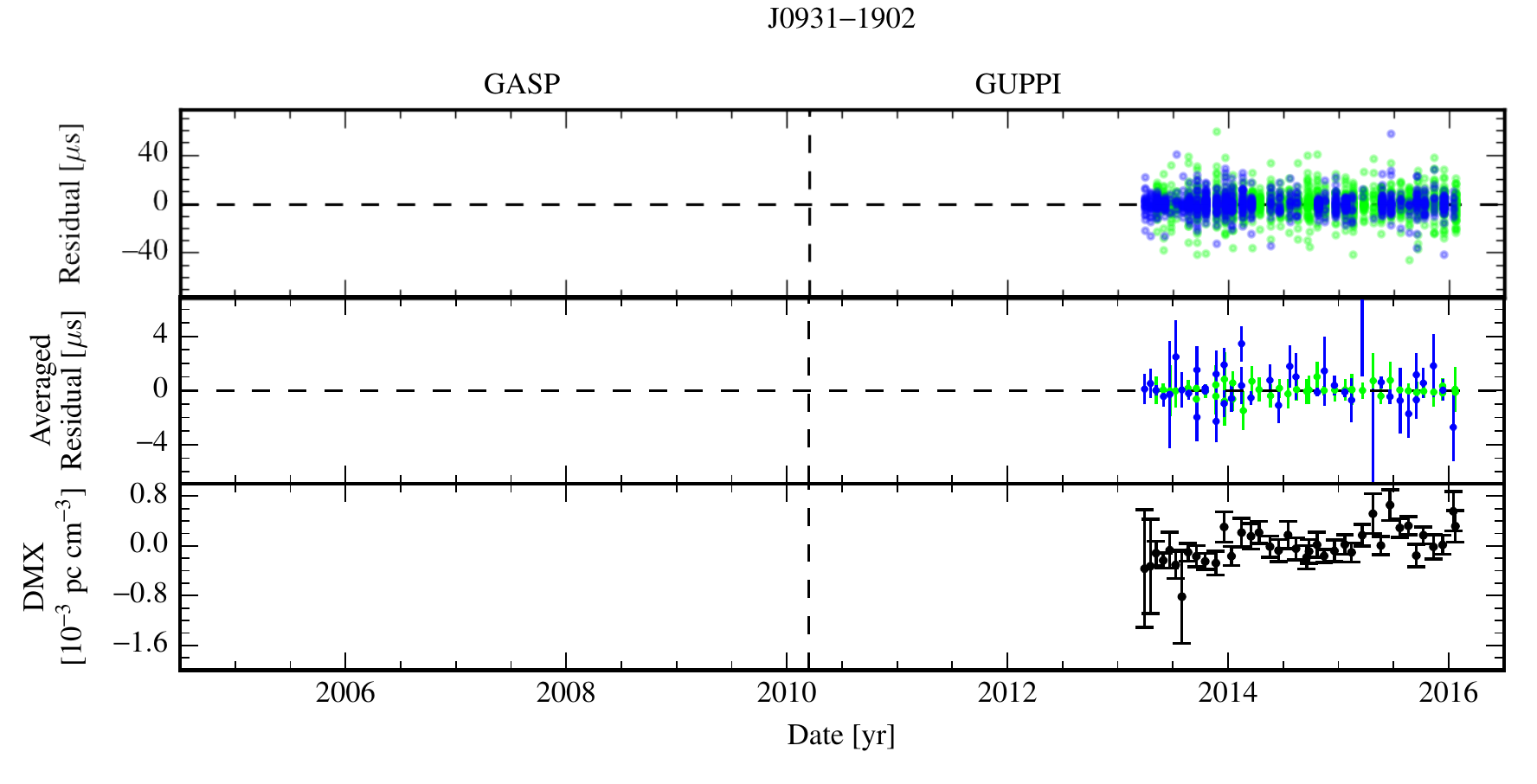}
\caption{Timing summary for PSR J0931-1902. Colors are: Blue: 1.4 GHz, Purple: 2.1 GHz, Green: 820 MHz, Orange: 430 MHz, Red: 327 MHz. In the top panel, individual points are semi-transparent; darker regions arise from the overlap of many points.}
\label{fig:summary-J0931-1902}
\end{figure*}

\begin{figure*}[p]
\centering
\includegraphics[scale=1.0]{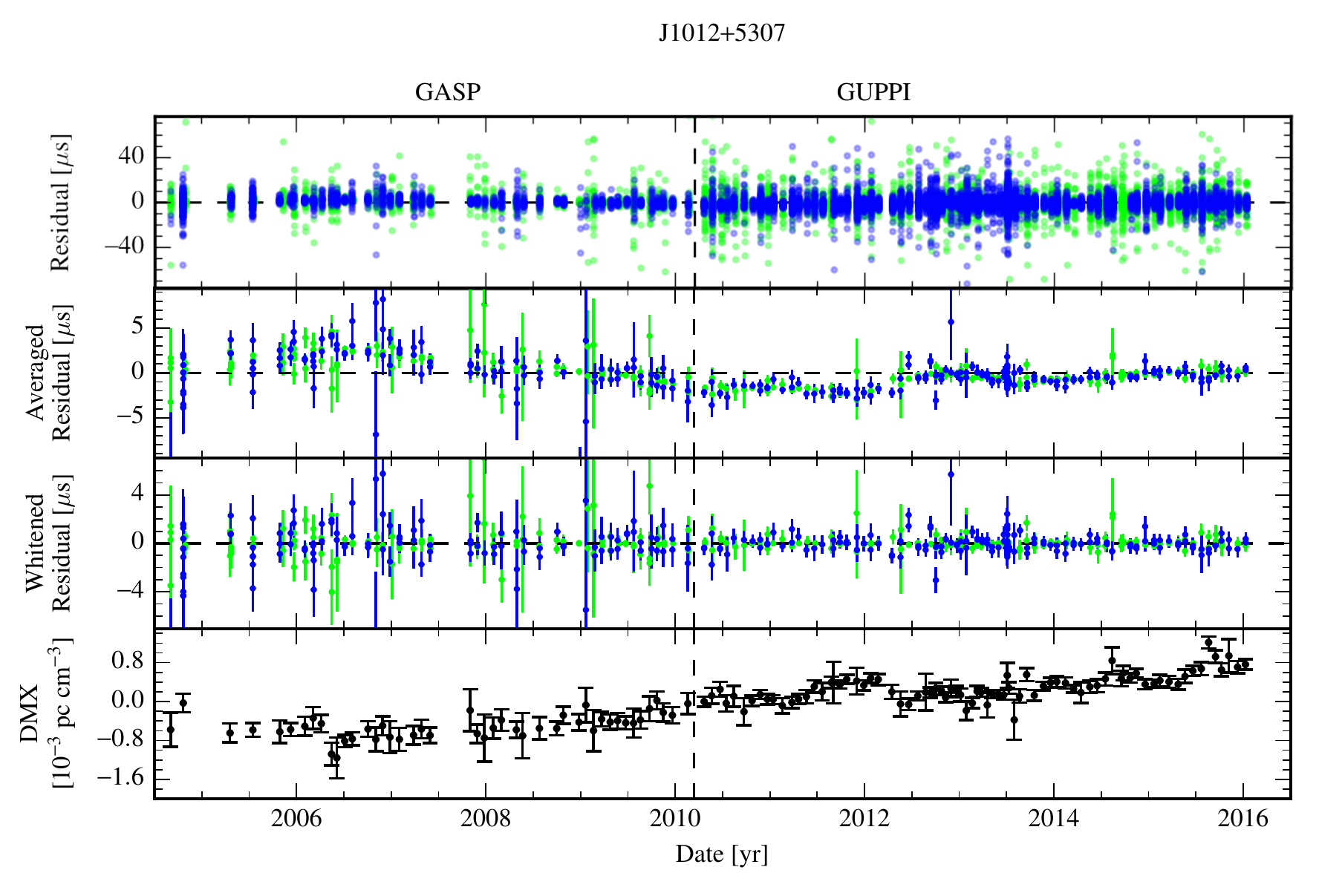}
\caption{Timing summary for PSR J1012+5307. Colors are: Blue: 1.4 GHz, Purple: 2.1 GHz, Green: 820 MHz, Orange: 430 MHz, Red: 327 MHz. In the top panel, individual points are semi-transparent; darker regions arise from the overlap of many points.}
\label{fig:summary-J1012+5307}
\end{figure*}

\begin{figure*}[p]
\centering
\includegraphics[scale=1.0]{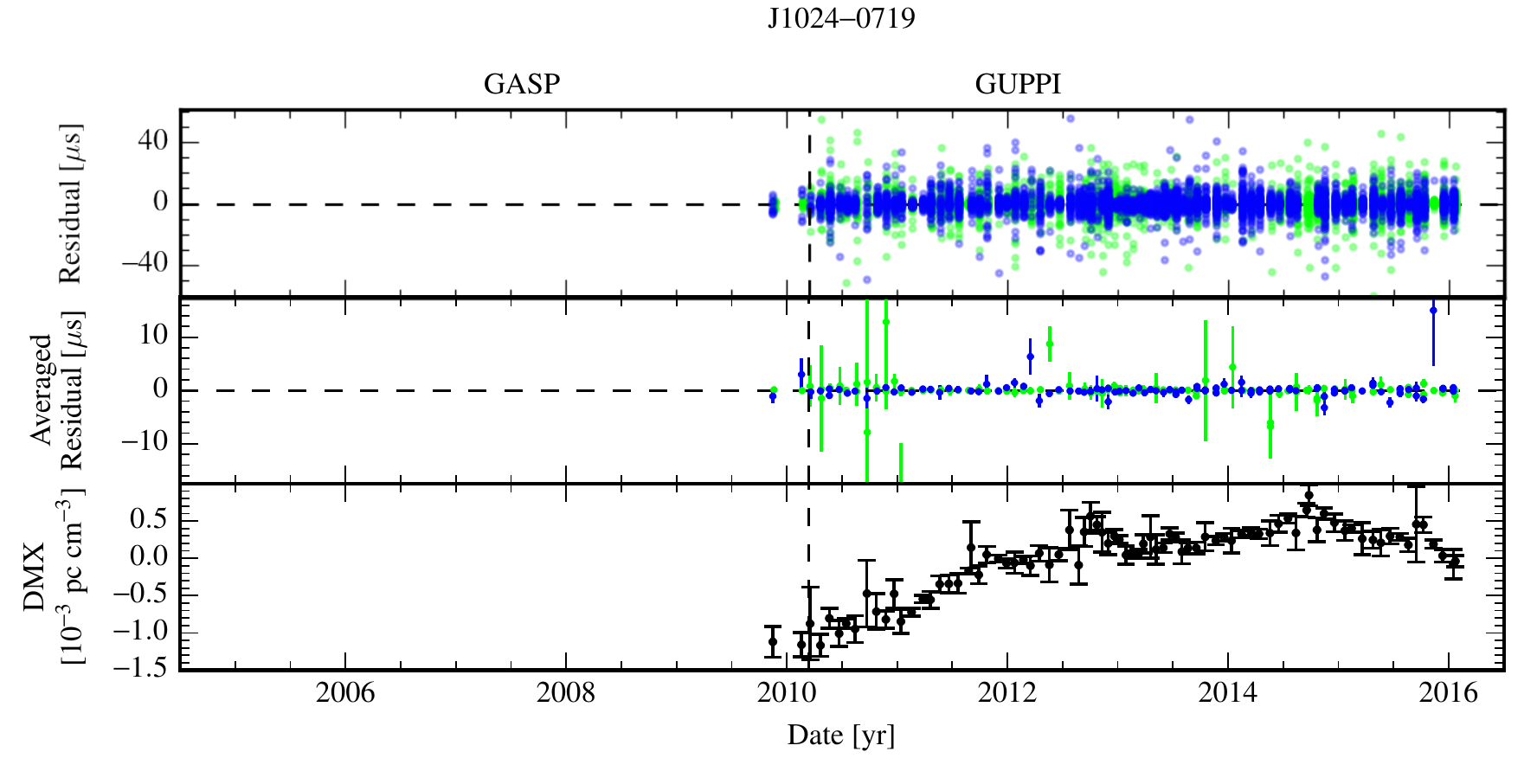}
\caption{Timing summary for PSR J1024-0719. Colors are: Blue: 1.4 GHz, Purple: 2.1 GHz, Green: 820 MHz, Orange: 430 MHz, Red: 327 MHz. In the top panel, individual points are semi-transparent; darker regions arise from the overlap of many points.}
\label{fig:summary-J1024-0719}
\end{figure*}

\begin{figure*}[p]
\centering
\includegraphics[scale=1.0]{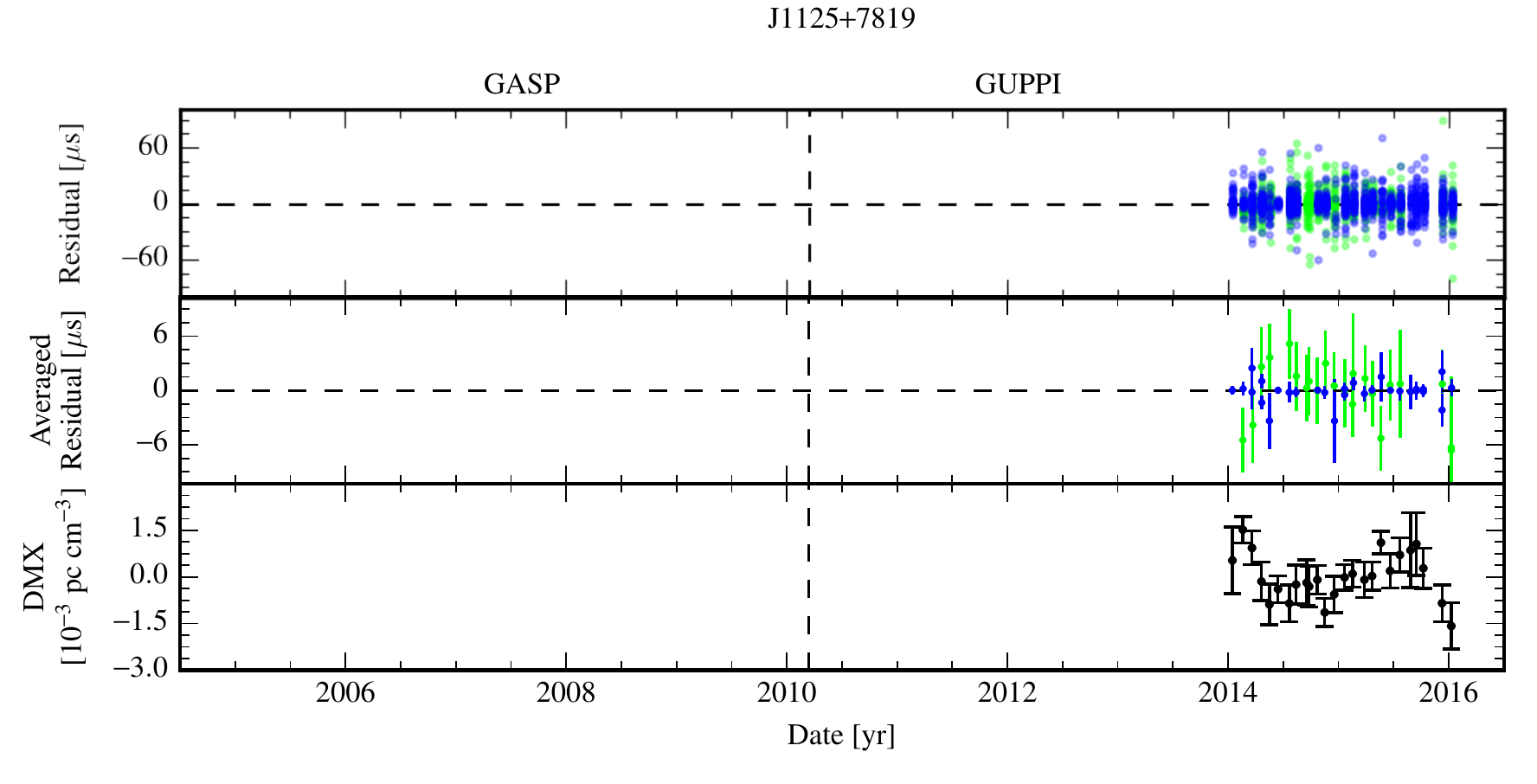}
\caption{Timing summary for PSR J1125+7819. Colors are: Blue: 1.4 GHz, Purple: 2.1 GHz, Green: 820 MHz, Orange: 430 MHz, Red: 327 MHz. In the top panel, individual points are semi-transparent; darker regions arise from the overlap of many points.}
\label{fig:summary-J1125+7819}
\end{figure*}

\begin{figure*}[p]
\centering
\includegraphics[scale=1.0]{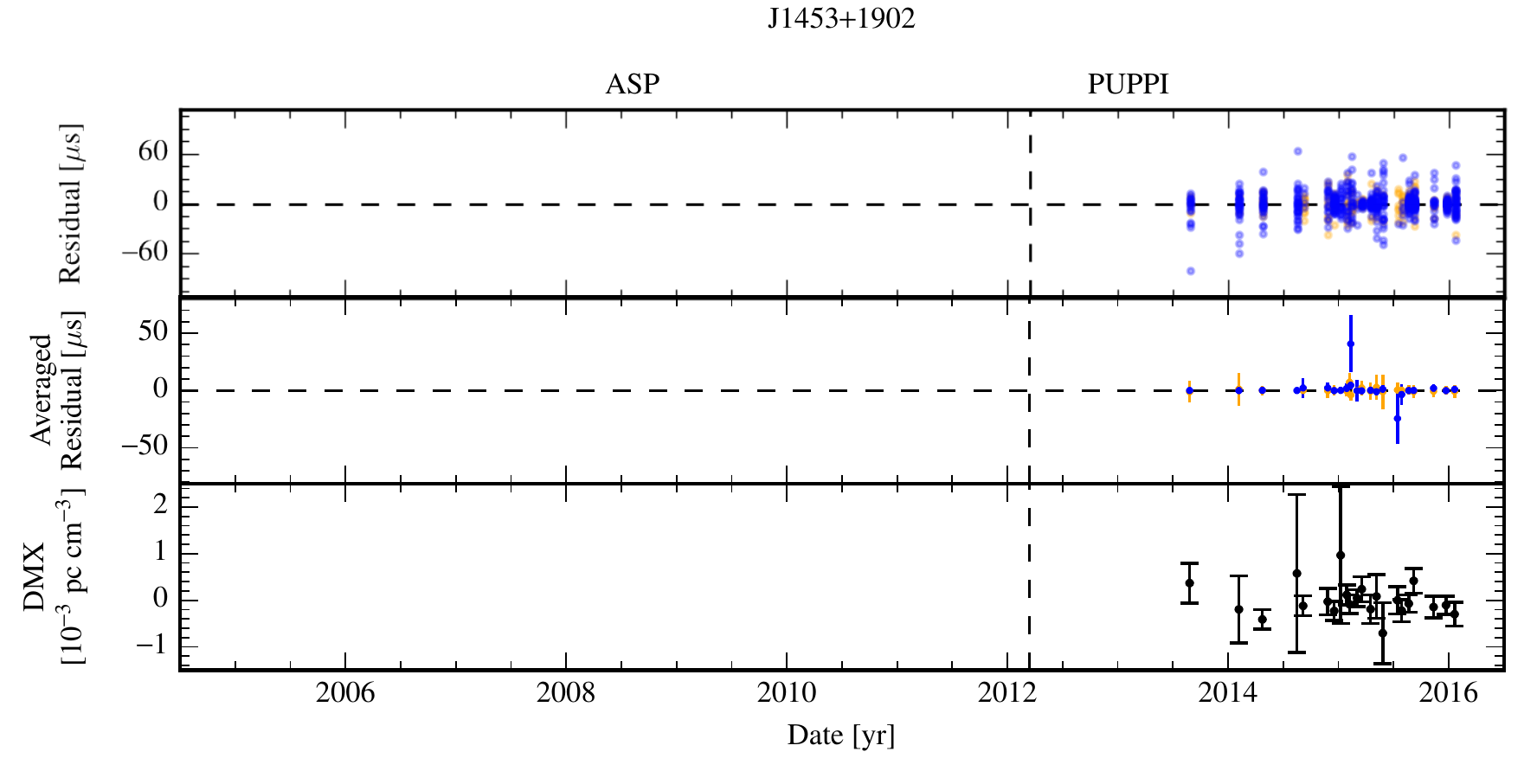}
\caption{Timing summary for PSR J1453+1902. Colors are: Blue: 1.4 GHz, Purple: 2.1 GHz, Green: 820 MHz, Orange: 430 MHz, Red: 327 MHz. In the top panel, individual points are semi-transparent; darker regions arise from the overlap of many points.}
\label{fig:summary-J1453+1902}
\end{figure*}

\begin{figure*}[p]
\centering
\includegraphics[scale=1.0]{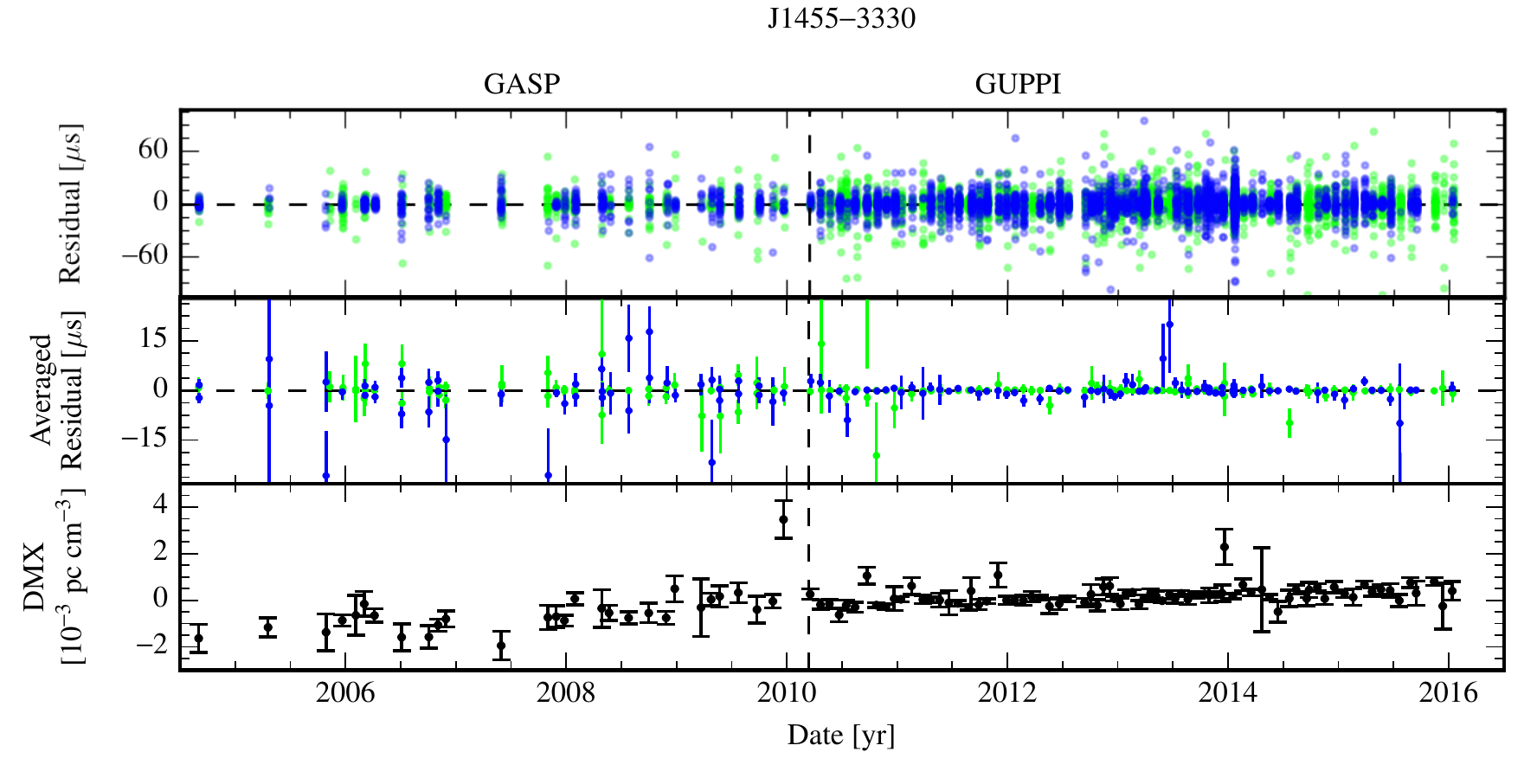}
\caption{Timing summary for PSR J1455-3330. Colors are: Blue: 1.4 GHz, Purple: 2.1 GHz, Green: 820 MHz, Orange: 430 MHz, Red: 327 MHz. In the top panel, individual points are semi-transparent; darker regions arise from the overlap of many points.}
\label{fig:summary-J1455-3330}
\end{figure*}

\begin{figure*}[p]
\centering
\includegraphics[scale=1.0]{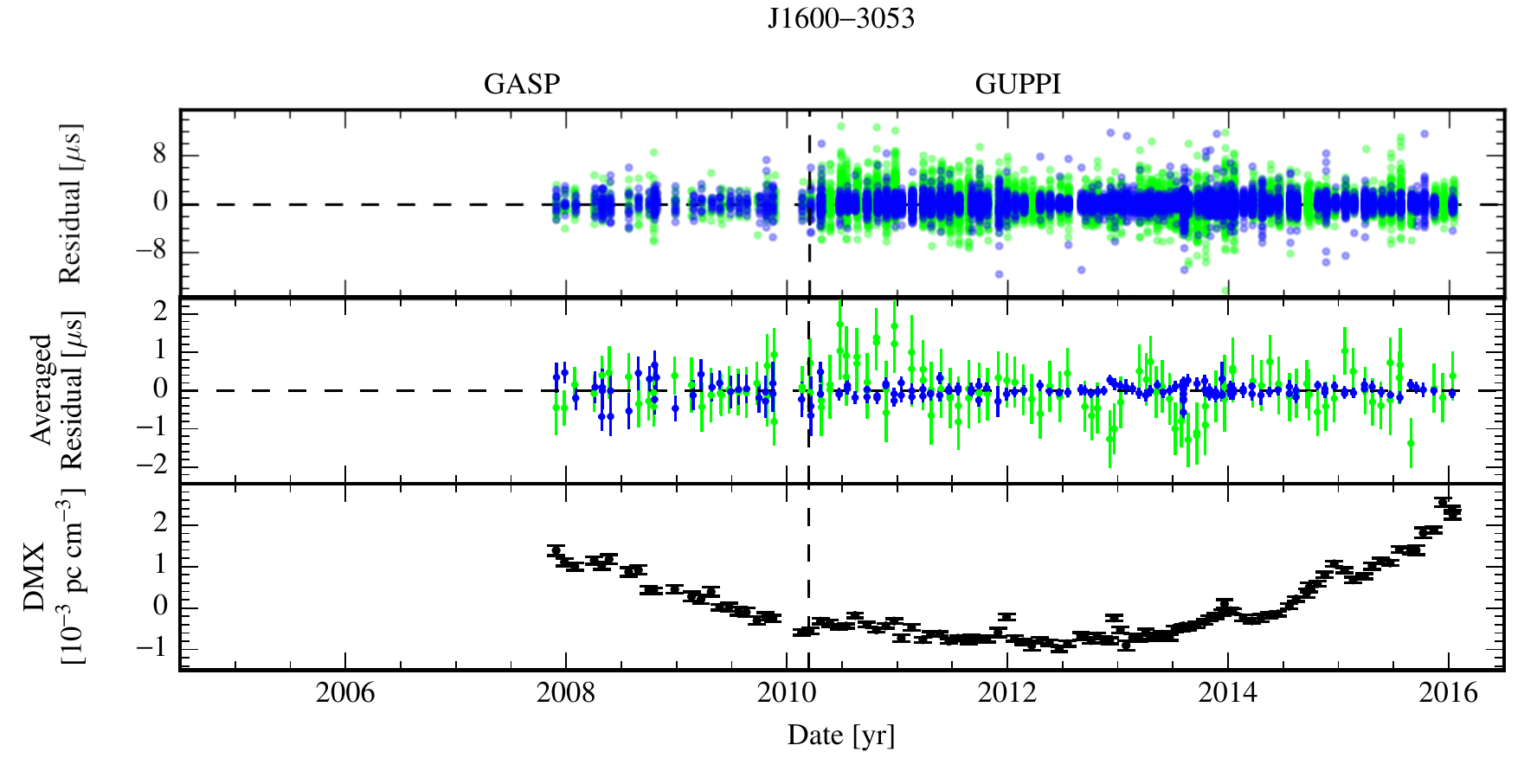}
\caption{Timing summary for PSR J1600-3053. Colors are: Blue: 1.4 GHz, Purple: 2.1 GHz, Green: 820 MHz, Orange: 430 MHz, Red: 327 MHz. In the top panel, individual points are semi-transparent; darker regions arise from the overlap of many points.}
\label{fig:summary-J1600-3053}
\end{figure*}

\begin{figure*}[p]
\centering
\includegraphics[scale=1.0]{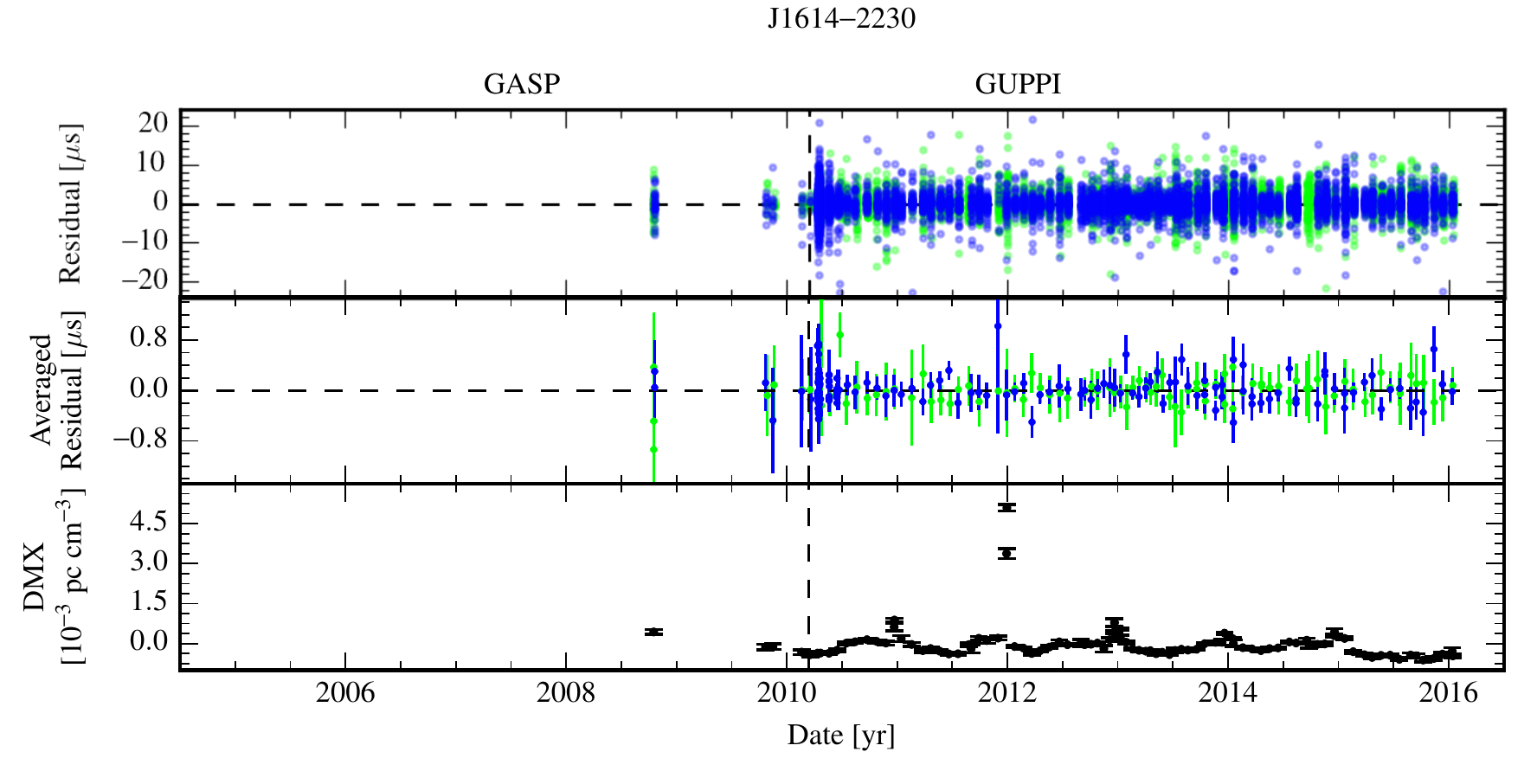}
\caption{Timing summary for PSR J1614-2230. Colors are: Blue: 1.4 GHz, Purple: 2.1 GHz, Green: 820 MHz, Orange: 430 MHz, Red: 327 MHz. In the top panel, individual points are semi-transparent; darker regions arise from the overlap of many points.}
\label{fig:summary-J1614-2230}
\end{figure*}

\begin{figure*}[p]
\centering
\includegraphics[scale=1.0]{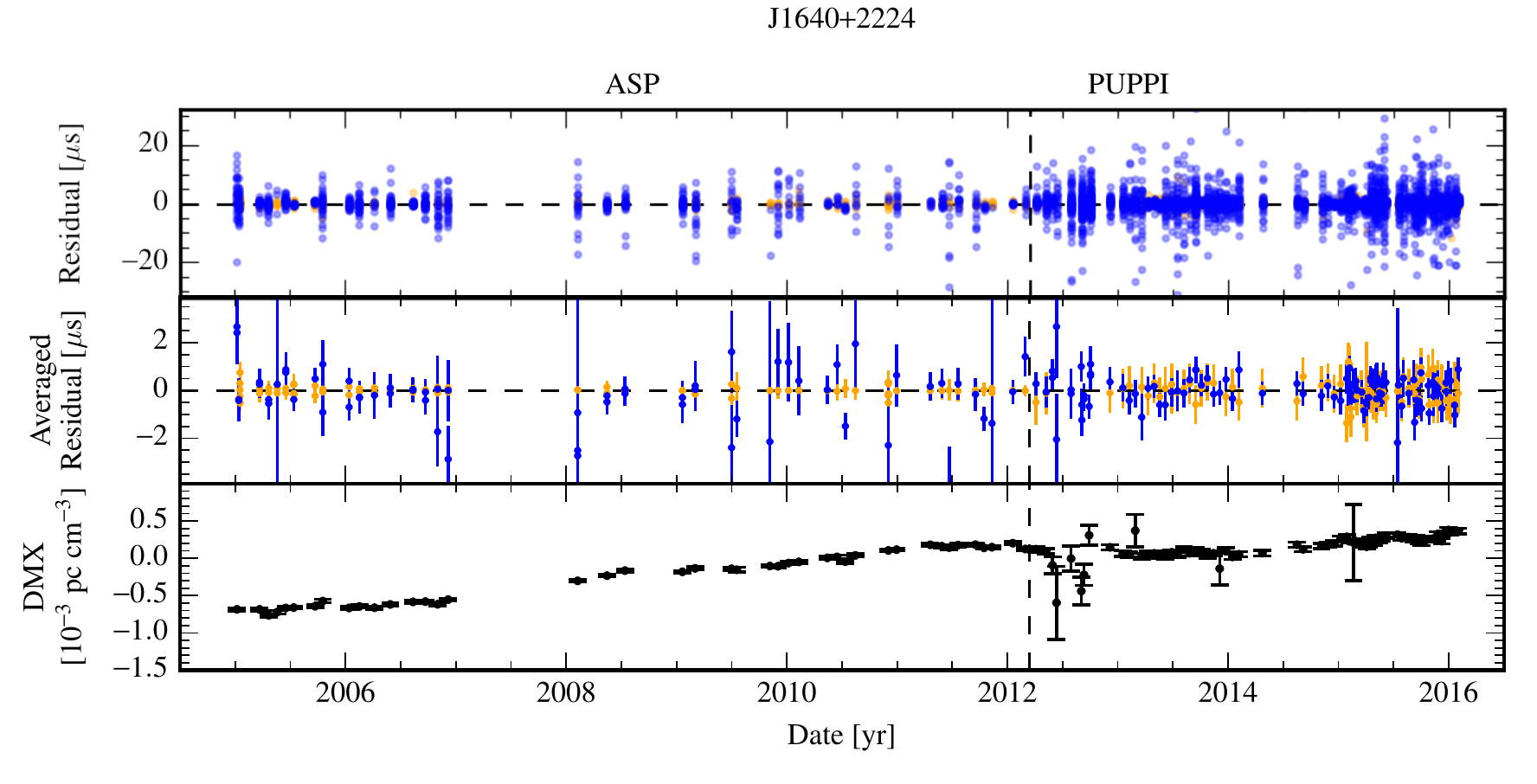}
\caption{Timing summary for PSR J1640+2224. Colors are: Blue: 1.4 GHz, Purple: 2.1 GHz, Green: 820 MHz, Orange: 430 MHz, Red: 327 MHz. In the top panel, individual points are semi-transparent; darker regions arise from the overlap of many points.}
\label{fig:summary-J1640+2224}
\end{figure*}

\begin{figure*}[p]
\centering
\includegraphics[scale=1.0]{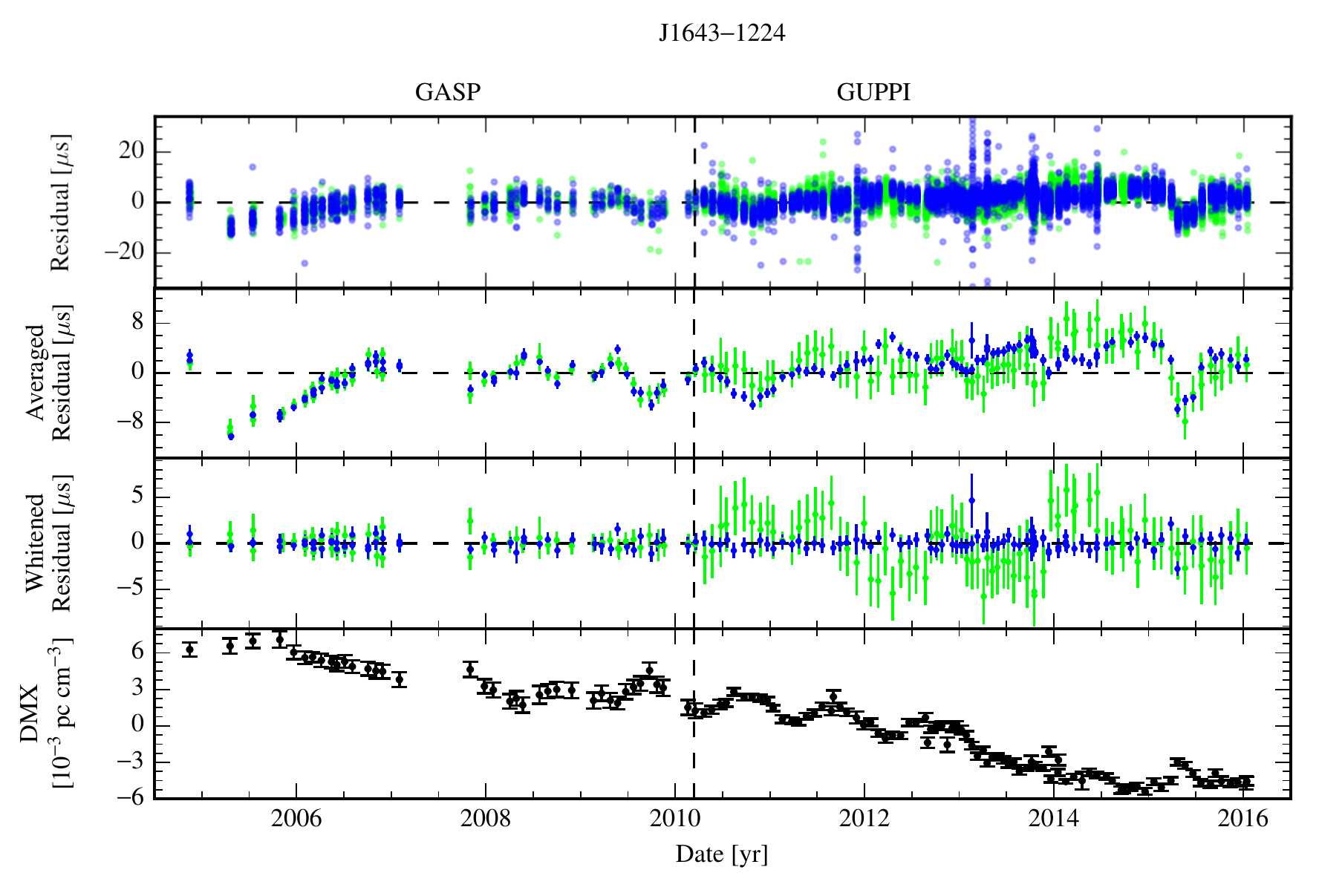}
\caption{Timing summary for PSR J1643-1224. Colors are: Blue: 1.4 GHz, Purple: 2.1 GHz, Green: 820 MHz, Orange: 430 MHz, Red: 327 MHz. In the top panel, individual points are semi-transparent; darker regions arise from the overlap of many points.}
\label{fig:summary-J1643-1224}
\end{figure*}

\begin{figure*}[p]
\centering
\includegraphics[scale=1.0]{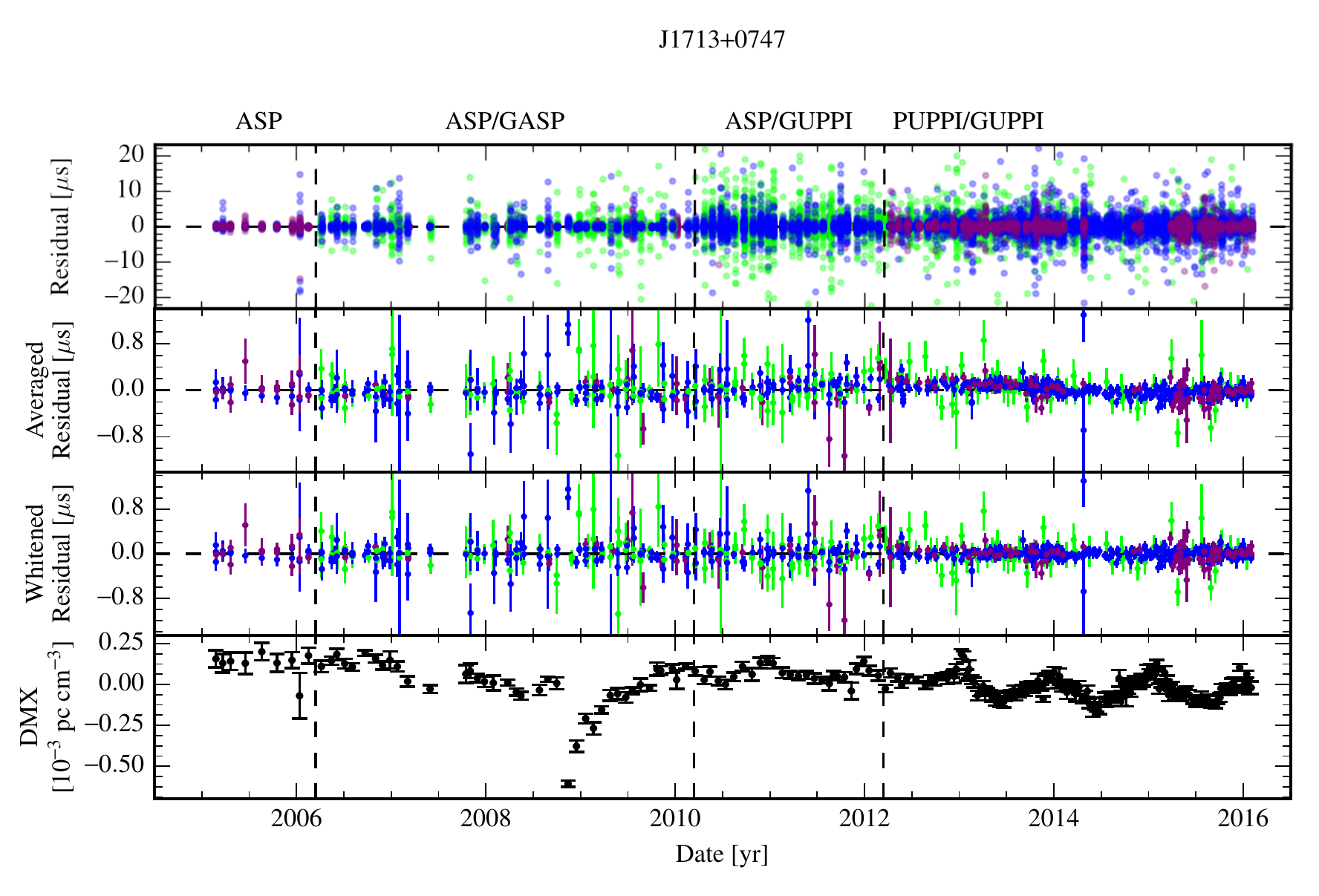}
\caption{Timing summary for PSR J1713+0747. Colors are: Blue: 1.4 GHz, Purple: 2.1 GHz, Green: 820 MHz, Orange: 430 MHz, Red: 327 MHz. In the top panel, individual points are semi-transparent; darker regions arise from the overlap of many points.}
\label{fig:summary-J1713+0747}
\end{figure*}

\begin{figure*}[p]
\centering
\includegraphics[scale=1.0]{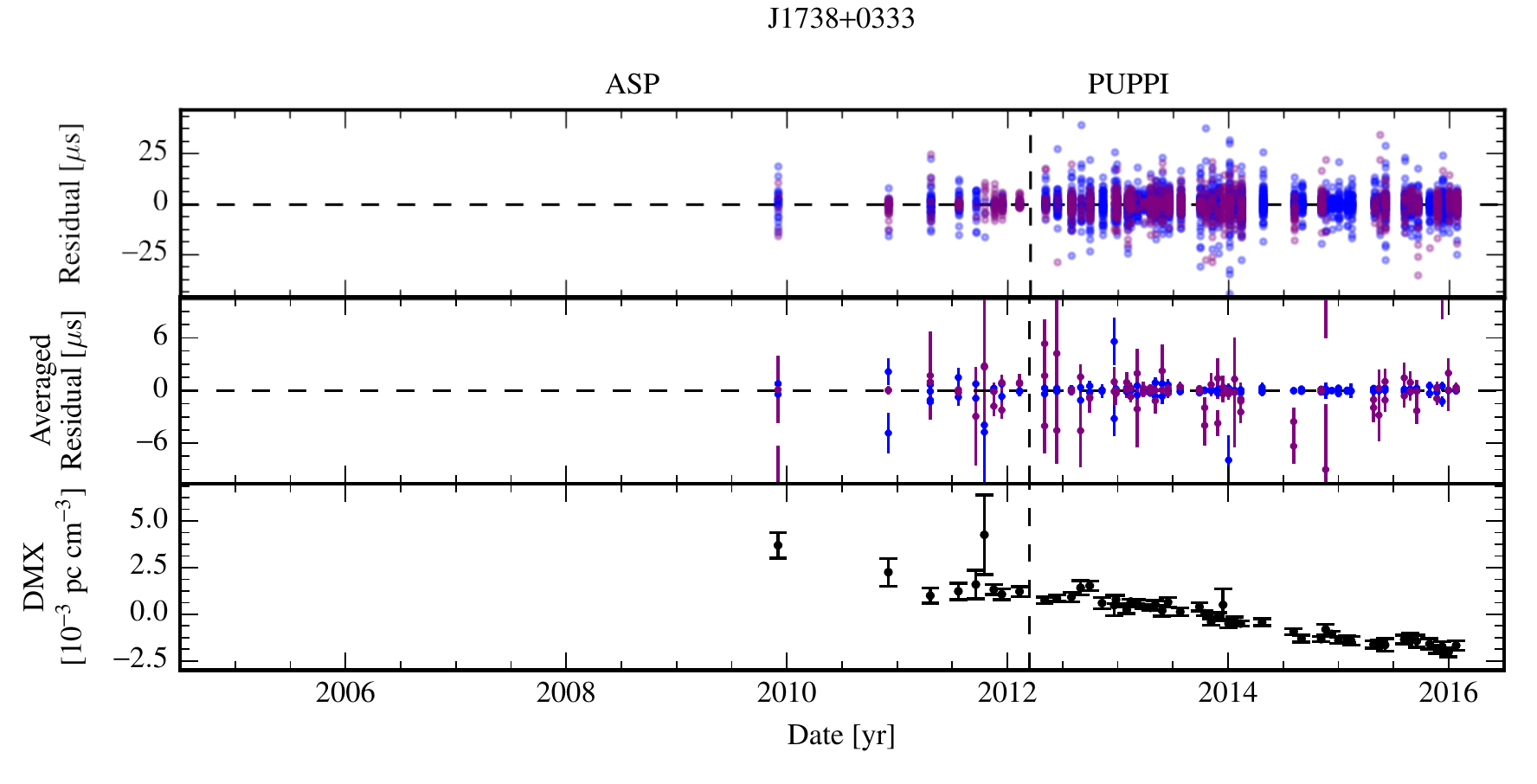}
\caption{Timing summary for PSR J1738+0333. Colors are: Blue: 1.4 GHz, Purple: 2.1 GHz, Green: 820 MHz, Orange: 430 MHz, Red: 327 MHz. In the top panel, individual points are semi-transparent; darker regions arise from the overlap of many points.}
\label{fig:summary-J1738+0333}
\end{figure*}

\begin{figure*}[p]
\centering
\includegraphics[scale=1.0]{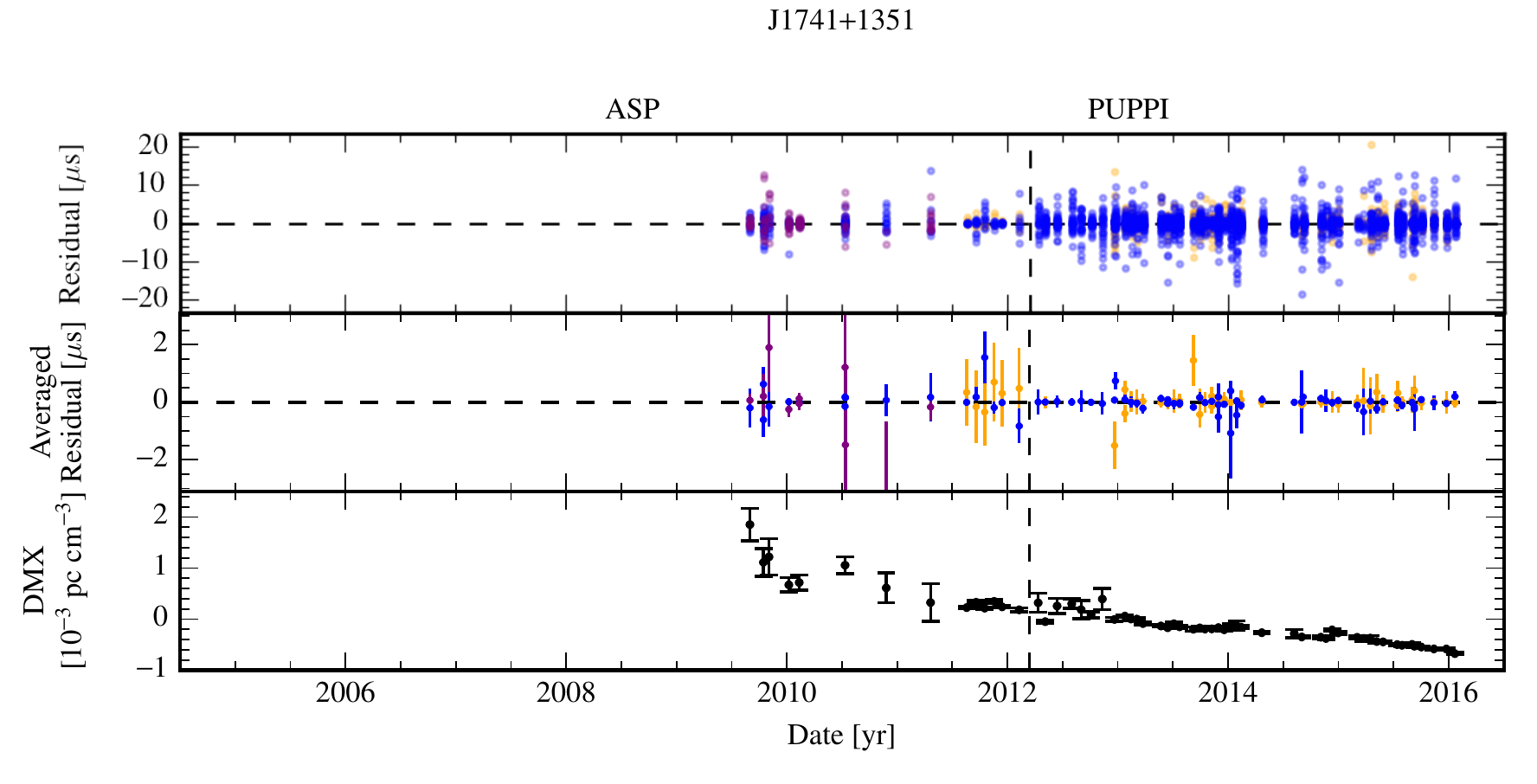}
\caption{Timing summary for PSR J1741+1351. Colors are: Blue: 1.4 GHz, Purple: 2.1 GHz, Green: 820 MHz, Orange: 430 MHz, Red: 327 MHz. In the top panel, individual points are semi-transparent; darker regions arise from the overlap of many points.}
\label{fig:summary-J1741+1351}
\end{figure*}

\begin{figure*}[p]
\centering
\includegraphics[scale=1.0]{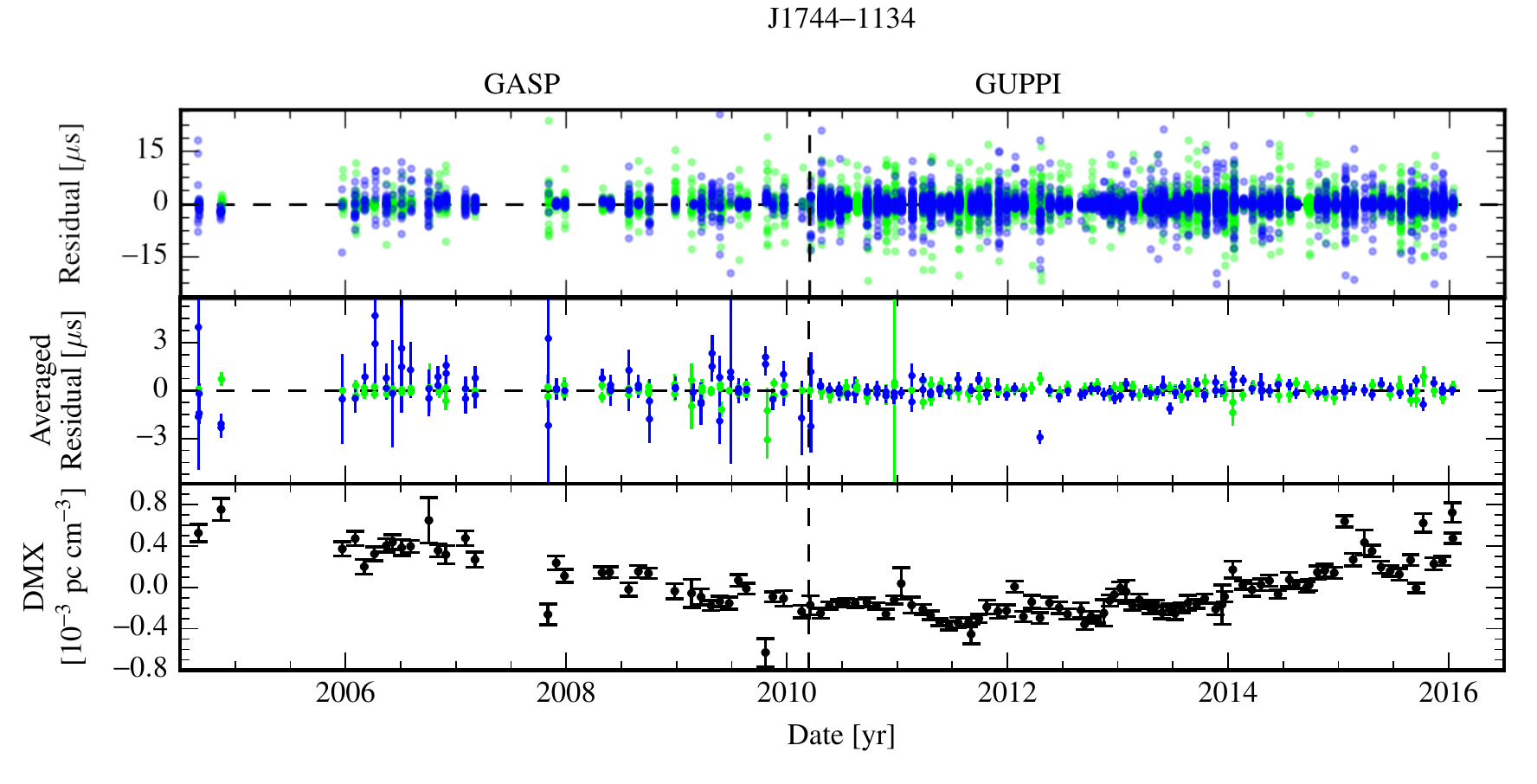}
\caption{Timing summary for PSR J1744-1134. Colors are: Blue: 1.4 GHz, Purple: 2.1 GHz, Green: 820 MHz, Orange: 430 MHz, Red: 327 MHz. In the top panel, individual points are semi-transparent; darker regions arise from the overlap of many points.}
\label{fig:summary-J1744-1134}
\end{figure*}

\begin{figure*}[p]
\centering
\includegraphics[scale=1.0]{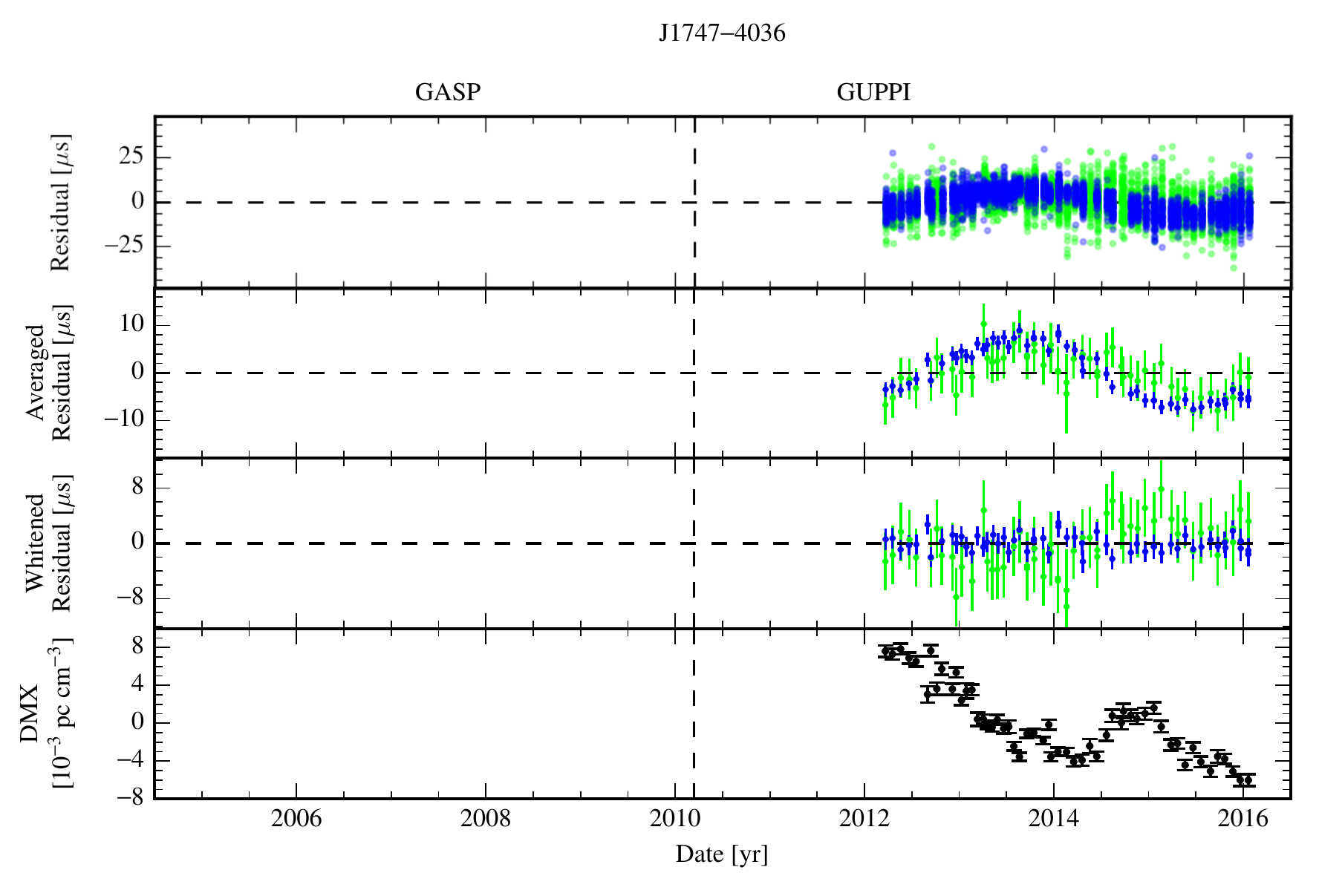}
\caption{Timing summary for PSR J1747-4036. Colors are: Blue: 1.4 GHz, Purple: 2.1 GHz, Green: 820 MHz, Orange: 430 MHz, Red: 327 MHz. In the top panel, individual points are semi-transparent; darker regions arise from the overlap of many points.}
\label{fig:summary-J1747-4036}
\end{figure*}

\begin{figure*}[p]
\centering
\includegraphics[scale=1.0]{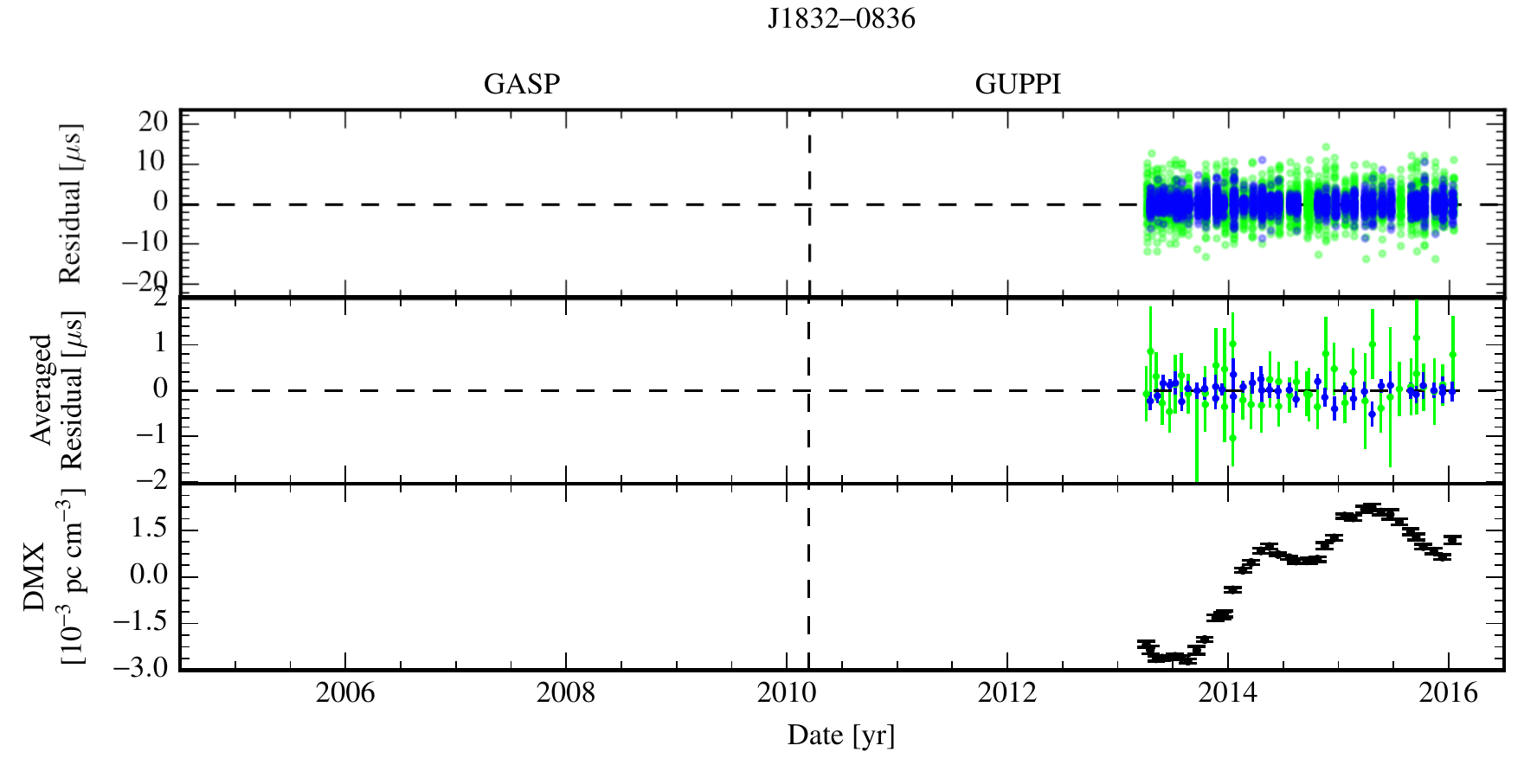}
\caption{Timing summary for PSR J1832-0836. Colors are: Blue: 1.4 GHz, Purple: 2.1 GHz, Green: 820 MHz, Orange: 430 MHz, Red: 327 MHz. In the top panel, individual points are semi-transparent; darker regions arise from the overlap of many points.}
\label{fig:summary-J1832-0836}
\end{figure*}

\begin{figure*}[p]
\centering
\includegraphics[scale=1.0]{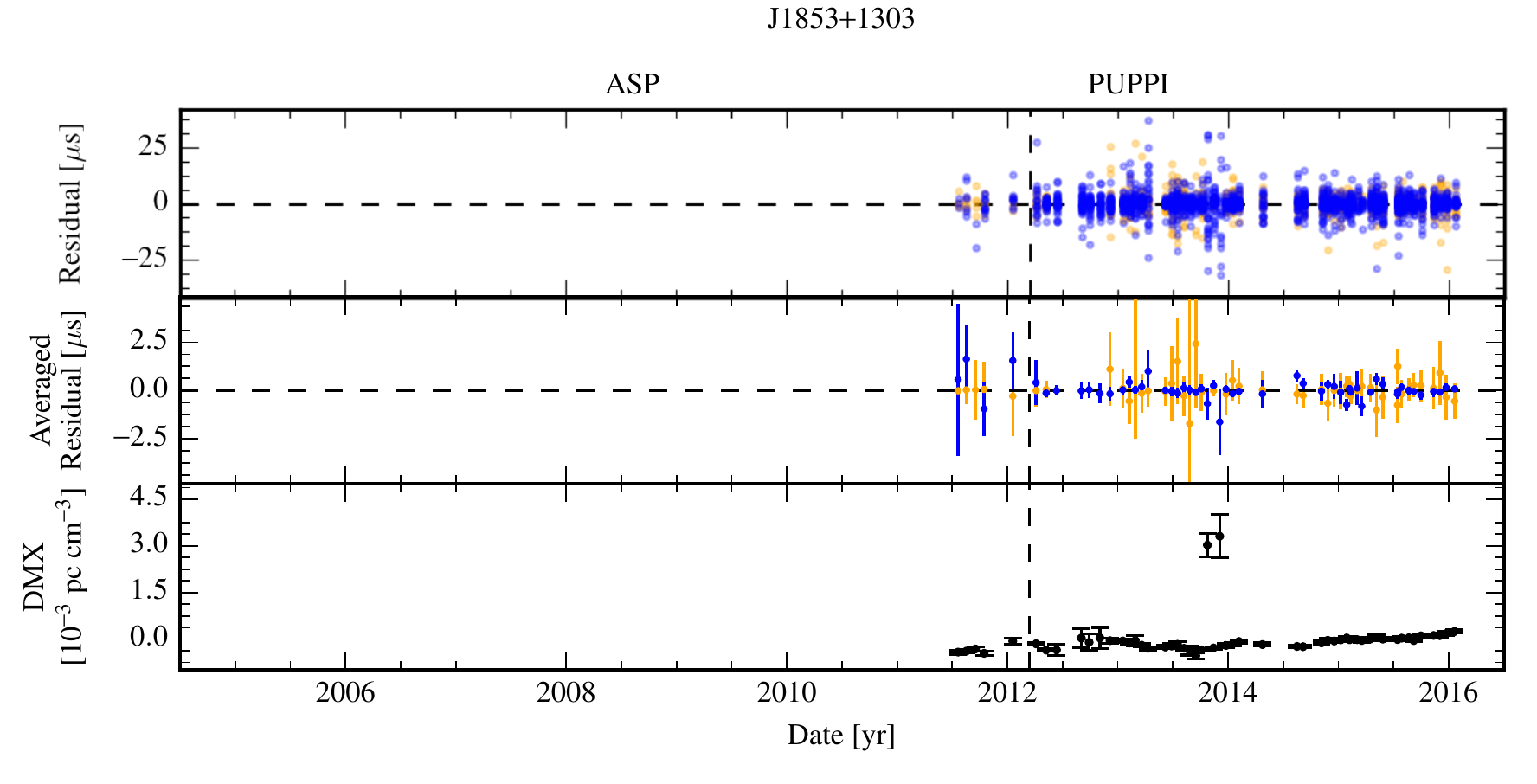}
\caption{Timing summary for PSR J1853+1303. Colors are: Blue: 1.4 GHz, Purple: 2.1 GHz, Green: 820 MHz, Orange: 430 MHz, Red: 327 MHz. In the top panel, individual points are semi-transparent; darker regions arise from the overlap of many points.}
\label{fig:summary-J1853+1303}
\end{figure*}

\begin{figure*}[p]
\centering
\includegraphics[scale=1.0]{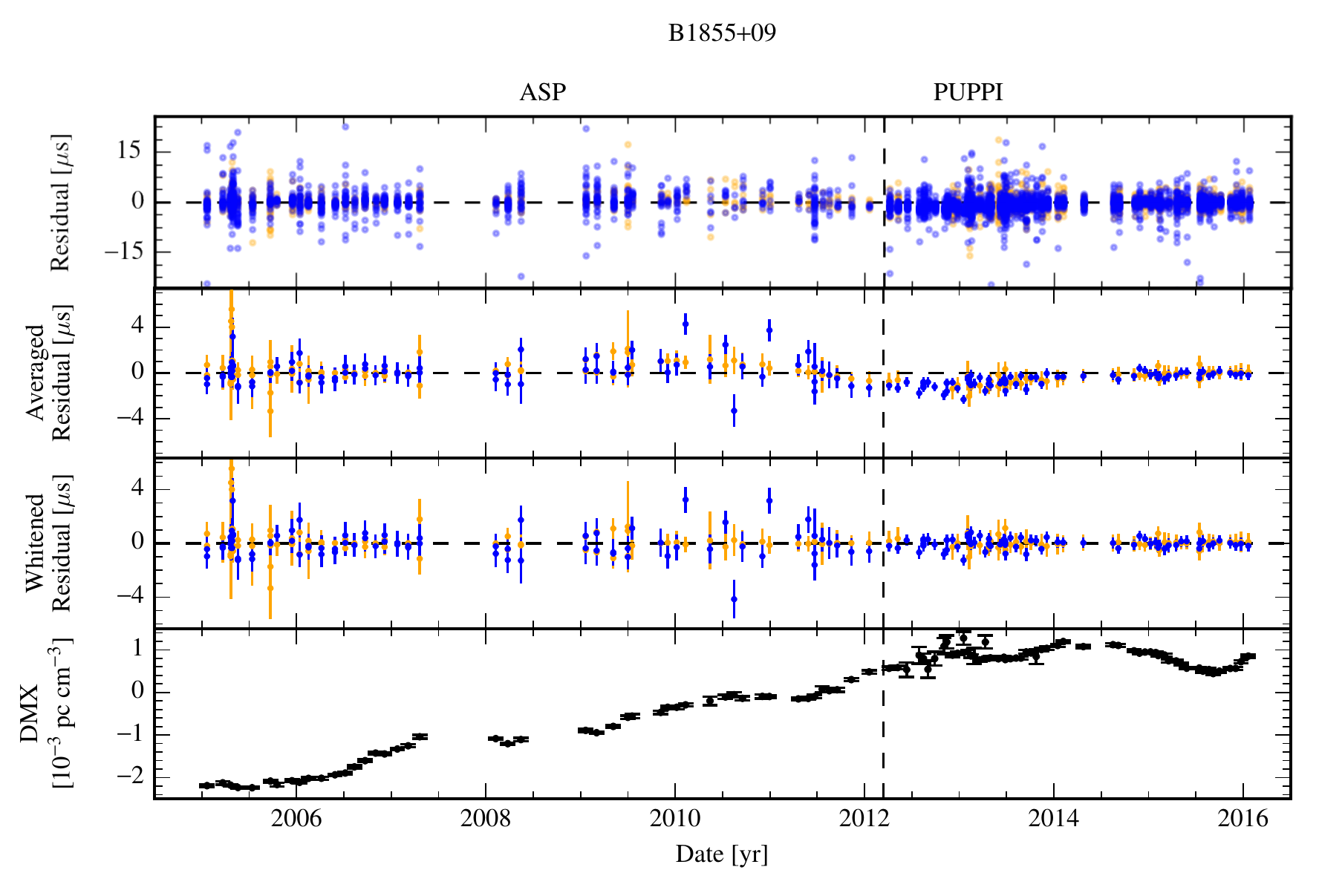}
\caption{Timing summary for PSR B1855+09. Colors are: Blue: 1.4 GHz, Purple: 2.1 GHz, Green: 820 MHz, Orange: 430 MHz, Red: 327 MHz. In the top panel, individual points are semi-transparent; darker regions arise from the overlap of many points.}
\label{fig:summary-B1855+09}
\end{figure*}

\begin{figure*}[p]
\centering
\includegraphics[scale=1.0]{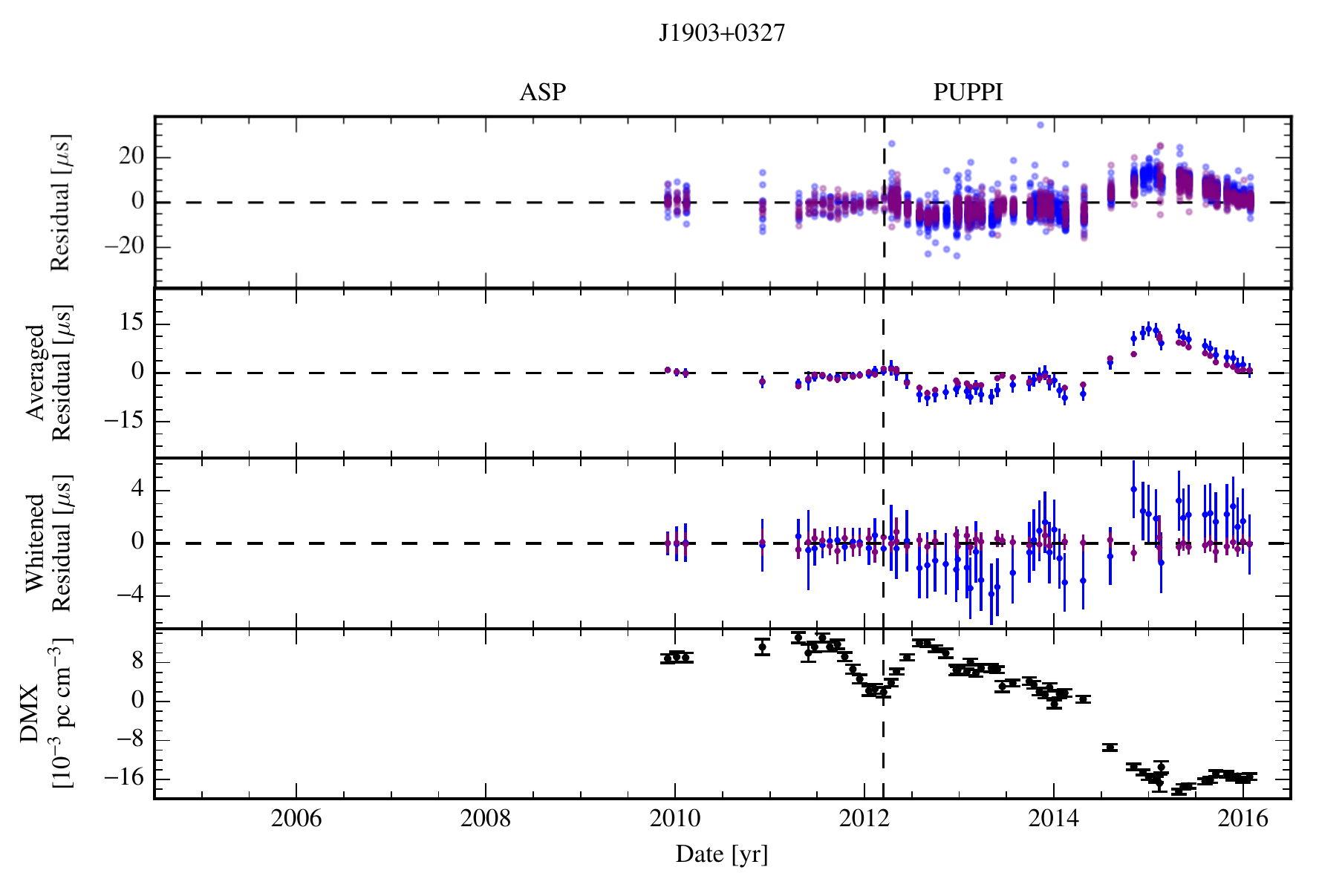}
\caption{Timing summary for PSR J1903+0327. Colors are: Blue: 1.4 GHz, Purple: 2.1 GHz, Green: 820 MHz, Orange: 430 MHz, Red: 327 MHz. In the top panel, individual points are semi-transparent; darker regions arise from the overlap of many points.}
\label{fig:summary-J1903+0327}
\end{figure*}

\begin{figure*}[p]
\centering
\includegraphics[scale=1.0]{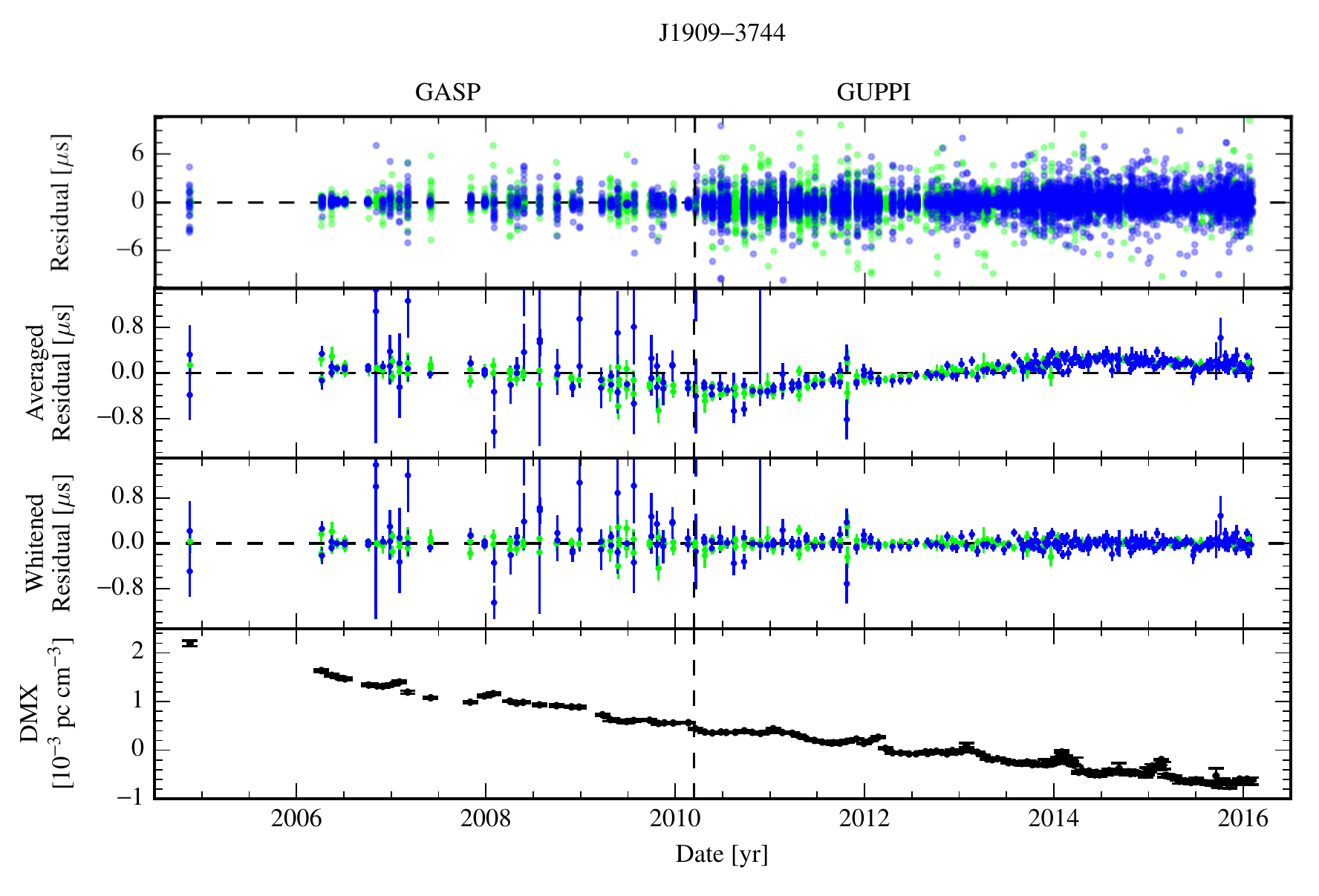}
\caption{Timing summary for PSR J1909-3744. Colors are: Blue: 1.4 GHz, Purple: 2.1 GHz, Green: 820 MHz, Orange: 430 MHz, Red: 327 MHz. In the top panel, individual points are semi-transparent; darker regions arise from the overlap of many points.}
\label{fig:summary-J1909-3744}
\end{figure*}

\begin{figure*}[p]
\centering
\includegraphics[scale=1.0]{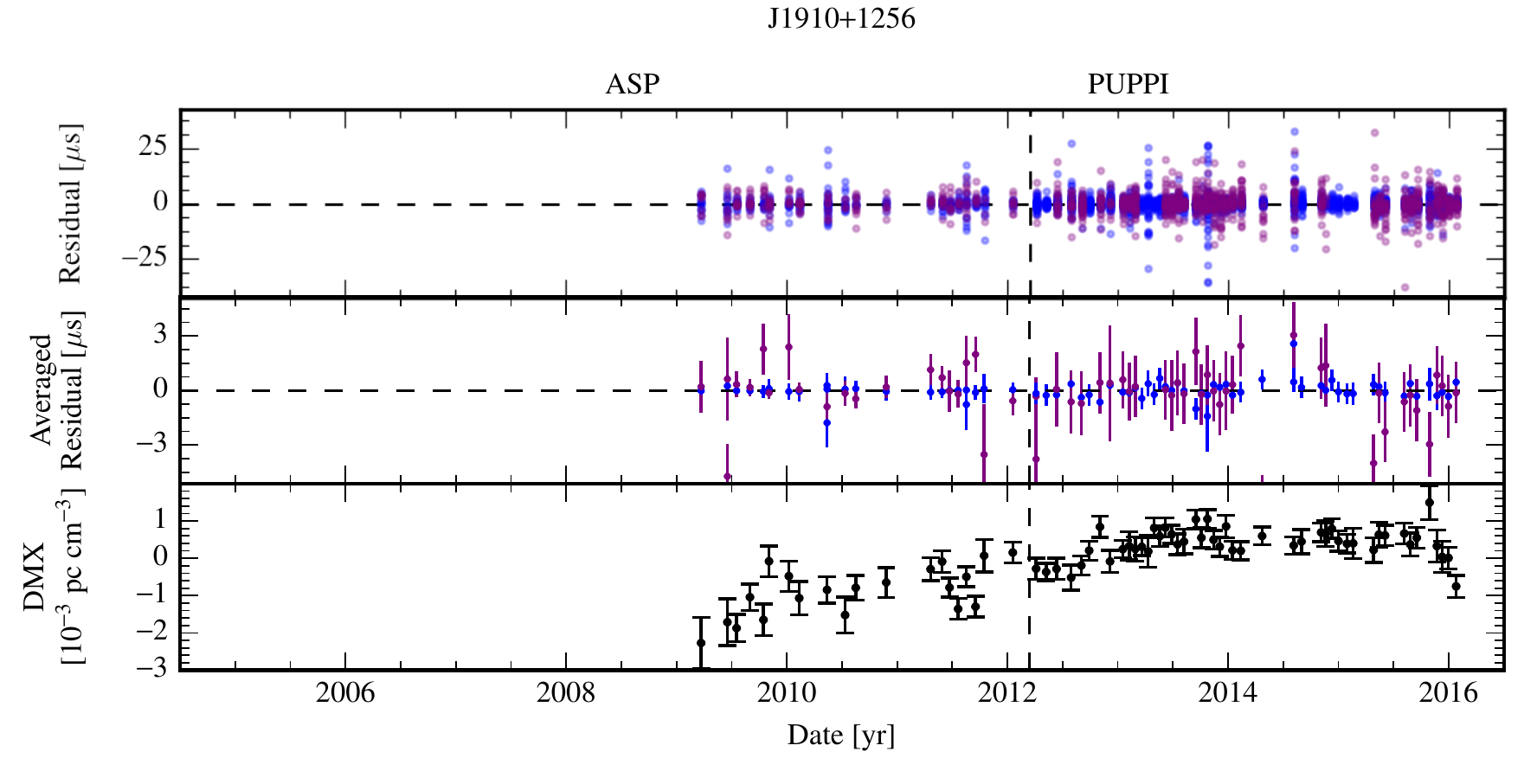}
\caption{Timing summary for PSR J1910+1256. Colors are: Blue: 1.4 GHz, Purple: 2.1 GHz, Green: 820 MHz, Orange: 430 MHz, Red: 327 MHz. In the top panel, individual points are semi-transparent; darker regions arise from the overlap of many points.}
\label{fig:summary-J1910+1256}
\end{figure*}

\begin{figure*}[p]
\centering
\includegraphics[scale=1.0]{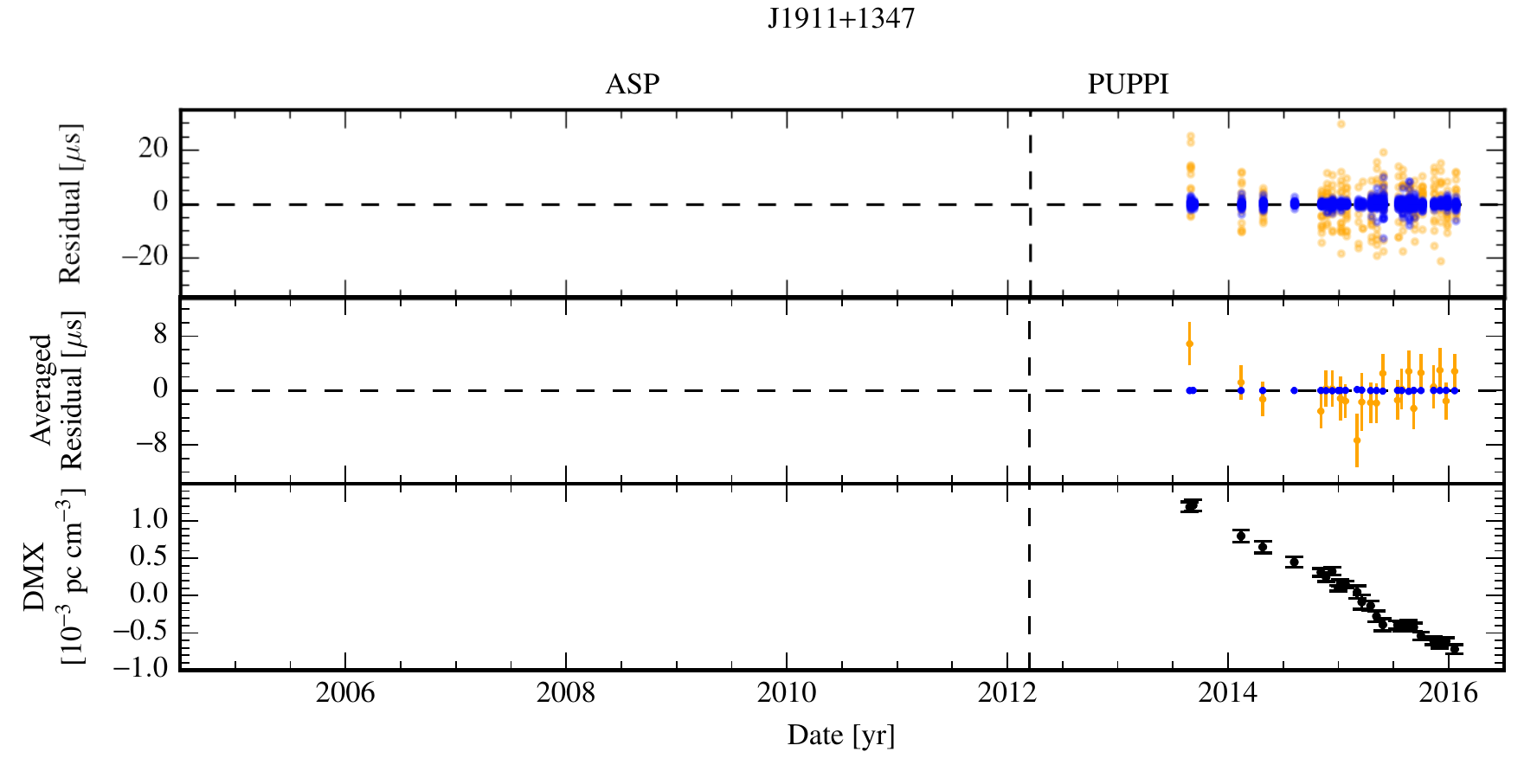}
\caption{Timing summary for PSR J1911+1347. Colors are: Blue: 1.4 GHz, Purple: 2.1 GHz, Green: 820 MHz, Orange: 430 MHz, Red: 327 MHz. In the top panel, individual points are semi-transparent; darker regions arise from the overlap of many points.}
\label{fig:summary-J1911+1347}
\end{figure*}

\begin{figure*}[p]
\centering
\includegraphics[scale=1.0]{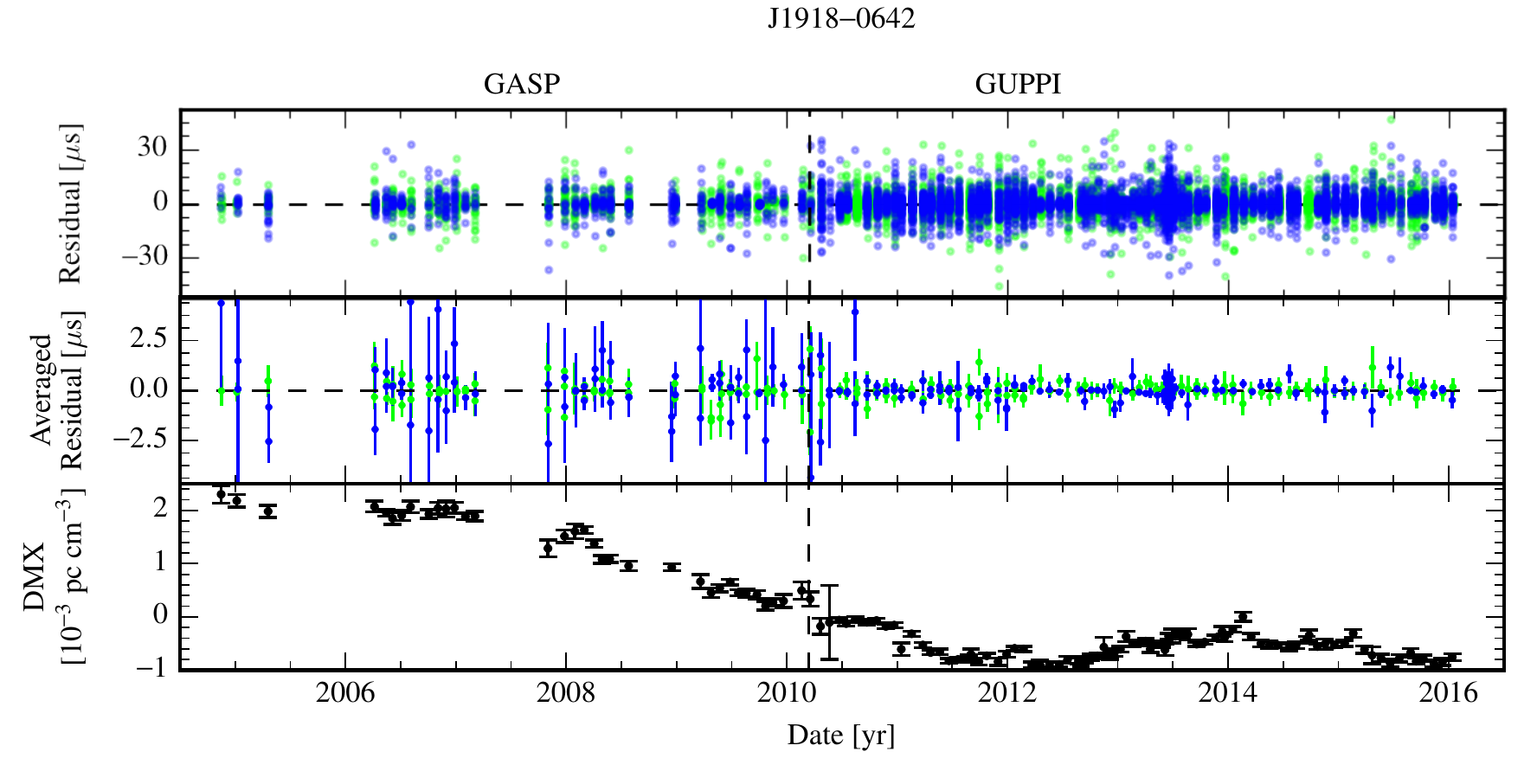}
\caption{Timing summary for PSR J1918-0642. Colors are: Blue: 1.4 GHz, Purple: 2.1 GHz, Green: 820 MHz, Orange: 430 MHz, Red: 327 MHz. In the top panel, individual points are semi-transparent; darker regions arise from the overlap of many points.}
\label{fig:summary-J1918-0642}
\end{figure*}

\begin{figure*}[p]
\centering
\includegraphics[scale=1.0]{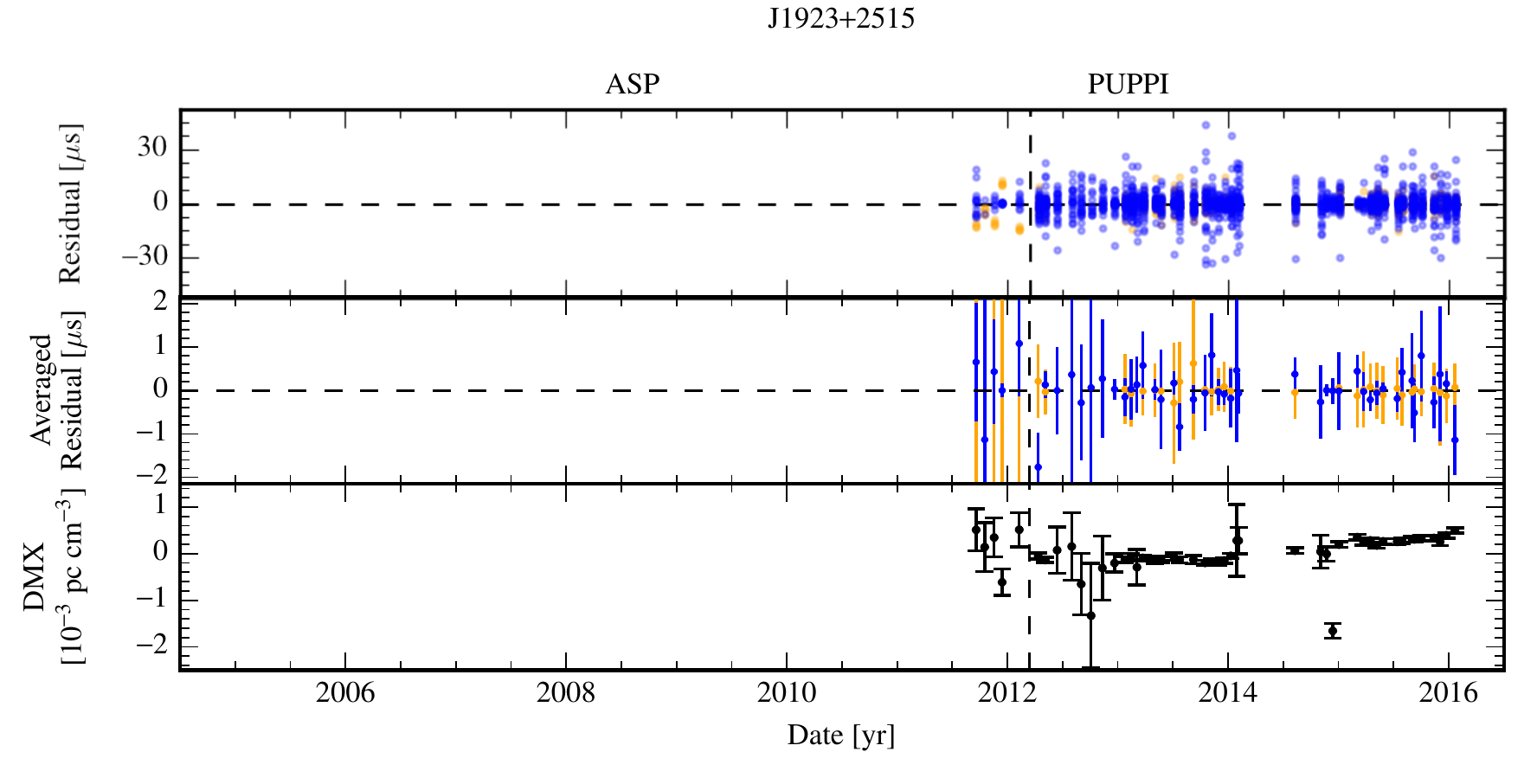}
\caption{Timing summary for PSR J1923+2515. Colors are: Blue: 1.4 GHz, Purple: 2.1 GHz, Green: 820 MHz, Orange: 430 MHz, Red: 327 MHz. In the top panel, individual points are semi-transparent; darker regions arise from the overlap of many points.}
\label{fig:summary-J1923+2515}
\end{figure*}

\begin{figure*}[p]
\centering
\includegraphics[scale=1.0]{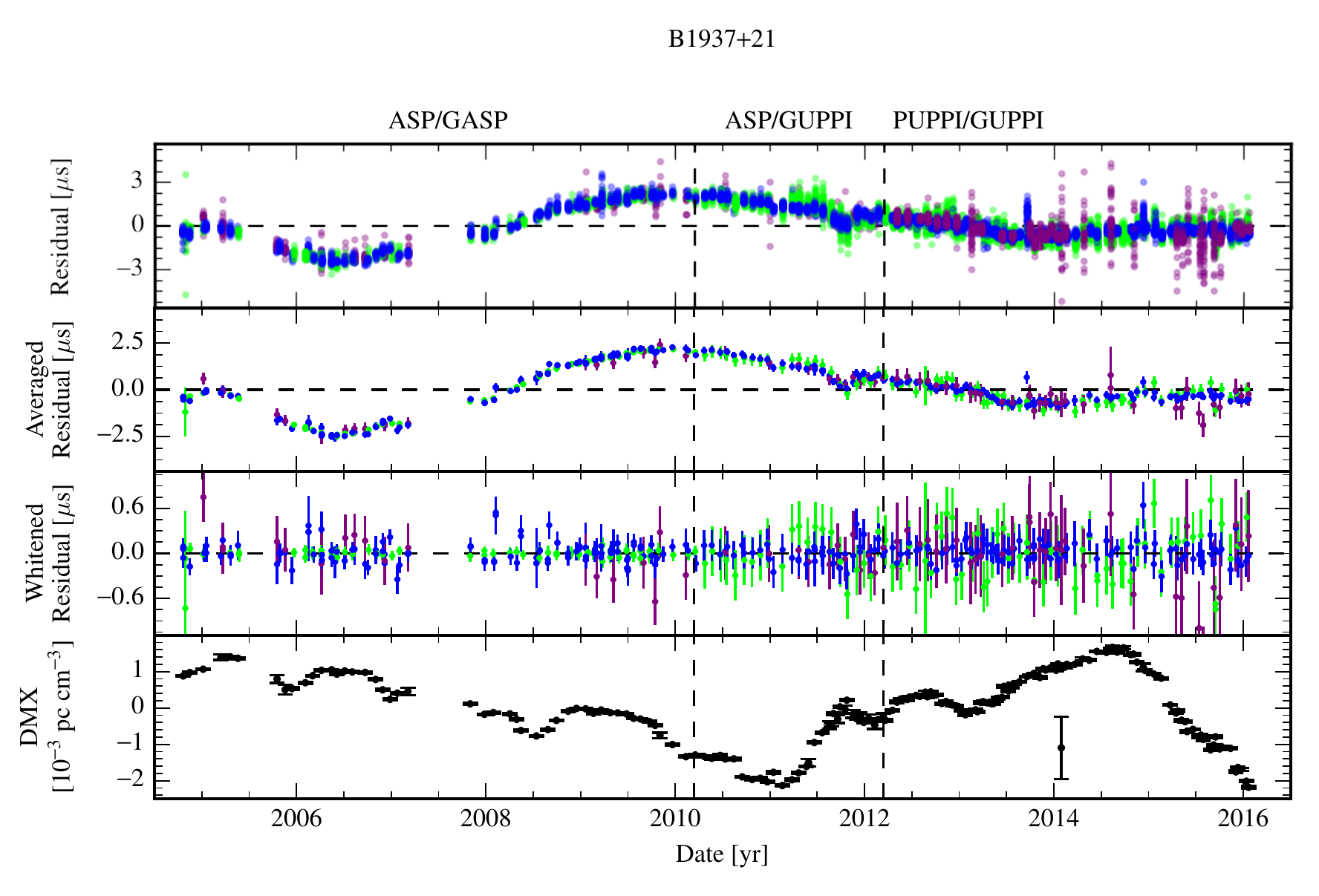}
\caption{Timing summary for PSR B1937+21. Colors are: Blue: 1.4 GHz, Purple: 2.1 GHz, Green: 820 MHz, Orange: 430 MHz, Red: 327 MHz. In the top panel, individual points are semi-transparent; darker regions arise from the overlap of many points.}
\label{fig:summary-B1937+21}
\end{figure*}

\begin{figure*}[p]
\centering
\includegraphics[scale=1.0]{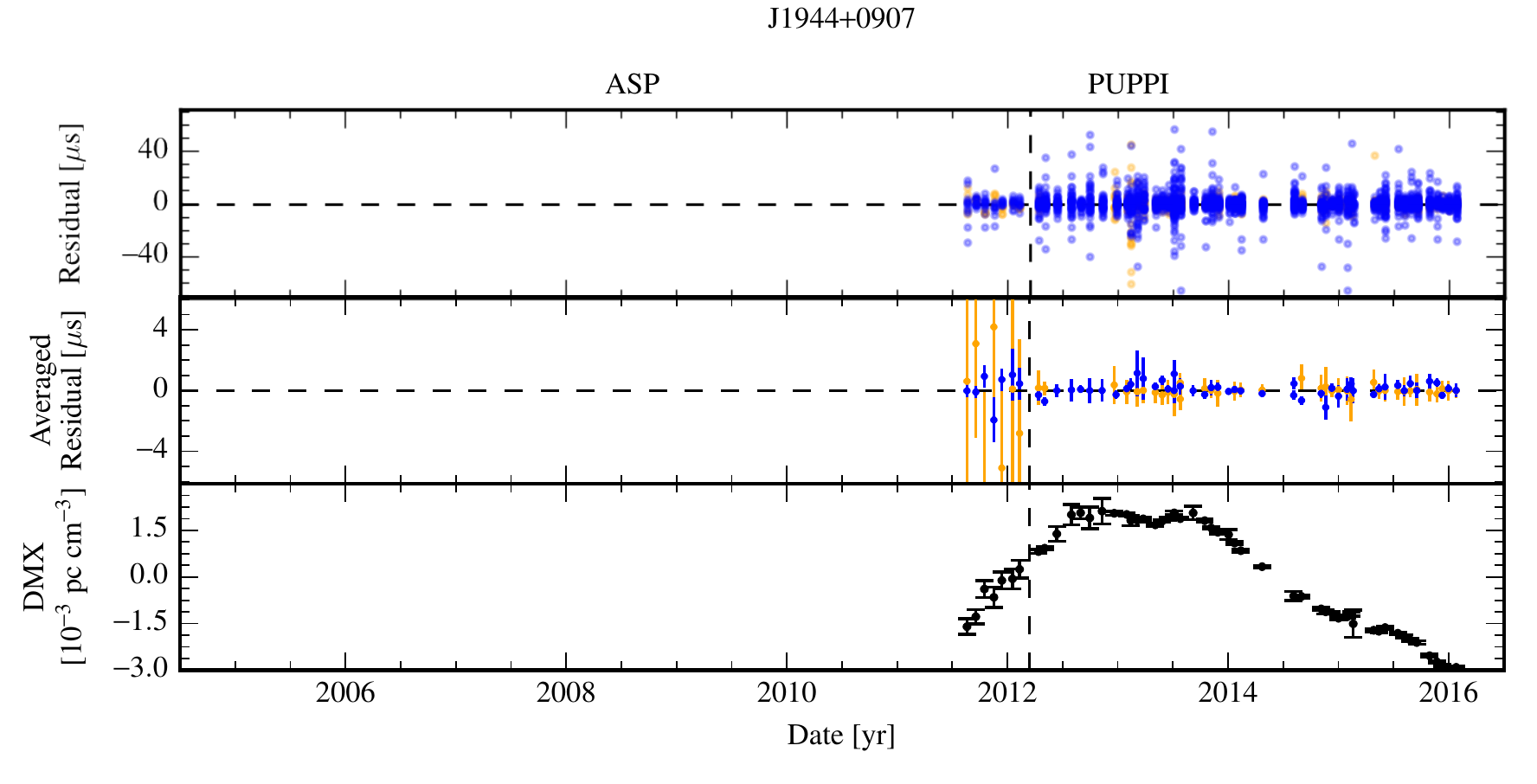}
\caption{Timing summary for PSR J1944+0907. Colors are: Blue: 1.4 GHz, Purple: 2.1 GHz, Green: 820 MHz, Orange: 430 MHz, Red: 327 MHz. In the top panel, individual points are semi-transparent; darker regions arise from the overlap of many points.}
\label{fig:summary-J1944+0907}
\end{figure*}

\begin{figure*}[p]
\centering
\includegraphics[scale=1.0]{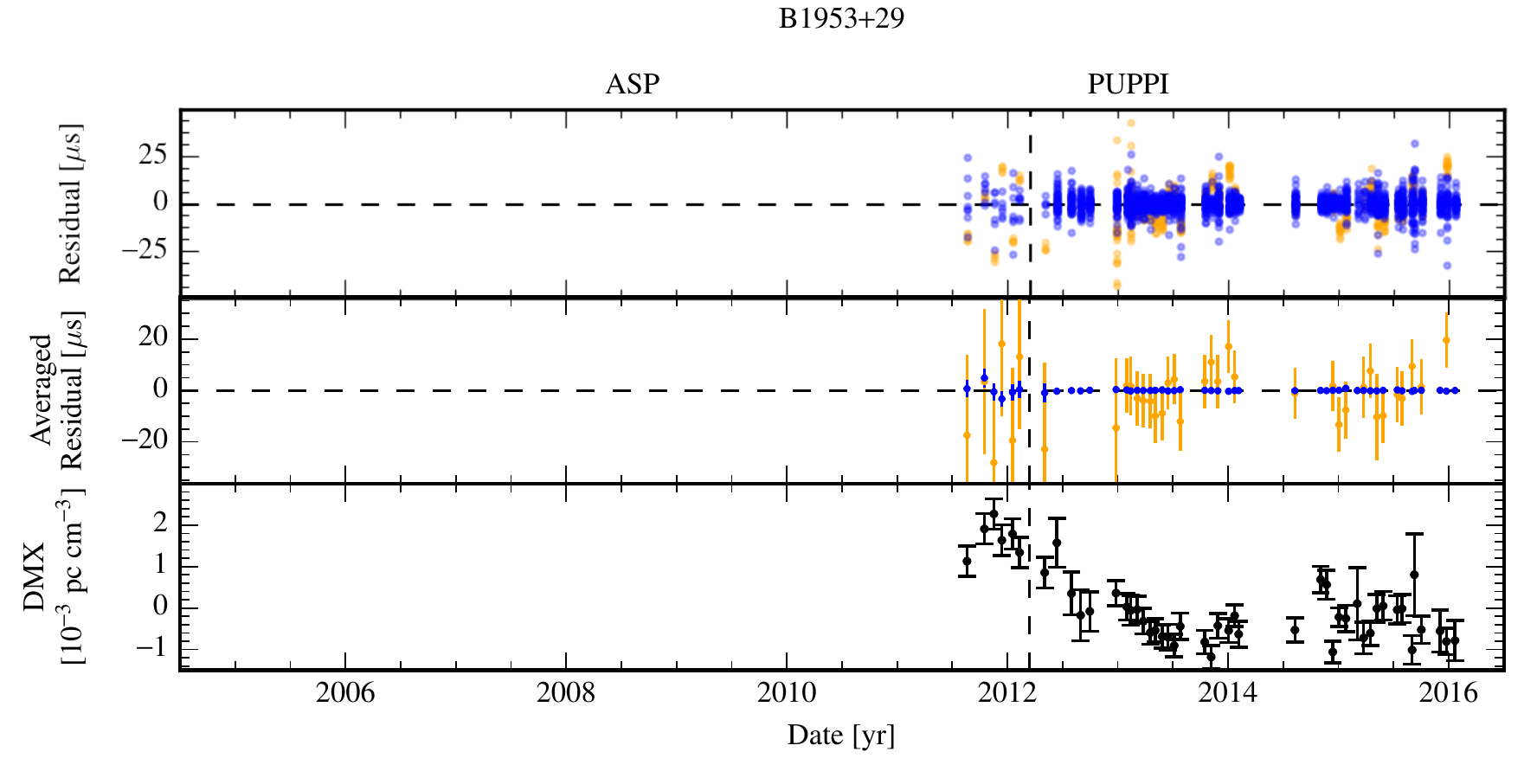}
\caption{Timing summary for PSR B1953+29. Colors are: Blue: 1.4 GHz, Purple: 2.1 GHz, Green: 820 MHz, Orange: 430 MHz, Red: 327 MHz. In the top panel, individual points are semi-transparent; darker regions arise from the overlap of many points.}
\label{fig:summary-B1953+29}
\end{figure*}

\begin{figure*}[p]
\centering
\includegraphics[scale=1.0]{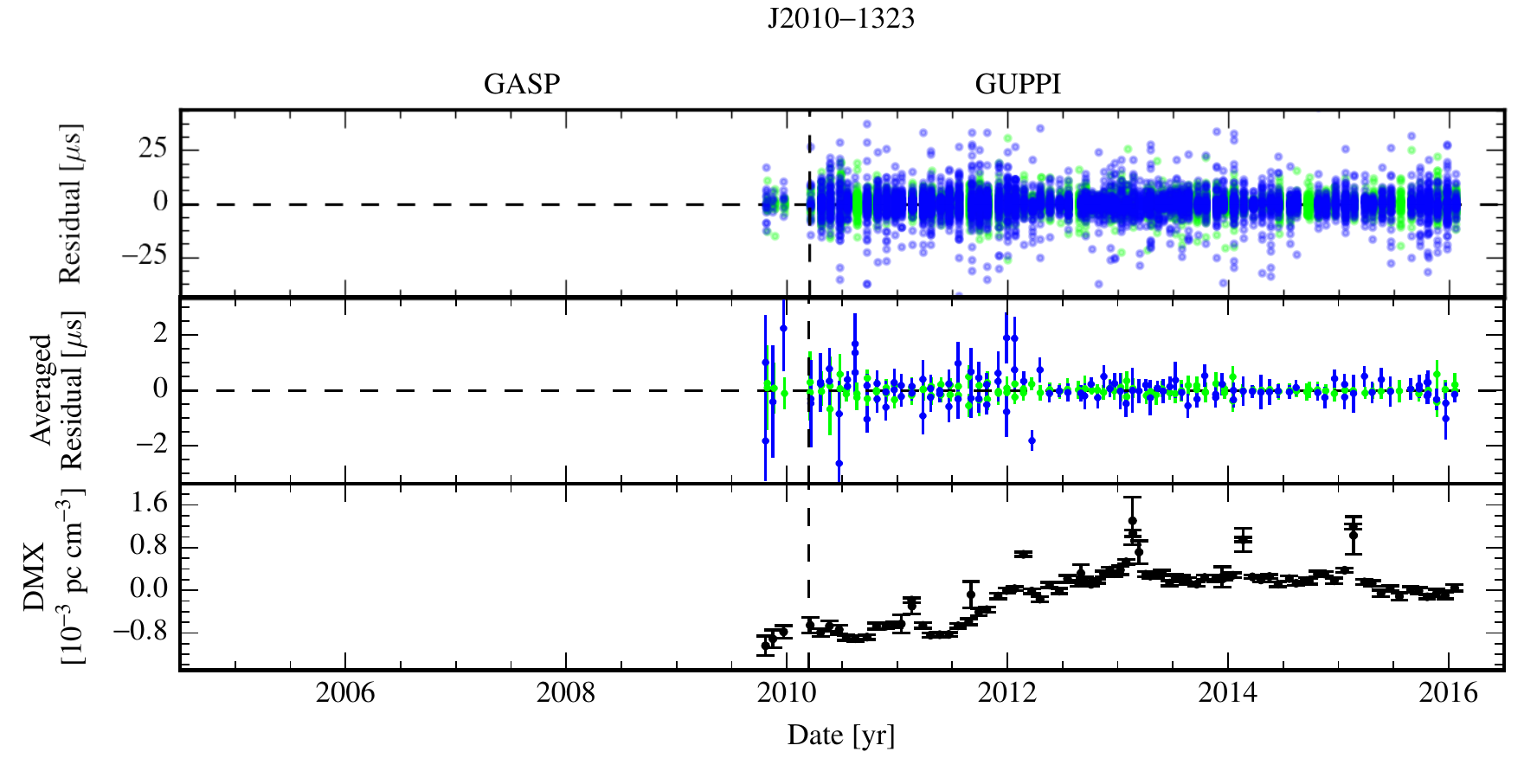}
\caption{Timing summary for PSR J2010-1323. Colors are: Blue: 1.4 GHz, Purple: 2.1 GHz, Green: 820 MHz, Orange: 430 MHz, Red: 327 MHz. In the top panel, individual points are semi-transparent; darker regions arise from the overlap of many points.}
\label{fig:summary-J2010-1323}
\end{figure*}

\begin{figure*}[p]
\centering
\includegraphics[scale=1.0]{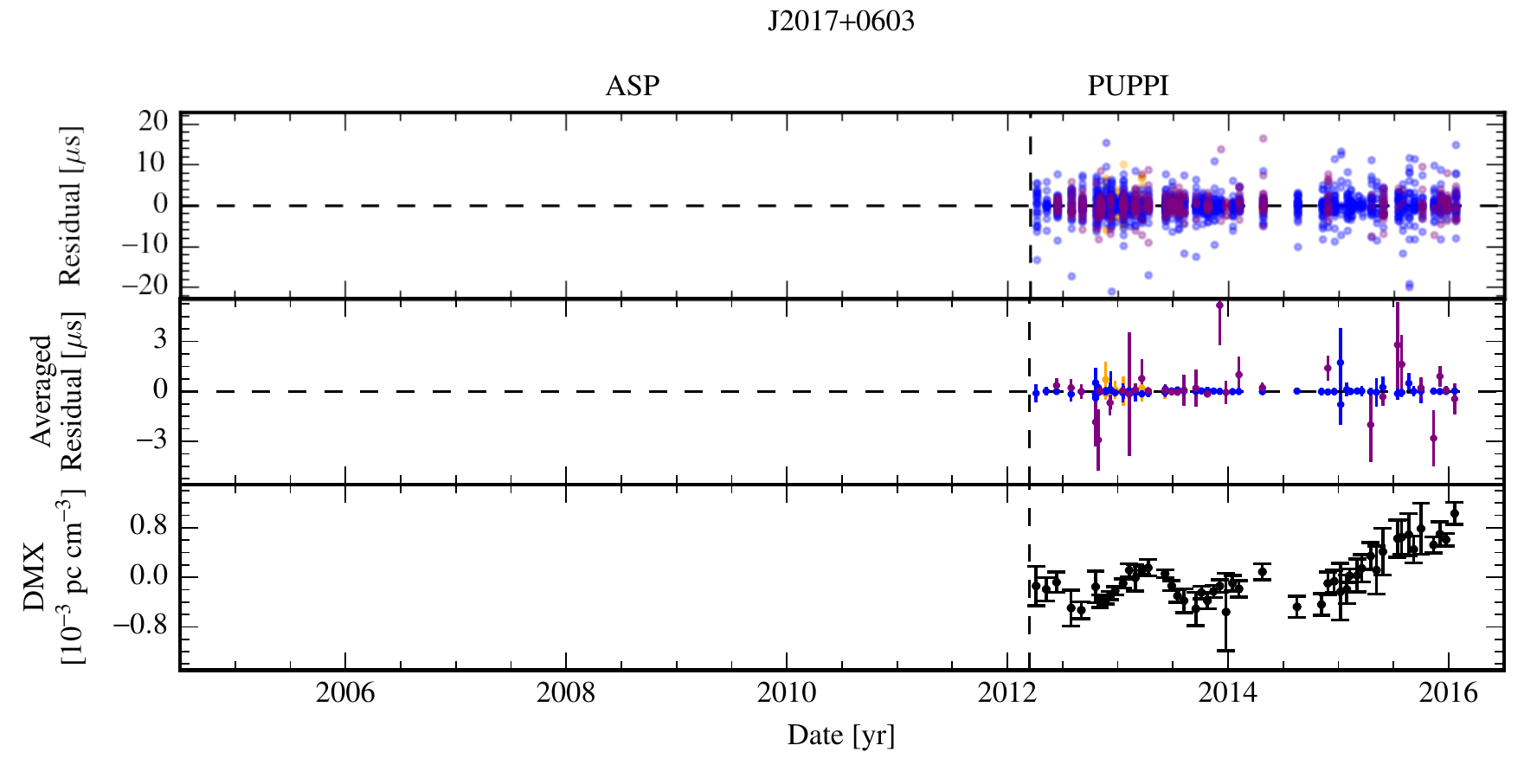}
\caption{Timing summary for PSR J2017+0603. Colors are: Blue: 1.4 GHz, Purple: 2.1 GHz, Green: 820 MHz, Orange: 430 MHz, Red: 327 MHz. In the top panel, individual points are semi-transparent; darker regions arise from the overlap of many points.}
\label{fig:summary-J2017+0603}
\end{figure*}

\begin{figure*}[p]
\centering
\includegraphics[scale=1.0]{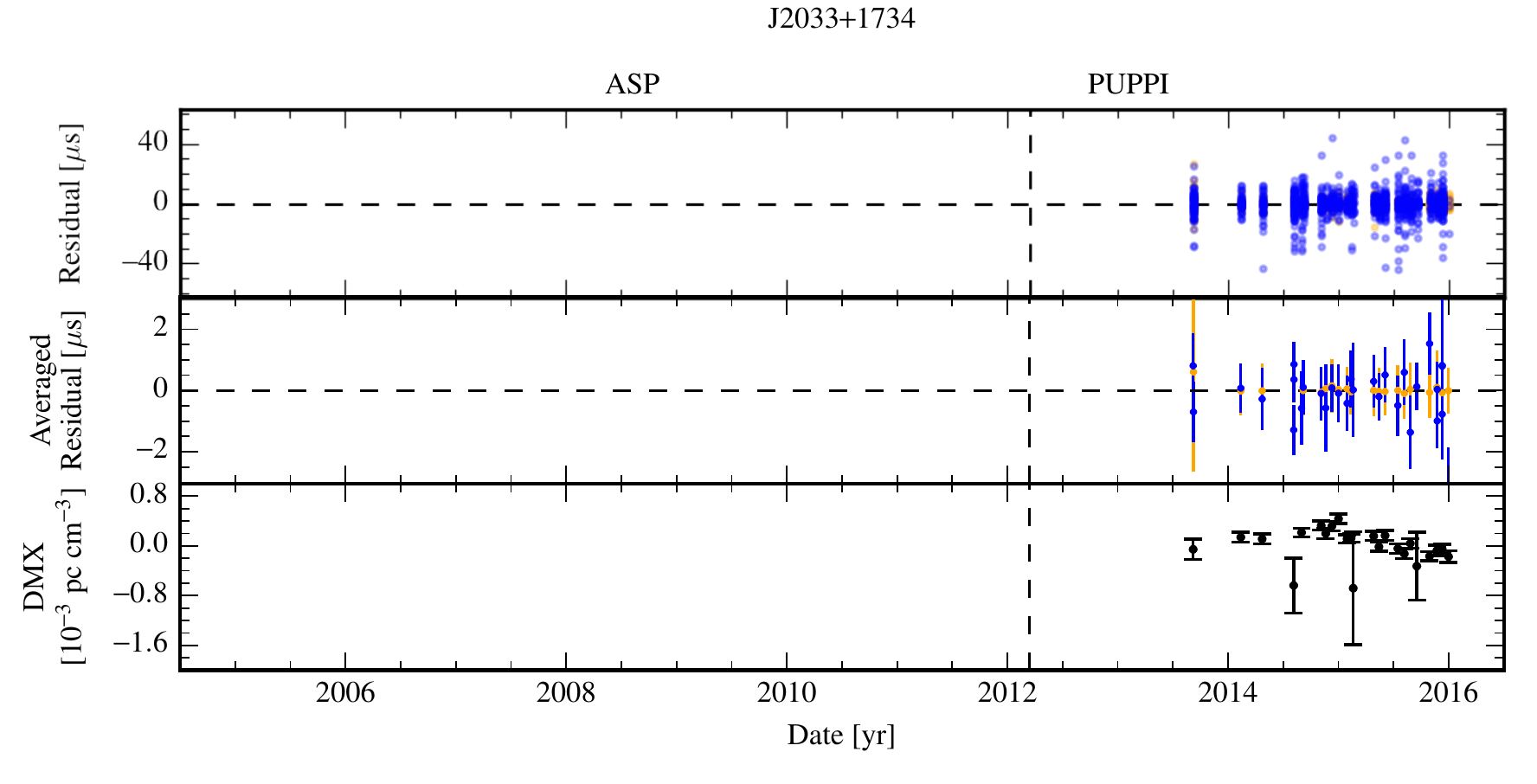}
\caption{Timing summary for PSR J2033+1734. Colors are: Blue: 1.4 GHz, Purple: 2.1 GHz, Green: 820 MHz, Orange: 430 MHz, Red: 327 MHz. In the top panel, individual points are semi-transparent; darker regions arise from the overlap of many points.}
\label{fig:summary-J2033+1734}
\end{figure*}

\begin{figure*}[p]
\centering
\includegraphics[scale=1.0]{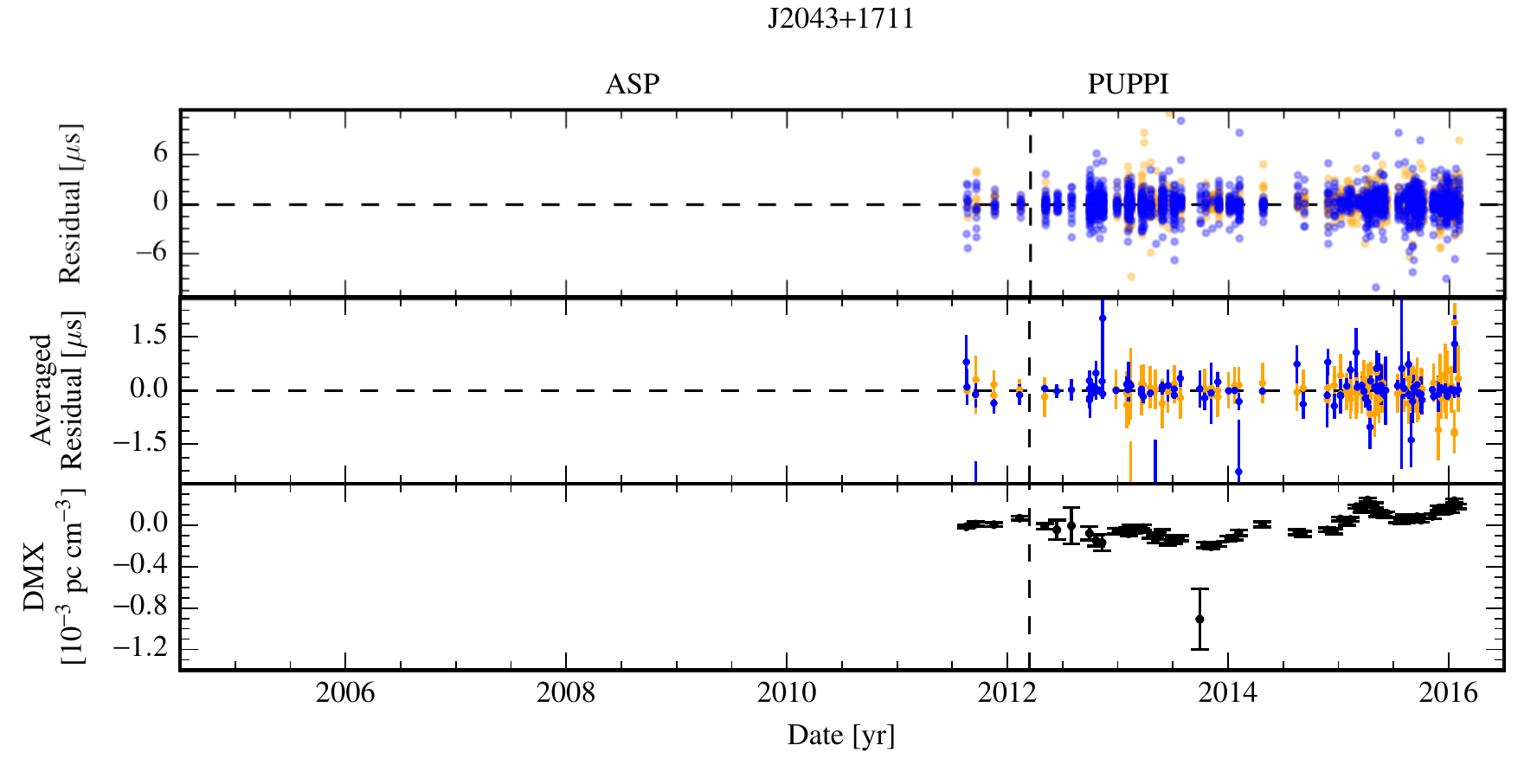}
\caption{Timing summary for PSR J2043+1711. Colors are: Blue: 1.4 GHz, Purple: 2.1 GHz, Green: 820 MHz, Orange: 430 MHz, Red: 327 MHz. In the top panel, individual points are semi-transparent; darker regions arise from the overlap of many points.}
\label{fig:summary-J2043+1711}
\end{figure*}

\begin{figure*}[p]
\centering
\includegraphics[scale=1.0]{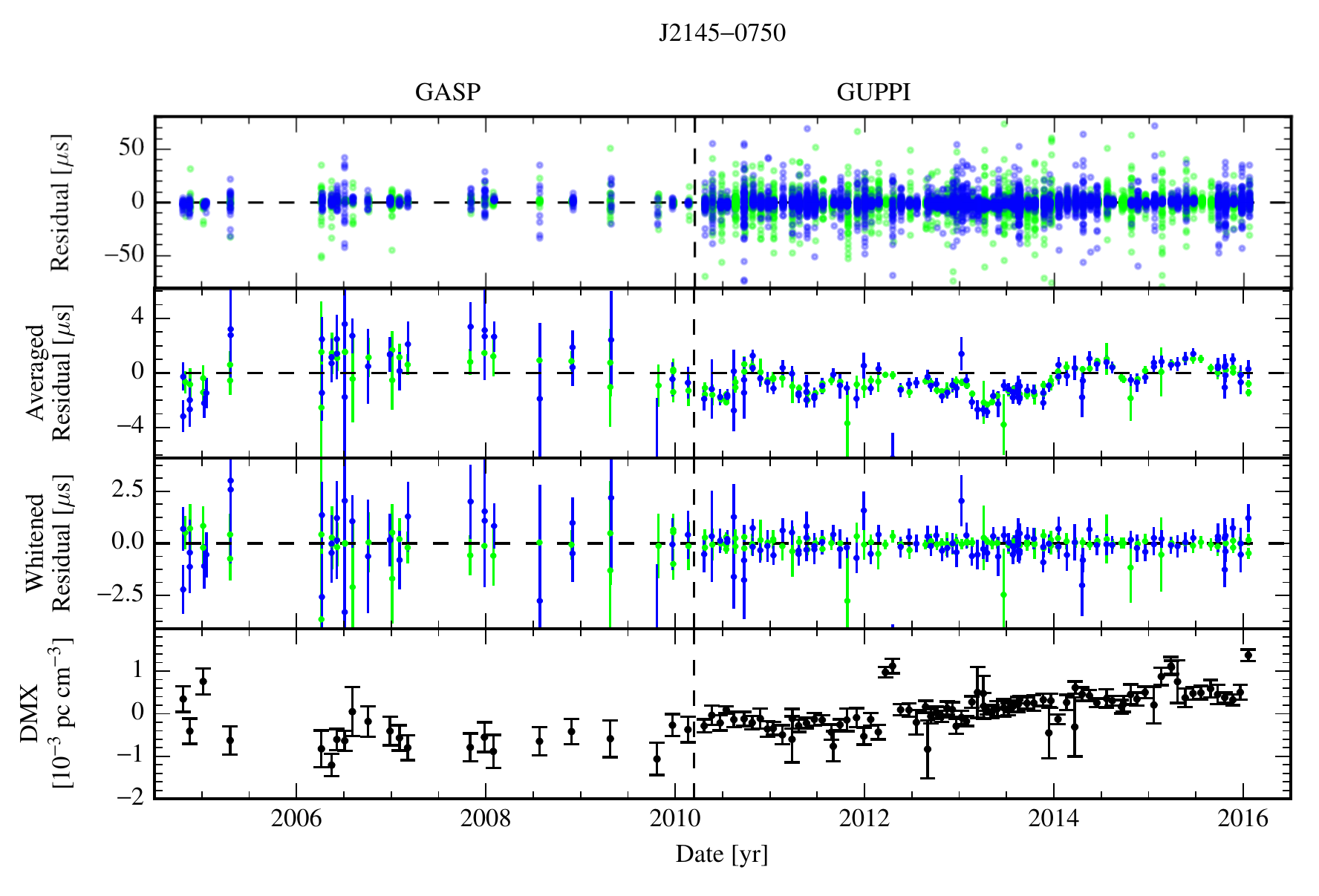}
\caption{Timing summary for PSR J2145-0750. Colors are: Blue: 1.4 GHz, Purple: 2.1 GHz, Green: 820 MHz, Orange: 430 MHz, Red: 327 MHz. In the top panel, individual points are semi-transparent; darker regions arise from the overlap of many points.}
\label{fig:summary-J2145-0750}
\end{figure*}

\begin{figure*}[p]
\centering
\includegraphics[scale=1.0]{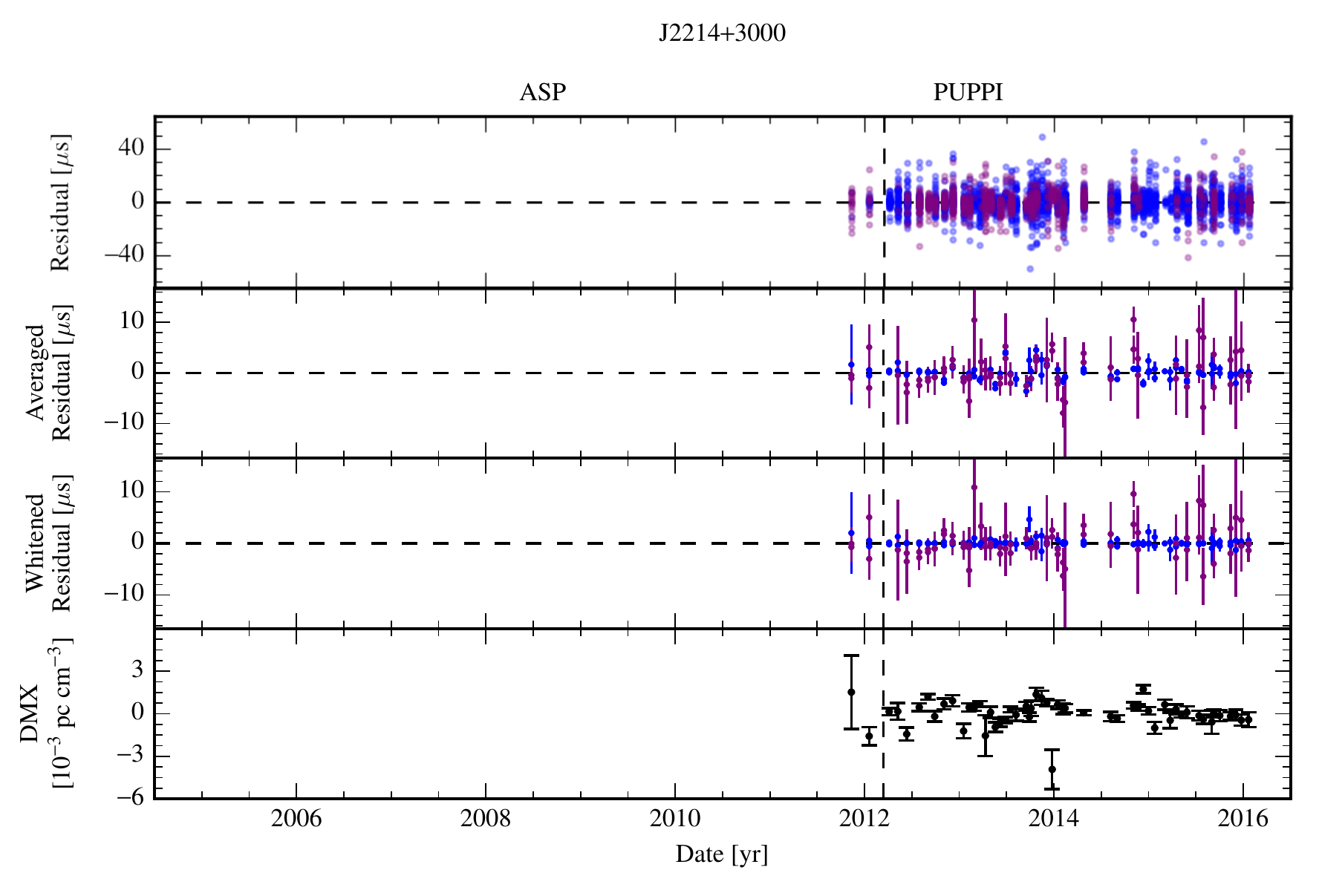}
\caption{Timing summary for PSR J2214+3000. Colors are: Blue: 1.4 GHz, Purple: 2.1 GHz, Green: 820 MHz, Orange: 430 MHz, Red: 327 MHz. In the top panel, individual points are semi-transparent; darker regions arise from the overlap of many points.}
\label{fig:summary-J2214+3000}
\end{figure*}

\begin{figure*}[p]
\centering
\includegraphics[scale=1.0]{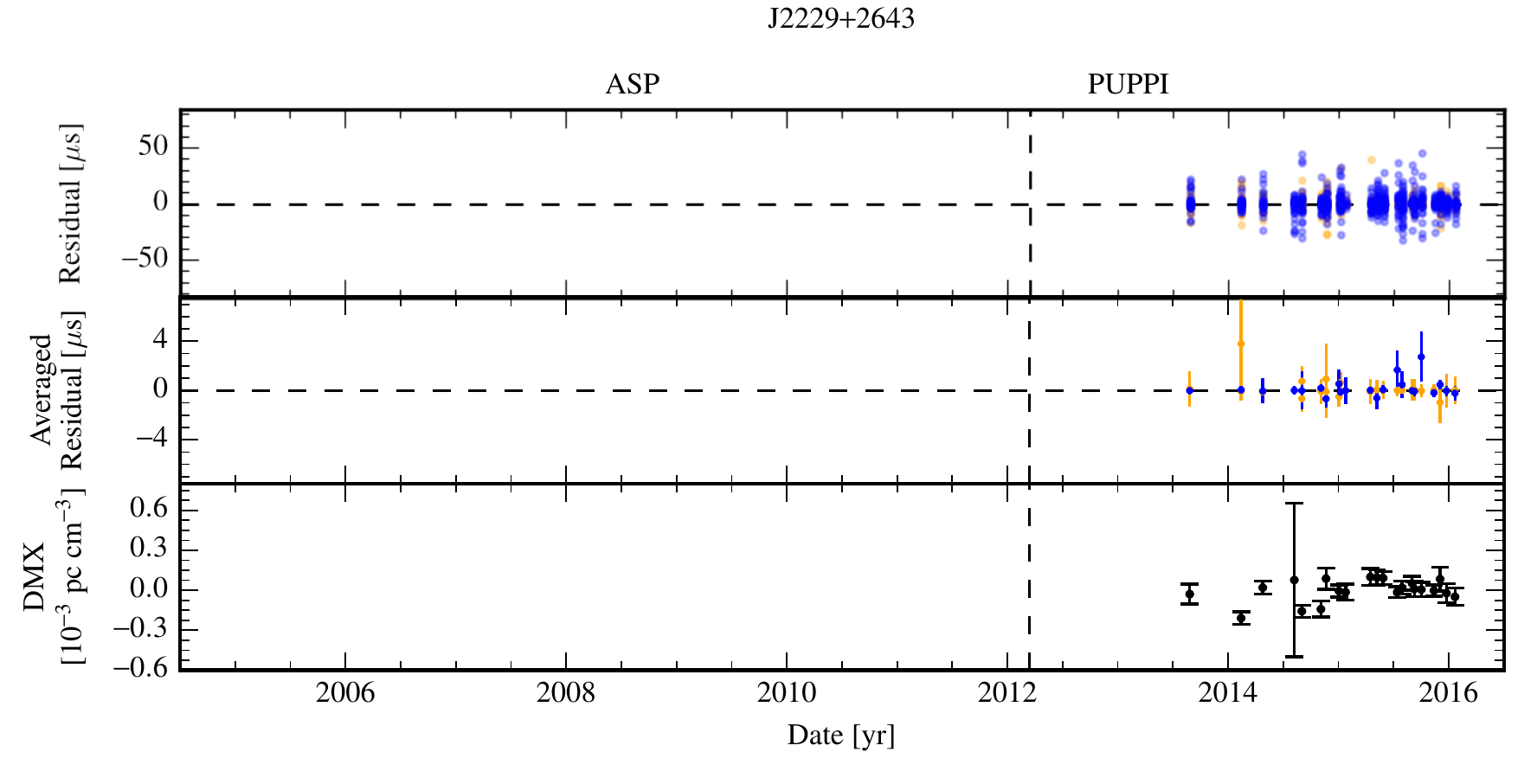}
\caption{Timing summary for PSR J2229+2643. Colors are: Blue: 1.4 GHz, Purple: 2.1 GHz, Green: 820 MHz, Orange: 430 MHz, Red: 327 MHz. In the top panel, individual points are semi-transparent; darker regions arise from the overlap of many points.}
\label{fig:summary-J2229+2643}
\end{figure*}

\begin{figure*}[p]
\centering
\includegraphics[scale=1.0]{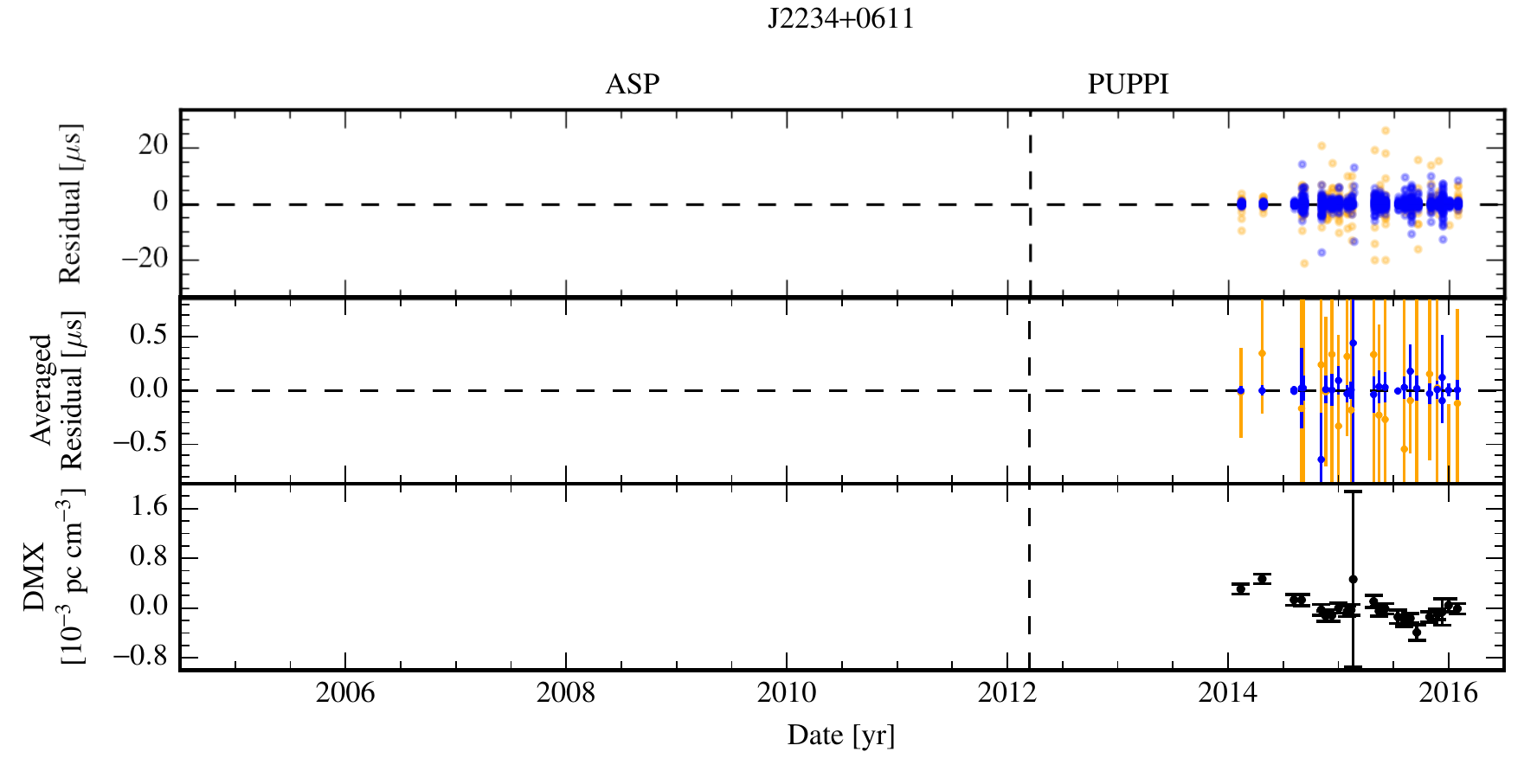}
\caption{Timing summary for PSR J2234+0611. Colors are: Blue: 1.4 GHz, Purple: 2.1 GHz, Green: 820 MHz, Orange: 430 MHz, Red: 327 MHz. In the top panel, individual points are semi-transparent; darker regions arise from the overlap of many points.}
\label{fig:summary-J2234+0611}
\end{figure*}

\begin{figure*}[p]
\centering
\includegraphics[scale=1.0]{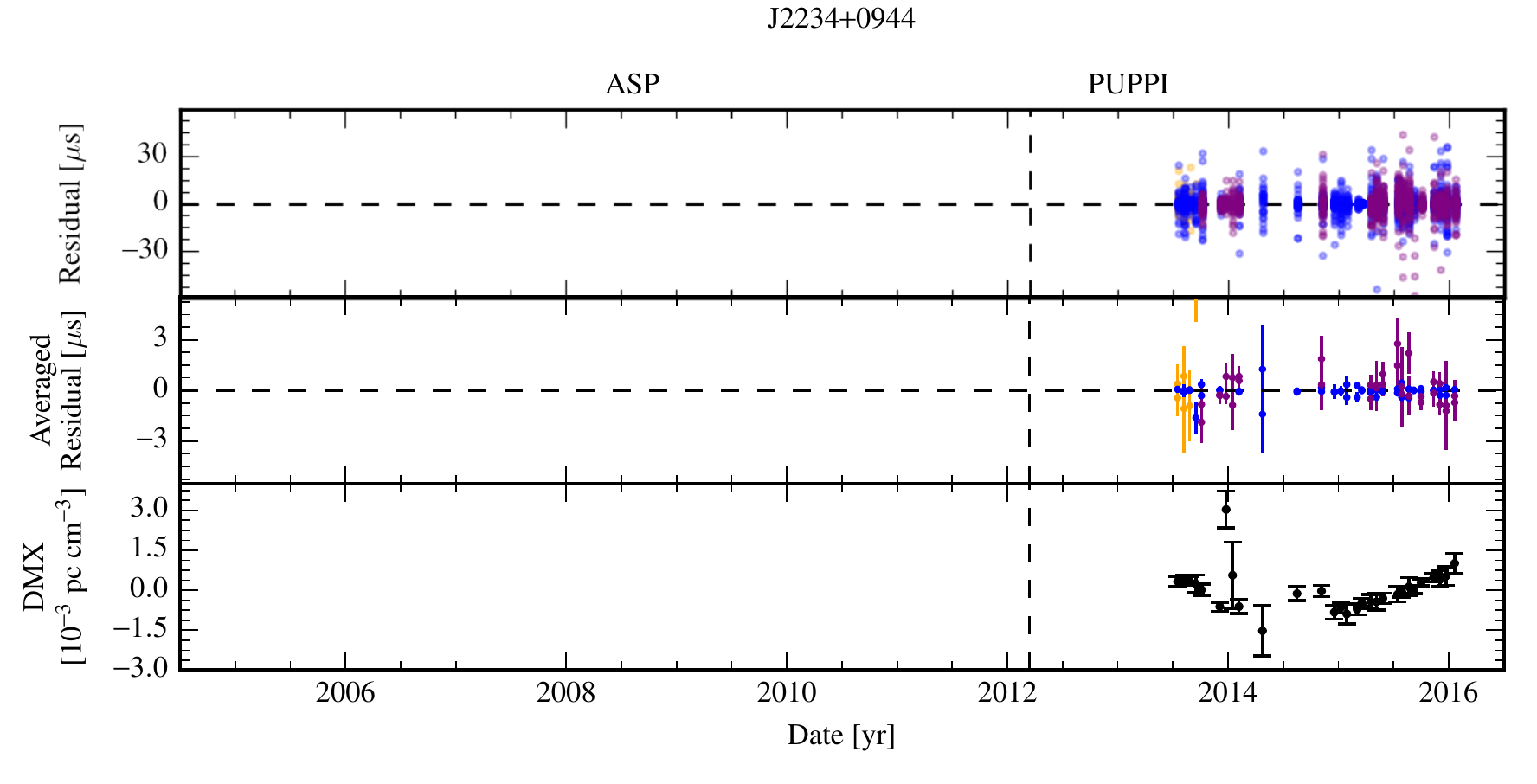}
\caption{Timing summary for PSR J2234+0944. Colors are: Blue: 1.4 GHz, Purple: 2.1 GHz, Green: 820 MHz, Orange: 430 MHz, Red: 327 MHz. In the top panel, individual points are semi-transparent; darker regions arise from the overlap of many points.}
\label{fig:summary-J2234+0944}
\end{figure*}

\begin{figure*}[p]
\centering
\includegraphics[scale=1.0]{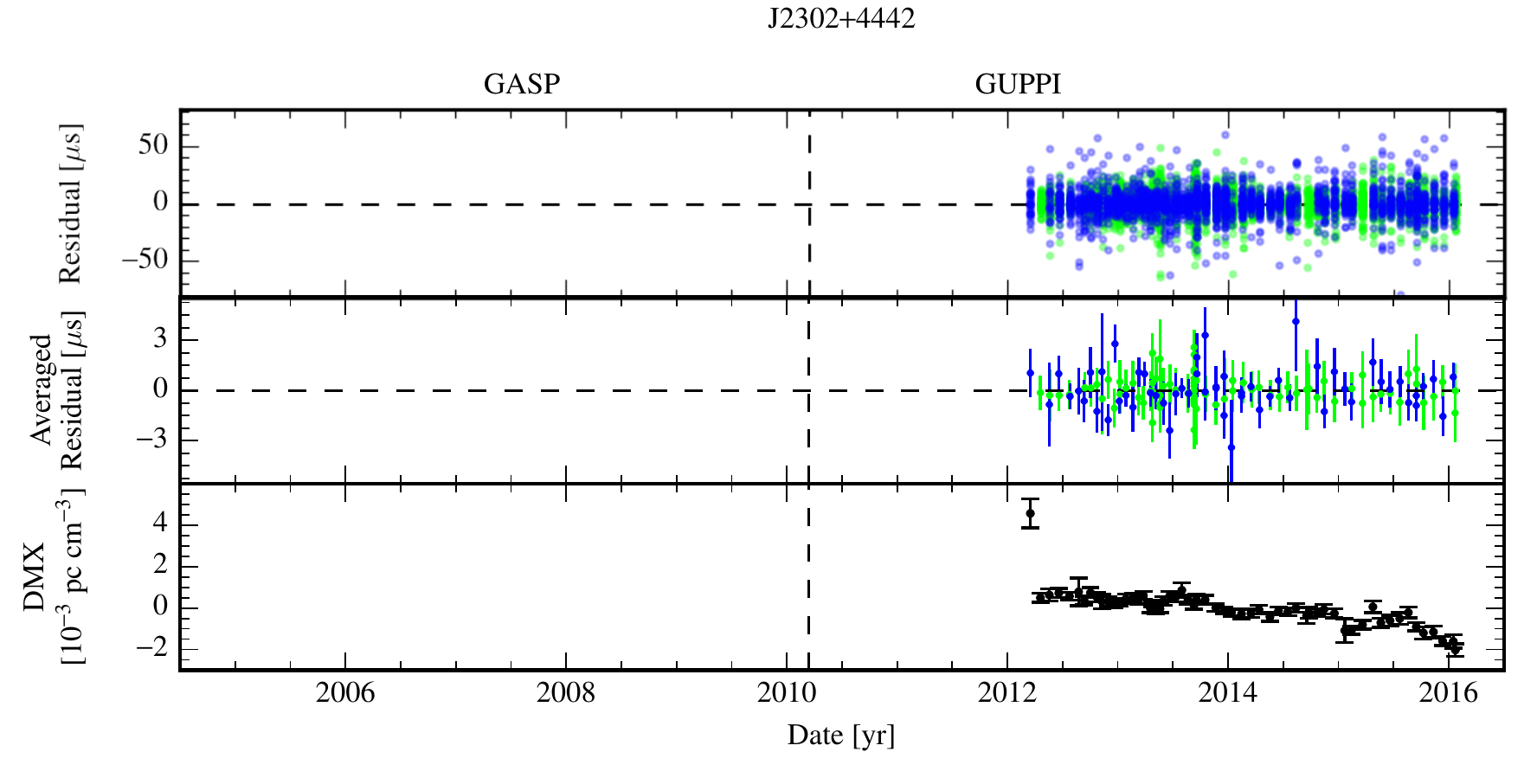}
\caption{Timing summary for PSR J2302+4442. Colors are: Blue: 1.4 GHz, Purple: 2.1 GHz, Green: 820 MHz, Orange: 430 MHz, Red: 327 MHz. In the top panel, individual points are semi-transparent; darker regions arise from the overlap of many points.}
\label{fig:summary-J2302+4442}
\end{figure*}

\begin{figure*}[p]
\centering
\includegraphics[scale=1.0]{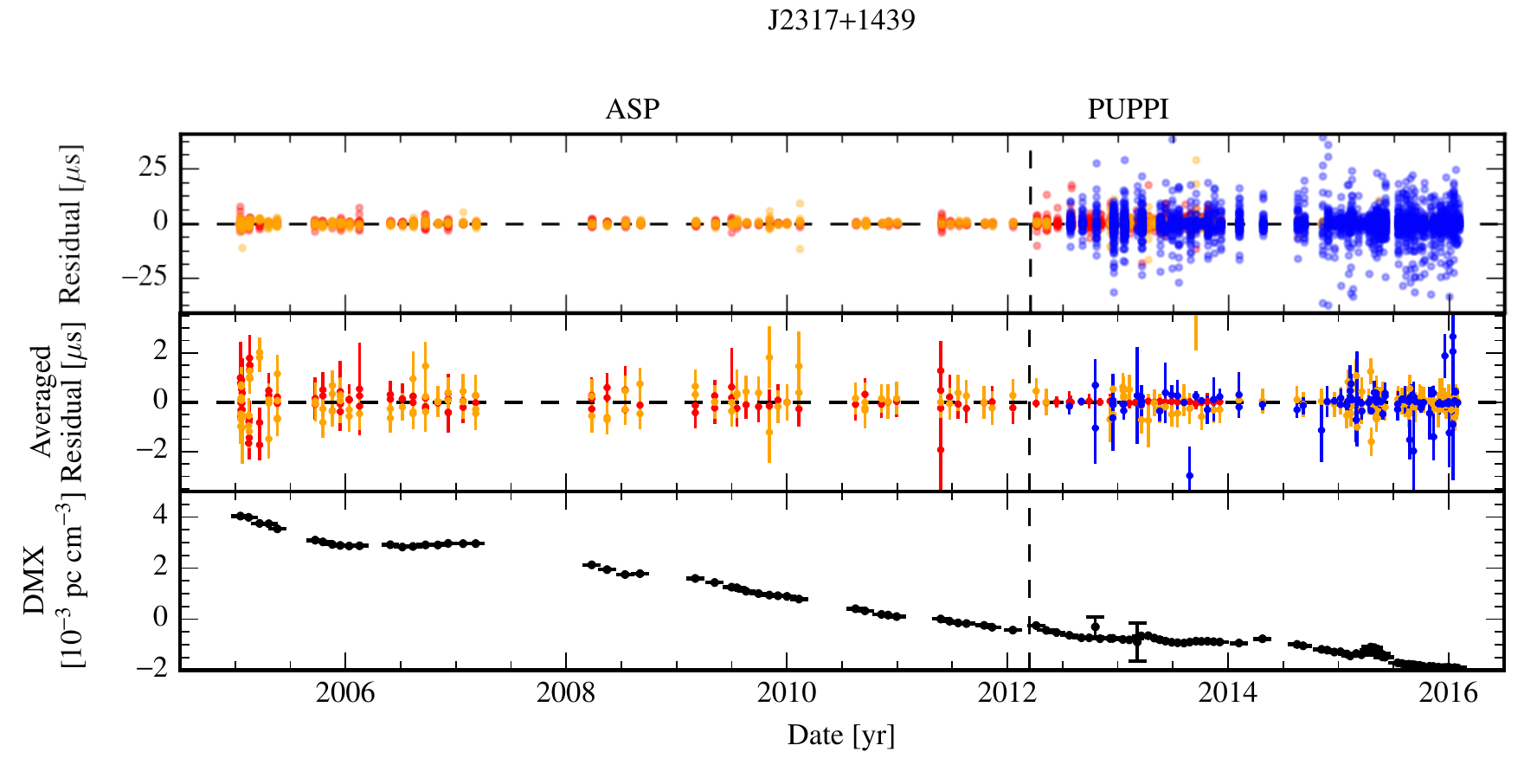}
\caption{Timing summary for PSR J2317+1439. Colors are: Blue: 1.4 GHz, Purple: 2.1 GHz, Green: 820 MHz, Orange: 430 MHz, Red: 327 MHz. In the top panel, individual points are semi-transparent; darker regions arise from the overlap of many points.}
\label{fig:summary-J2317+1439}
\end{figure*}

\end{document}